%% file: Arxiv-version.tex
\documentclass[nohyperref]{article} 

\usepackage{geometry}

\usepackage[round]{natbib}

\usepackage{hyperref}

\usepackage{algorithm2e}
\RestyleAlgo{ruled}

\usepackage{booktabs}       
\usepackage{amsfonts}       
\usepackage{nicefrac}       
\usepackage{microtype}      
\usepackage{enumerate}
\usepackage{caption}
\usepackage{subcaption}
\usepackage{bm}
\usepackage{textcomp}
\usepackage{graphicx}
\usepackage{epstopdf}
\usepackage{amssymb}
\usepackage{amsthm}
\usepackage{mathtools}
\usepackage{bbm,dsfont}
\usepackage{enumitem}
\allowdisplaybreaks
\hypersetup{final}
\input{Math-Commands}
\input{preamble}

\usepackage{sansmath}

\DeclareSymbolFont{extraup}{U}{zavm}{m}{n}
\DeclareMathSymbol{\varheart}{\mathord}{extraup}{86}
\DeclareMathSymbol{\vardiamond}{\mathalpha}{extraup}{87}

\graphicspath{{../figs/}}

\usepackage{mathrsfs}
\usepackage{psfrag}
\usepackage{tikz}
\usepackage{url}            

\usepackage{xcolor}
\usepackage{graphicx}
\usepackage{amsmath,amsthm,amssymb,amsfonts}

\makeatletter
\newtheorem*{rep@theorem}{\rep@title}
\newcommand{\newreptheorem}[2]{%
\newenvironment{rep#1}[1]{%
 \def\rep@title{#2 \ref{##1}}%
 \begin{rep@theorem}}%
 {\end{rep@theorem}}}
\makeatother

\theoremstyle{plain}
\newreptheorem{theorem}{Theorem}
\newreptheorem{proposition}{Proposition}

\theoremstyle{definition}
\theoremstyle{remark}
\setlist[itemize]{leftmargin=*}
\setlist[enumerate]{leftmargin=*}

\usepackage[capitalize]{cleveref}
\crefname{section}{Sec.}{Secs.}
\Crefname{section}{Section}{Sections}
\crefname{table}{Tab.}{Tabs.}
\Crefname{table}{Table}{Tables}
\crefname{algocf}{alg.}{algs.}
\Crefname{algocf}{Algorithm}{Algorithms}
\allowdisplaybreaks

\begin{document}
%

%


\title{How Does Pseudo-Labeling Affect the Generalization Error of the Semi-Supervised  Gibbs Algorithm?}


\author{Haiyun He$^\dagger$, $\quad$ Gholamali Aminian$^\ddagger$,$\quad$ Yuheng Bu$^\mathsection$, \\ Miguel Rodrigues$^\mathparagraph$, $\quad$ Vincent Y. F. Tan$^{\dagger\dagger}$\\ \\ 
\normalsize{$^\dagger$Cornell University, $^\ddagger$The Alan Turing Institute}\\
\normalsize{$^\mathsection$University of Florida, $^\mathparagraph$University College London, $^{\dagger\dagger}$National University of Singapore}\\ \\ 
\normalsize{Emails: \href{mailto:haiyun.he@u.nus.edu}{haiyun.he@u.nus.edu}, \href{mailto:gaminian@turing.ac.uk}{gaminian@turing.ac.uk},
\href{mailto:buyuheng@ufl.edu}{buyuheng@ufl.edu}},\\ 
\normalsize{\href{mailto:m.rodrigues@ucl.ac.uk}{m.rodrigues@ucl.ac.uk},
\href{mailto:vtan@nus.edu.sg}{vtan@nus.edu.sg} }}

\date{}
\maketitle

\begin{abstract}
We provide an exact characterization of the expected generalization error (gen-error) for semi-supervised learning (SSL) with pseudo-labeling via the Gibbs algorithm. The gen-error is expressed in terms of the symmetrized KL information between the output hypothesis, the pseudo-labeled dataset, and the labeled dataset. Distribution-free upper and lower bounds on the gen-error can also be obtained. Our findings offer new insights that the generalization performance of SSL with pseudo-labeling is affected not only by the information between the output hypothesis and input  training data but also by the information {\em shared} between the {\em labeled} and {\em pseudo-labeled} data samples. This serves as a guideline to choose an appropriate pseudo-labeling method from a given family of methods. To deepen our understanding, we further explore two examples---mean estimation and logistic regression. In particular, we analyze how the ratio of the number of unlabeled to labeled  data $\lambda$ affects the gen-error under both scenarios. As $\lambda$ increases, the gen-error for mean estimation decreases and then saturates at a value larger than when all the samples are labeled, and the gap can be quantified {\em exactly} with our analysis, and is dependent on the \emph{cross-covariance} between the labeled and pseudo-labeled data samples. For logistic regression, the gen-error and the variance component of the excess risk also decrease as $\lambda$ increases.

\end{abstract}

\section{Introduction}

There are several areas, like natural language processing, computer vision, and finance, where labeled data are rare but unlabeled data are abundant. In these situations, semi-supervised learning (SSL) techniques enable us to utilize both labeled and unlabeled data.
 
Self-training algorithms~\citep{ouali2020overview} are a subcategory of SSL techniques. These algorithms use the supervised-learned model's confident predictions to predict the labels of unlabeled data. Entropy minimization and pseudo-labeling are two basic approaches used in self-training-based SSL. The entropy function may be viewed as a regularization term that penalizes uncertainty in the label prediction of unlabeled data in entropy minimization approaches~\citep{grandvalet2005semi}. The \emph{manifold assumption} \citep{iscen2019label}---where it is assumed that labeled and unlabeled features are drawn from a common data manifold---or the \emph{cluster assumption} \citep{chapelle2003cluster}---where it is assumed that similar data features have a similar label---are assumptions for adopting the entropy minimization algorithm.
In contrast, in pseudo-labeling, which is the focus of this work, the model is trained using labeled data and then used to produce a pseudo-label for the unlabeled data~\citep{lee2013pseudo}. These pseudo-labels are then utilized to build another model, which is trained using both labeled and pseudo-labeled data in a supervised manner. Studying the generalization error (gen-error) of this procedure is critical to understanding and improving pseudo-labeling performance.

There have been various efforts to characterize the gen-error of SSL algorithms. In \citet{rigollet2007generalization}, an upper bound on the gen-error of binary classification under the cluster assumption is derived. \citet{niu2013squared} provides an upper bound on gen-error based on the Rademacher complexity for binary classification with squared-loss mutual information regularization.  \citet{gopfert2019can} employs the VC-dimension method to characterize the SSL gen-error.  In \citet{gopfert2019can} and \citet{zhu2020semi}, upper bounds for SSL gen-error using Bayes classifiers are provided. 
\citet{zhu2020semi} also provides an upper bound on the excess risk of SSL algorithm by assuming an exponentially concave function\footnote{A function $f(x)$ is called $\beta$-exponentially concave function if $\exp(-\beta f(x))$ is concave} based on the conditional mutual information. \citet{he2022information} investigates the gen-error of iterative SSL techniques based on pseudo-labels.   An information-theoretic gen-error upper bound on self-training algorithms under the covariate-shift assumption is proposed by \cite{aminian2022information}. These upper bounds on excess risk and gen-error do not entirely  capture the impact of SSL, in particular pseudo-labeling and the relative number of labeled and unlabeled data, and thus constrain our ability to fully comprehend the performance of SSL.

In this paper, we are interested in characterizing the expected gen-error of pseudo-labeling-based SSL---using an appropriately-designed Gibbs algorithm---and studying how it depends on the output hypothesis, the labeled, and the pseudo-labeled data. Moreover, we intend to understand the effect of the ratio between the numbers of the unlabeled and labeled training data examples on the gen-error in different scenarios. 

Our main contributions in this paper are as follows:
\begin{itemize}
    \item We provide an exact characterization of the expected gen-error of Gibbs algorithm that models pseudo-labeling-based SSL. This characterization can be applied to obtain novel and informative upper and lower bounds. 
    
    \item The characterization and bounds offer an insight that reducing the shared information between the labeled and pseudo-labeled samples can help to improve the generalization performance of pseudo-labeling-based SSL.
    
    \item We analyze the effect of the ratio  of the number of unlabeled data to labeled data $\lambda$ on the gen-error of Gibbs algorithm using a mean estimation example.
    
    \item Finally, we study the asymptotic behaviour of the Gibbs algorithm and analyze the effect of $\lambda$ on the  gen-error and the excess risk, applying our results  to logistic regression.
\end{itemize}

\section{Related Works}\label{app: related works}
\textbf{Semi-supervised learning:}
The SSL approaches can be partitioned into six main classes: generative models, low-density separation methods, graph-based methods, self-training and co-training \citep{zhu2008semi}. Among all these, SSL first appeared as self-training in which the model is first trained on the labeled data and annotates the unlabeled data to improve the intial model \citep{books/mit/06/CSZ2006}. SSL has gradually gained more attention after the well-known expectation-minimization (EM) algorithm \citet{moon1996expectation} was proposed in 1996. One key problem of interest in SSL is whether the unlabeled data can help to improve the performance of learning algorithms.  Many existing works have studied this problem either theoretically (e.g., providing bounds) or empirically (e.g., proposing new algorithms).
One classical work by \citet{castelli1996relative} set out to study SSL under a traditional setup with unknown mixture of known distributions and characterized the error probability by the fisher information of the labeled and unlabeled data.
\citet{szummer2002information} proposed an algorithm that utilizes the unlabeled data to learn the marginal data distribution to augment learning the class conditional distribution.
\citet{amini2002semi} empirically showed that semi-supervised logistic regression based on EM algorithm has higher accuracy than the naive Bayes classifier.
\citet{singh2008unlabeled} studied the benefit of unlabeled data on the excess risk based on the number of unlabeled data and the margin between classes. 
\citet{ji2012simple} developed a simple algorithm based on top eigenfunctions of integral operator derived from both labeled and unlabeled examples that can improve the regression error bound.
\citet{li2019multi} showed how the unlabeled data can improve the Rademacher-complexity-based generalization error bound of a multi-class classification problem.  In deep learning, \citet{berthelot2019mixmatch} introduced an effective algorithm that generates low-entropy labels for unlabeled data and then mixes them up with the labeled  data to train the model. \citet{sohn2020fixmatch} showed that augmenting the confidently pseudo-labeled images can help to improve the accuracy of their model.
For a more comprehensive overview of SSL, one can refer to the report by \citet{seeger2000learning} and the book by \citet{books/mit/06/CSZ2006}. 

\textbf{Pseudo-labeling:}
Pseudo-labeling is one of the approaches in self-training \citep{zhu2009introduction}.
Due to the reliance of pseudo-labeling on the quality of the pseudo labels, pseudo-labeling approach might perform poorly. \citet{seeger2000learning} stated that there exists a tradeoff between robustness of the learning algorithm and the information gain from the pseudo-labels.
\citet{rizve2020defense} offered an uncertainty-aware pseudo-labeling strategy to circumvent this difficulty. In \citet{wei2020theoretical}, a theoretical framework for combining input-consistency regularization with self-training algorithms in deep neural networks is provided. \citet{dupre2019improving} empirically showed that iterative pseudo-labeling with a confidence threshold can improve the test accuracy in early stage. \citet{arazo2020pseudo} showed that the method of generating soft labels for unlabeled data plus mixup augmentation can outperform consistency regularization methods.

Despite the plenty of works on SSL and pseudo-labeling, our work provides a new viewpoint of understanding the effect of pseudo-labeling method on the generalization error using information-theoretic quantities.

\textbf{Information-theoretic upper bounds:} 
\citet{russo2019much,xu2017information} proposed to use the mutual information between the input training set and the output hypothesis to upper bound the expected generalization error.  This paves a new way of understanding and improving the generalization performance of a learning algorithm from an information-theoretic viewpoint. Tighter upper bounds by considering the individual sample mutual information is proposed by \citet{bu2020tightening}.  \citet{asadi2018chaining} proposed using chaining mutual information, and \citet{hafez2020conditioning,haghifam2020sharpened,steinke2020reasoning} advocated the conditioning and processing techniques. 
Information-theoretic generalization error bounds using other information quantities are also studied, e.g., $f$-divergences, $\alpha$-R\'enyi divergence and generalized Jensen-Shannon divergence~\citet{esposito2019generalization,aminian2022learning,aminian2021information}. 

\section{Semi-Supervised Learning Via The Gibbs Algorithm }

In this section, we formulate our problem using the Gibbs algorithm based on both the labeled and unlabeled training data with pseudo-labels. The Gibbs algorithm is a tractable and idealized model for learning algorithms with various approaches, e.g., stochastic optimization methods or relative entropy regularization \citep{raginsky2017non}.

\subsection{Problem Formulation}\label{Sec:prob setup}

Let $S_{\rml}=\{\bZ_i\}_{i=1}^n=\{(\bX_i,Y_i)\}_{i=1}^n$ be the labeled training dataset, where $\bX_i\in\calX=\bbR^d$ is the data feature, $Y_i\in\calY=[K]$ is the class label and each pair of $\bZ_i=(\bX_i,Y_i) \in \calX \times \calY=\calZ$ is independently and identically distributed (i.i.d.) from $P_{\bZ}=P_{\bX,Y}\in\calP(\calZ)$.  Conditioned on $\bX_i$, each label $Y_i$ is i.i.d.\ from $P_{Y|\bX}$. 
Let $S_{\rmu}=\{\bX_i\}_{i=n+1}^{n+m}$ be the unlabeled training dataset. For all $i\in[n+m]$, each $\bX_i$ is i.i.d.\ from $P_{\bX}\in\calP(\calX)$. 
Based on the labeled dataset $S_\rml$,  a  pseudo-labeling method assigns a pseudo-label $\hatY_i$ to each $\bX_i \in S_\rmu$ and $\hatY_i$ is drawn conditionally i.i.d.\ from $P_{\hatY|\bX,S_\rml}$. Let the pseudo-labeled data point be $\hat{\bZ}_i=(\bX_i,\hatY_i)$ and the pseudo-labeled dataset be $\hatS_\rmu=\{\hat{\bZ}_i\}_{i=n+1}^{n+m}$. For any labeled and pseudo-label datasets, we use $P_{S_\rml}$, $P_{\hatS_\rmu}$ and $P_{S_\rml,\hatS_\rmu}$ to denote the joint distributions of the data samples in $S_\rml$, $\hatS_\rmu$, and $S_\rml \cup \hatS_\rmu$, respectively. Note that $P_{S_\rml}=P_{\bZ}^{\otimes n}$.

Let $\bw\in\calW$ denote the output hypothesis. 
In semi-supervised learning, one considers the empirical risk based on both the labeled and unlabeled data. In this paper,  by fixing a mixing weight $\eta\in\bbR_+$, for any loss function $l:\calW \times \calZ \to \bbR_+$, the \emph{total empirical risk} is the $\eta$-weighted sum of the empirical risks of the labeled and pseudo-labeled data \citep{mclachlan2005discriminant,books/mit/06/CSZ2006}
\begin{align}
	L_{\rmE}(\bw,S_\rml,\hatS_\rmu):=L_{\rmE}(\bw,S_\rml)+\eta L_{\rmE}(\bw,\hatS_\rmu),
\end{align}
where $L_{\rmE}(\bw,S_\rml):=\frac{1}{n}\sum_{i=1}^n l(\bw,\bZ_i)$ and $L_{\rmE}(\bw,\hatS_\rmu):=\frac{1}{m}\sum_{i=n+1}^{n+m} l(\bw,\hat{\bZ}_i)$.
Note that, by normalizing the weight $\eta$, minimizing the empirical risk $L_{\rmE}(\bw,S_\rml,\hatS_\rmu)$  is equivalent to minimizing 
\begin{align}
	\!\!\barL_{\rmE}(\bw,S_\rml,\hatS_\rmu)\!=\!\frac{1}{1+\eta}L_{\rmE}(\bw,S_\rml)\!+\!\frac{\eta}{1+\eta}L_{\rmE}(\bw,\hatS_\rmu). \label{Eq:normalize emp risk}
\end{align}
The \emph{population risk} under the true data distribution is
\begin{align}
	L_{\rm P}(\bw,P_{S_\rml}):=\bbE_{S_\rml\sim P_{S_\rml}}\big[L_{\rmE}(\bw,S_\rml)\big]. \label{Eq:population risk}
\end{align}
Under the i.i.d.\ assumption, such a definition reduces to $\bbE_{S_\rml\sim P_{S_\rml}}\big[L_{\rmE}(\bw,S_\rml)\big]=\bbE_{\bZ\sim P_{\bZ}}\big[l(\bw,\bZ)\big] $.

Any SSL algorithm can be characterized by a conditional distribution 
$P_{\bW|S_\rml,\hatS_\rmu}$, which is a stochastic map from the labeled dataset $S_\rml$, the pseudo-labeled dataset $\hatS_{\rmu}$ to the output hypothesis $\bW$. For any training datasets $S_\rml,\hatS_\rmu$ and any SSL algorithm $P_{\bW|S_\rml,\hatS_\rmu}$, the \emph{expected gen-error} is defined as the expected gap between the population risk and the empirical risk of the \emph{labeled data} $S_\rml$, i.e.,
\begin{align}
	&\overline{\mathrm{gen}}(P_{\bW|S_\rml,\hatS_\rmu}, P_{S_\rml,\hatS_\rmu})\!\!:=\!\bbE_{\bW,S_\rml}\![ L_{\rm P}(\bW,P_{S_\rml}) \!\!-\!\! L_{\rmE}(\bW,S_\rml)], \label{Eq:def gen}
\end{align}
which measures the extent to which the algorithm overfits to the labeled training data.

In particular, we consider the Gibbs algorithm (also known as the  Gibbs posterior \citep{catoni2007pac}) to model a pseudo-labeling-based SSL algorithm.
Given any $(S_\rml,\hatS_\rmu)$, \emph{the $(\alpha,\pi(\bw),\barL_{\rmE}(\bw,S_\rml,\hatS_\rmu))$-Gibbs algorithm} \citep{gibbs1902elementary,jaynes1957information} is
\begin{align}
	&P_{\bW|S_\rml,\hatS_\rmu}^\alpha(\bw|S_\rml,\hatS_\rmu)
	=\frac{\pi(\bw)\exp\big(-\alpha\barL_{\rmE}(\bw,S_\rml,\hatS_\rmu)\big)}{\Lambda_{\alpha,\eta}(S_\rml,\hatS_\rmu)},\nn
\end{align} 
where $\alpha\geq 0$ is the ``inverse temperature'', $\Lambda_{\alpha,\eta}(S_\rml,\hatS_\rmu)\!=\!\int \pi(\bw) \exp(-\alpha\barL_{\rmE}(\bw,S_\rml,\hatS_\rmu))\, \rmd\bw$ is the partition function and $\pi(\bw)$ is the prior of $\bw$. We provide more motivations for the Gibbs algorithm model in Appendix~\ref{App: motivation of Gibbs}.

Our goal---relying on the characterization of $P_{\bW|S_\rml,\hatS_\rmu}^\alpha$---is to precisely quantify the gen-error in \eqref{Eq:def gen} as a function of various information-theoretic quantities.

If $P$ is absolutely continuous with respect to $Q$ and vice versa, let the \emph{symmetrized KL-divergence} (also known as \emph{Jeffrey's divergence} \citep{jeffreys1946invariant}) be defined as $D_{\mathrm{SKL}}(P\|Q):=D(P\|Q)+D(Q\|P)$, where $D$ is the Kullback--Leibler (KL) divergence \citep{cover1999elements}.

For  random variables $X$ and $Y$ with joint distribution $P_{X,Y}$, the mutual information is $I(X;Y)=D(P_{X,Y}\| P_X\otimes P_Y )$ and the {\em Lautum information} \citep{palomar2008lautum} is $L(X;Y)=D(P_X\otimes P_Y \| P_{X,Y})$.
Similarly, the \emph{symmetrized KL information} between $X$ and $Y$ \citep{aminian2015capacity} is defined as $I_{\mathrm{SKL}}(X;Y):=D_{\mathrm{SKL}}(P_{X,Y}\|P_X \otimes P_Y)=I(X;Y)+L(X;Y)$. We define the \emph{conditional symmetrized KL information} as $I_{\mathrm{SKL}}(X;Y|Z):=I(X;Y|Z)+L(X;Y|Z)$.

\subsection{Main Results}\label{Sec:main result}
One of our main results offers an exact closed-form  information-theoretic 
expression for the gen-error of the $(\alpha,\pi(\bw), \barL_{\rmE}(\bw,S_\rml,\hatS_\rmu))$-Gibbs algorithm in terms of the symmetrized KL information defined above.
\begin{theorem}\label{Thm:gen SSL PL}
	Under the $(\alpha,\pi(\bw), \barL_{\rmE}(\bw,S_\rml,\hatS_\rmu))$-Gibbs algorithm, the expected gen-error is
	\begin{align}
		&\overline{\mathrm{gen}}(P_{\bW|S_\rml,\hatS_\rmu}^\alpha, P_{S_\rml,\hatS_\rmu})\!=\!\frac{1+\eta}{\alpha}\big( \bbE_{\Delta_{S_\rml,\hatS_\rmu}}[\log\Lambda_{\alpha,\eta}(S_\rml,\hatS_\rmu)] \nn\\
		&\quad +I_{\mathrm{SKL}}(\bW,\hatS_\rmu;S_\rml)-I_{\mathrm{SKL}}(\hatS_\rmu;S_\rml) \big), \label{Eq: gen main thm}
	\end{align}
	where $\Lambda_{\alpha,\eta}(S_\rml,\hatS_\rmu):=\bbE_{\bW\sim\pi}[\exp(-\alpha\barL_{\rmE}(\bW,S_\rml,\hatS_\rmu))]$ and $\bbE_{\Delta_{S_\rml,\hatS_\rmu}}[\cdot]:=\bbE_{S_\rml,\hatS_\rmu}[\cdot]-\bbE_{S_\rml}\bbE_{\hatS_\rmu}[\cdot]$.
\end{theorem}
The proof of Theorem \ref{Thm:gen SSL PL} is provided in Appendix \ref{App:proof of Thm:gen SSL PL}, where we also show that by letting $\eta\to 0$, the result reduces to that for supervised learning (SL). 
In addition, we can even extend  \cref{Thm:gen SSL PL} to SSL based on other methods, e.g., entropy minimization \citep{amini2002semi,grandvalet2005semi}. This corollary is provided in Section \ref{Sec: SSL entropy minimization}.

Theorem \ref{Thm:gen SSL PL} can also be applied to  derive novel bounds on the expected gen-error of the Gibbs algorithm as follows.

    \begin{proposition}\label{Prop: upper and lower bound}
        Assume that  the loss function $l(\bw,\bZ)$ is bounded in $[a,b]\subset \bbR_+$. Then, the expected gen-error of the $(\alpha,\pi(\bw), \barL_{\rmE}(\bw,S_\rml,\hatS_\rmu))$-Gibbs algorithm satisfies
        \begin{align}\label{Eq: gen UL bd}
        \!\!\left| \overline{\mathrm{gen}}(P_{\bW|S_\rml,\hatS_\rmu}^\alpha, P_{S_\rml,\hatS_\rmu}) \!-\! \text{SKL}\right|\!\le\! c(\eta,a,b)\sqrt{\!I(\hatS_\rmu;S_\rml)} ,
\end{align}
where  $c(\eta,a,b):= \frac{1}{\sqrt{2}}(1+\eta)(b-a)$ and   $\text{SKL}:=\frac{1+\eta}{\alpha}(I_{\mathrm{SKL}}(\bW,\hatS_\rmu;S_\rml)-I_{\mathrm{SKL}}(\hatS_\rmu;S_\rml))$.
    \end{proposition}

The proof and more discussions are provided in~\cref{App pf of Prop: upper and lower bound}.
In fact, we can further explore how the gen-error depends on the information that the output hypothesis contains about the labeled and unlabeled datasets by writing (see Appendix~\ref{App:rewrite I_SKL-I_SKL})
    \begin{align}\label{Eq:rewrite I_SKL-I_SKL}
        &I_{\mathrm{SKL}}(\bW,\hatS_\rmu;S_\rml)-I_{\mathrm{SKL}}(\hatS_\rmu;S_\rml) \nn
       \\& =I(\bW;S_\rml|\hatS_{\rmu})+D(P_{\bW|\hatS_{\rmu}} \| P_{\bW|S_\rml,\hatS_{\rmu}} |P_{S_\rml} P_{\hatS_{\rmu}} ).  
    \end{align}
    From \cref{Thm:gen SSL PL} and \eqref{Eq:rewrite I_SKL-I_SKL}, we observe that for any  $(\eta,\alpha)$, given $\hatS_\rmu$, as the dependency (or the information shared) between $\bW$ and $S_\rml$ decreases, the expected gen-error decreases, and the algorithm is less likely to overfit to the training data. This intuition dovetails with the results for supervised learning in \citet{xu2017information}, \citet{russo2019much} and \citet{aminian2021exact}. However, the difference here is that the quantities are {\em conditioned on} the pseudo-labeled data $\hatS_\rmu$, which reflects the impact of SSL. 
    
     Together with \cref{Prop: upper and lower bound}, we can observe that the gen-error is also dependent on the {\em information shared between the input labeled and pseudo-labeled data}. If  the mutual information $I(\hatS_\rmu;S_\rml)$ decreases, the expected gen-error is likely to decrease as well. This implies that pseudo-labels highly dependent on the labeled dataset may not be beneficial in terms of the generalization performance of an algorithm. In fact, in our subsequent example in Section \ref{Sec: mean est eg}, we exactly quantify the shared information using the cross-covariance between the labeled and pseudo-labeled data.
     This result sheds light on the future design of pseudo-labeling methods.

\subsubsection{Special Cases of Theorem \ref{Thm:gen SSL PL}}
It is instructive to elaborate how Theorem \ref{Thm:gen SSL PL} specializes in some well-known learning settings such as transfer learning and SSL not reusing labeled data.
	\begin{itemize}[itemsep=0pt, topsep=0pt]
		\item {\bf Case 1 ($\hatS_{\rmu}$ independent of $S_\rml$)}: If the pseudo-labels $\{\hatY_i \}_{i=n+1}^{n+m}$ are not generated based on the labeled dataset $S_\rml$, e.g. randomly assigned or generated from another domain (similar to transfer learning), the pseudo-labeled dataset $\hatS_{\rmu}$ is independent of $S_\rml$.
		According to the basic properties of mutual and Lautum information \citep{palomar2008lautum} and~\eqref{Eq:rewrite I_SKL-I_SKL}, we have
		\begin{align}
		   & I_{\mathrm{SKL}}(\bW,\hatS_\rmu;S_\rml)-I_{\mathrm{SKL}}(\hatS_\rmu;S_\rml)=I_{\mathrm{SKL}}(\bW;S_\rml|\hatS_\rmu), \nn
		\end{align}
		and $\bbE_{\Delta_{S_\rml,\hatS_\rmu}}[\log\Lambda_{\alpha,\eta}(S_\rml,\hatS_\rmu)]=0$. That is, 
		\begin{align}
		    \!\!\! \overline{\mathrm{gen}}(P_{\bW|S_\rml,\hatS_\rmu}^\alpha, P_{S_\rml,\hatS_\rmu})\!=\!\frac{(1+\lambda)I_{\mathrm{SKL}}(\bW;S_\rml|\hatS_\rmu)}{\alpha}, \label{Eq:gen indep}
		\end{align}
		which corresponds to the result of transfer learning in \citet[Theorem 1]{bu2022characterizing}. 
		
		\item {\bf Case 2 ($S_\rml-\hatS_{\rmu}-\bW$)}: When this Markov chain holds, meaning that the output hypothesis $\bW$ is independent of $S_\rml$ conditioned on  $\hatS_\rmu$, we have
		\begin{align}
		    &I(\bW;S_\rml|\hatS_\rmu)=0, ~P_{\bW|\hatS_\rmu}=P_{\bW|S_\rml,\hatS_\rmu} \nn\\* &~\text{and}~ D(P_{\bW|\hatS_{\rmu}} \| P_{\bW|S_\rml,\hatS_{\rmu}} |P_{S_\rml} P_{\hatS_{\rmu}} )=0. \nn
		\end{align}
        Thus, the expected gen-error $\overline{\mathrm{gen}}(P_{\bW|S_\rml,\hatS_\rmu}^\alpha, P_{S_\rml,\hatS_\rmu})=\frac{1+\eta}{\alpha}\bbE_{\Delta_{S_\rml,\hatS_\rmu}}[\log\Lambda_{\alpha,\eta}(S_\rml,\hatS_\rmu)]$. However, this is a degenerate case. For example, it only occurs when $\eta\to\infty$, i.e., the labeled dataset $S_\rml$ might be used for generating pseudo-labels for $S_\rmu$ but not used in the Gibbs algorithm to learn the output hypothesis $\bW$. In this case, $\overline{\mathrm{gen}}(P_{\bW|S_\rml,\hatS_\rmu}^\alpha, P_{S_\rml,\hatS_\rmu})=0$.
	\end{itemize}

\subsubsection{SSL vs. SL with $n+m$ labeled data}
    It is also instructive to elaborate on how SSL compares to SL with $n+m$ labeled data based on \cref{Thm:gen SSL PL}. In particular, assume that 
	that the labeled training dataset $S_\rml^{(n+m)}=\{\bZ'_i\}_{i=1}^{n+m}$ contains $n+m$ samples drawn i.i.d.\ from $P_{\bZ}$.
	Then for any output hypothesis $\bw_{\mathrm{SL}}^{(n+m)}\in\calW$, the population risk is given by $L_\rmP(\bw_{\mathrm{SL}}^{(n+m)},P_{S_\rml^{(n+m)}})=\bbE_{\bZ'\sim P_\bZ}[l(\bw_{\mathrm{SL}}^{(n+m)},\bZ')]$ (cf.~\eqref{Eq:population risk}), and 
	the empirical risk of the labeled data samples is given by
	\begin{align}
		L_{\rmE}(\bw_{\mathrm{SL}}^{(n+m)},S_\rml^{(n+m)})&=\frac{1}{n+m}\sum_{i=1}^{n+m}l(\bw_{\mathrm{SL}}^{(n+m)},\bZ'_i). \nn
	\end{align}
	From \citet[Theorem 1]{aminian2021exact}, under the $(\alpha,\pi(\bw_{\mathrm{SL}}^{(n+m)}), L_{\rmE}(\bw_{\mathrm{SL}}^{(n+m)},S_\rml^{(n+m)}))$-Gibbs algorithm, the expected gen-error is
	\begin{align}\label{Eq:gen n add m label}
		\overline{\mathrm{gen}}_{\mathrm{SL}}^{(n+m)} 
		&:=\bbE[L_{\rm P}(\bW_{\mathrm{SL}}^{(n+m)},P_{S_\rml^{(n+m)}})-L_{\rmE}(\bW_{\mathrm{SL}}^{(n+m)},S_\rml^{(n+m)})]  \nn\\
		&=\frac{I_{\mathrm{SKL}}(\bW_{\mathrm{SL}}^{(n+m)};S_\rml^{(n+m)})}{\alpha}. 
	\end{align}
	In the SSL setup, in comparison with \eqref{Eq:gen n add m label}, under the $(\alpha,\pi(\bw), \barL_{\rmE}(\bw,S_\rml,\hatS_\rmu))$-Gibbs algorithm, we let the mixing weight $\eta$ of the empirical risk in \eqref{Eq:normalize emp risk}  be the \emph{ratio} of the number of unlabeled to labeled data, i.e., $\eta=\lambda:=m/n$.
	We similarly define another expected gen-error $\overline{\mathrm{gen}}_{\mathrm{all}}$ which is also evaluated over $n+m$ data points as follows:
	\begin{align}
		&\!\overline{\mathrm{gen}}_{\mathrm{all}}(P_{\bW|S_\rml,\hatS_\rmu}^\alpha, P_{S_\rml,\hatS_\rmu})
		\!:=\!\bbE[L_{\rm P}(\bW,P_{S_\rml})\!-\!L_{\rmE}(\bW,S_\rml,\hatS_\rmu)]. \nn
	\end{align}
	Consider the situation where the pseudo-labeling method is perfect such that $P_{\hatY_i|\bX_i}=P_{Y|\bX}$ for all $i\in[n+1:n+m]$. Then any $\hat{\bZ}_i \in \hatS_\rmu$ has the same distribution as that of any $\bZ \in S_\rml$.  By applying the same technique in the proof of Theorem \ref{Thm:gen SSL PL}, we obtain (details provided in Appendix \ref{App:proof of Eq:gen_all})
	\begin{align}
		\overline{\mathrm{gen}}_{\mathrm{all}}(P_{\bW|S_\rml,\hatS_\rmu}^\alpha, P_{S_\rml,\hatS_\rmu})=\frac{I_{\mathrm{SKL}}(\bW;S_\rml,\hatS_\rmu)}{\alpha}. \label{Eq:gen_all}
	\end{align}
	However, only when $P_{\hatY_i|\bX_i,S_\rml}=P_{\hatY_i|\bX_i}=P_{Y|\bX}$ for any $S_\rml$,
	$I_{\mathrm{SKL}}(\bW;S_\rml,\hatS_\rmu)=I_{\mathrm{SKL}}(\bW_{\mathrm{SL}}^{(n+m)};S_\rml^{(n+m)})$ and $\overline{\mathrm{gen}}_{\mathrm{all}}=\overline{\mathrm{gen}}_{\mathrm{SL}}^{(n+m)}$; otherwise, the gen-errors of SL in \eqref{Eq:gen n add m label}  and of SSL in \eqref{Eq:gen_all} may not be equal. This is because the pseudo-label $\hatY_i$'s are assumed to be generated based on the labeled dataset $S_\rml$ and this dependence affects the gen-error. As we have discussed following \cref{Prop: upper and lower bound}, if the information shared between $S_\rml$ and $\hatS_\rmu$ is large, the learning algorithm tends to overfit to the labeled training data.
	
	If the $\hatY_i$'s are independent of $S_\rml$, then  $P_{\hatY_i|\bX_i,S_\rml}=P_{\hatY_i|\bX_i}$ for any $S_\rml$.  Consider the situation where the pseudo-labeling method is \emph{close} to perfect, i.e.,  $P_{\hat{\bZ}}=P_{\bZ}+\epsilon\Delta$, where $\epsilon> 0$ is   small and $\int_{\calZ}\Delta(\bz)\rmd \bz =0$ such that $P_{\hat{\bZ}}\geq 0$. 
	We have
	\begin{align}
	    &\overline{\mathrm{gen}}_{\mathrm{all}}(P_{\bW|S_\rml,\hatS_\rmu}^\alpha, P_{S_\rml,\hatS_\rmu})\!=\!\frac{I_{\mathrm{SKL}}(\bW_{\mathrm{SL}}^{(n+m)};S_\rml^{(n+m)})}{\alpha}\!+\!\tilde{\epsilon}, \nn
	\end{align}
	where $\tilde{\epsilon}$ is proportional to $\epsilon$ and $\tilde{\epsilon}\to 0$ as $\epsilon\to 0$ (details are provided in Appendix \ref{App: gen_all close to perfect}). Therefore, in this special case, we see that the gap between $\overline{\mathrm{gen}}_{\mathrm{all}}$ of SSL and $\overline{\mathrm{gen}}_{\mathrm{SL}}^{(n+m)}$ of SL can be directly quantified by the gap between the labeled and pseudo-labeled distributions.


\subsection{Distribution-Free Upper and Lower Bounds}
We can also build upon Theorem \ref{Thm:gen SSL PL} and \citet[Theorem 1]{xu2017information} to provide distribution-free upper and lower bounds of the gen-error in~\eqref{Eq: gen main thm}. We next state an informal version of such bounds (the formal version and the proof  are provided in Appendix \ref{App:pf of Prop:dis-free ub}). Throughout this section, we let the mixing weight $\eta=\lambda=m/n$.
\begin{proposition}[Informal Version]\label{Prop:dis-free ub}
    Assume $I(\hatS_\rmu;S_\rml) =\gamma_{\alpha,\lambda} I(\bW,\hatS_\rmu;S_\rml)$, for some $\gamma_{\alpha,\lambda}\in[0,1]$ that depends on $\lambda$ and $\alpha$.
    Assume that $l(\bw,\bZ)$ is bounded in $[a,b]\subset\bbR_+$. Let $\gamma'_{\alpha,\lambda,a,b}=\frac{\alpha(b-a)^2\sqrt{\gamma_{\alpha,\lambda}}}{2(1-\gamma_{\alpha,\lambda})}$. Then the expected gen-error can be bounded as follows
    \begin{align}
        &\!\!-\gamma'_{\alpha,\lambda,a,b} \!\bigg(\frac{1}{\sqrt{n}}\!+\!\frac{\sqrt{\gamma_{\alpha,\lambda}}}{1+\lambda} \bigg)\leq \overline{\mathrm{gen}}(P_{\bW|S_\rml,\hatS_\rmu}^\alpha, P_{S_\rml,\hatS_\rmu}) \leq \gamma'_{\alpha,\lambda,a,b}
        \bigg(\frac{1}{\sqrt{n}}+\frac{1}{(n+m)\sqrt{\gamma_{\alpha,\lambda}}}\bigg).
    \end{align}
\end{proposition}

Recall that for  SL with $n$ labeled data, the expected gen-error is of order $O(\frac{1}{n})$  \citet[Theorem~2]{aminian2021exact}. This can be naturally extended to SL  with $n+m$ labeled data, leading to the expected gen-error behaving as $O(\frac{1}{n+m})$.  Compared to the SL case,  we  see that using the pseudo-labeled data may degrade the convergence rate of the expected gen-error and the extent to which it degrades depends on $\gamma_{\alpha,\lambda}$, which captures the impact of the amount of information shared among the output hypothesis, the input labeled and unlabeled data as well as the ratio $\lambda$.  If $\gamma_{\alpha,\lambda}$ is small, it implies $I(\hatS_\rmu;S_\rml)$ is small and $I(\bW;S_\rml|\hatS_\rmu)$ is large, which leads to a tight upper bound.

If we fix $0<\lambda<\infty$ and assume $\hatS_\rmu$ is independent of $S_\rml$, then $\gamma_{\alpha,\lambda}=0$ and
\begin{align}
     0\leq \overline{\mathrm{gen}}(P_{\bW|S_\rml,\hatS_\rmu}^\alpha, P_{S_\rml,\hatS_\rmu})\leq \frac{\alpha(b-a)^2}{2(n+m)}, \nn
\end{align}
which coincides with the result of transfer learning in \citet[Remark 3]{bu2022characterizing} and corresponds to the analysis of~\eqref{Eq:gen indep}.  Although it appears that using independent pseudo-labeled data $\hatS_\rmu$ does not degrade the convergence rate of the expected gen-error, it may affect the excess risk or the test accuracy, which taken in unison, represents the performance of a learning algorithm. The definition and further analysis of the excess risk is provided in Section \ref{Sec: excess risk}.

When $\lambda \to 0$, the gen-error converges to that of SL with $n$ labeled data and the distribution-free upper bound is of order $O(\frac{1}{n})$, given in \citet[Theorem 2]{aminian2021exact}. 

\section{An Application to Mean Estimation }\label{Sec: mean est eg}
To deepen our understanding of the expected gen-error in Theorem \ref{Thm:gen SSL PL},  we study a mean estimation example and analyze how  $\lambda=m/n$ affects the gen-error. 

\subsection{Problem Setup}
For any $(\bX_i,Y_i)\in S_\rml$ and $i\in[n]$, we assume that $Y_i\in\{-1,+1\}$, $\bbE[\bX_i|Y_i=1]=\bmu$, $\bbE[\bX_i|Y_i=-1]=-\bmu$, $\|\bmu\|_2=1$ and $\Cov[Y_i\bX_i]=\sigma^2 \bI_d$. Any $\bX\in S_\rmu$ is drawn i.i.d.\ from the same distribution as $\bX_i$. Consider the problem of learning $\bmu$ using $S_\rml$ and $\hatS_\rmu$. We adopt the mean-squared loss $l(\bw,\bz)=\|\bw- y\bx\|_2^2$ and assume the prior $\pi(\bw)$  is uniform on $\calW$. Based on $\bW_0$ learned from the labeled dataset $S_\rml$ (e.g., $\bW_0=\frac{1}{n}\sum_{i=1}^nY_i\bX_i$), we assign a pseudo-label to each $X_i\in S_\rmu$ as $\hatY_i$ (e.g. $\hatY_i=\sgn(\bW_0^\top \bX_i)$) and let $\bmu'=\bbE[\hatY_i\bX_i]$. Let us construct $S_\rml'=\{Y_i\bX_i\}_{i=1}^n$ and $\hatS_\rmu'=\{\hatY_i\bX_i\}_{i=n+1}^{n+m}$.
The {\em empirical risk} is given by (cf.~\eqref{Eq:normalize emp risk} by replacing $S_\rml,\hatS_\rmu$ with $S_\rml',\hatS_\rmu'$)
\begin{align}
	\barL_{\rmE}(\bw,S_\rml',\hatS_\rmu') 
	&=\frac{1}{(1+\lambda)n}\sum_{i=1}^n \|Y_i\bX_i-\bw \|_2^2 +\frac{\lambda}{(1+\lambda)m}\sum_{j=n+1}^{n+m} \|\hatY_j\bX_j-\bw \|_2^2. \nn
\end{align}
The expected gen-error is equal to the right-hand side of \eqref{Eq: gen main thm} by replacing $S_\rml,\hatS_\rmu$ with $S_\rml',\hatS'_\rmu$.

The $(\alpha,\pi(\bW), \barL_{\rmE}(\bW,S_\rml',\hatS_\rmu'))$-Gibbs algorithm is given by the following Gibbs posterior distribution
\begin{align}
	P_{\bW|S_\rml',\hatS_\rmu'}^\alpha (\bW|S_\rml',\hatS'_\rmu)= \calN(\bmu_{n+m}, \sigma^2_{\rml,\rmu}\bI_d),\nn
\end{align}
where $\sigma^2_{\rml,\rmu}=\frac{1}{2\alpha}$ and
\begin{align}
	\!\bmu_{n,m}\!=\!\frac{1}{(1+\lambda)n}\sum_{i=1}^n  Y_i\bX_i + \frac{\lambda}{(1+\lambda)m}\sum_{j=n+1}^{n+m} \hatY_j\bX_j . \nn
\end{align}
Details are provided in Appendix \ref{App:proofs for mean est}.
With this posterior Gaussian distribution $P_{\bW|S_\rml',\hatS_\rmu'}^\alpha$,  the expected gen-error in Theorem \ref{Thm:gen SSL PL} can be exactly computed as follows
\begin{align}
    &\!\!\overline{\mathrm{gen}}(P_{\bW|S_\rml,\hatS_\rmu}^\alpha, P_{S_\rml,\hatS_\rmu})=\frac{2 \sigma^2 d}{n+m}+\frac{2m}{n+m}\bbE\big[ (Y_i\bX_i-\bmu)^\top(\hatY_j\bX_j-\bmu')\big], \label{Eq: mean est gen}
\end{align}
where $i\in [n]$ and $j \in [n+1:n+m]$  run over the labeled and unlabeled datasets  respectively. 
From the definition of the expected gen-error in \eqref{Eq:def gen}, we  obtain the same result, which corroborates the characterization of the expected gen-error in Theorem~\ref{Thm:gen SSL PL}. See Appendix \ref{App:proofs for mean est} for details.

Note that the second term in \eqref{Eq: mean est gen} is the trace of the cross-covariance between the labeled and pseudo-labeled data sample, i.e., for any  $i\in[n] $ and $ j\in[n+1:n+m]$,
\begin{align}
    \!\bbE\big[ (Y_i\bX_i-\bmu)^\top  (\hatY_j\bX_j-\bmu')\big]\!=\!\tr(\Cov[Y_i\bX_i,\hatY_j\bX_j]). \nn
\end{align}
This result  shows that   for any fixed $(n,m)$, the expected gen-error decreases when the trace of the cross-covariance decreases, which corroborates our analyses in  Section \ref{Sec:main result}.

\vspace{-5pt}
\subsection{Effect of the Pseudo-labeling and the Ratio of Unlabeled to Labeled Samples~$\lambda$}
To futher study the effect of the pseudo-labeling method and the ratio $\lambda=m/n$, in this mean estimation example, we consider the class-conditional feature distribution of $\bX_i|Y_i \sim \calN(Y_i\bmu,\sigma^2 \bI_d)$ for any $i\in[n+m]$. 
Let $S_\rml'^{(n+m)}$ be a dataset with $n+m$ independent copies of  $Y\bX\in S_\rml'$. Similarly to~\eqref{Eq: mean est gen}, for the supervised $(\alpha,\pi(\bw_{\mathrm{SL}}^{(n)}), L_{\rmE}(\bw_{\mathrm{SL}}^{(n)},S_\rml'))$-Gibbs algorithm  and the supervised $(\alpha,\pi(\bw_{\mathrm{SL}}^{(n+m)}), L_{\rmE}(\bw_{\mathrm{SL}}^{(n+m)},S_\rml'^{(n+m)}))$-Gibbs algorithm, the expected gen-errors are respectively given by (see details in Appendix \ref{App:mean est gen SL})
\begin{align}
	&\overline{\mathrm{gen}}_{\mathrm{SL}}^{(n)} 
	=\frac{2\sigma^2 d}{n} \quad \mbox{and}\quad\overline{\mathrm{gen}}_{\mathrm{SL}}^{(n+m)}
	=\frac{2\sigma^2 d}{n+m}.\nn
\end{align}
Let $\overline{\mathrm{gen}}_{\mathrm{SSL}}$ denote $\overline{\mathrm{gen}}(P_{\bW|S_\rml,\hatS_\rmu}^\alpha, P_{S_\rml,\hatS_\rmu})$ for brevity.
By comparing  $\overline{\mathrm{gen}}_{\mathrm{SL}}^{(n)}$ and $\overline{\mathrm{gen}}_{\mathrm{SL}}^{(n+m)}$, we observe that: 
\begin{enumerate}[itemsep=0pt, topsep=0pt]
    \item If $\bbE\big[ (Y_i\bX_i-\bmu)^\top(\hatY_j\bX_j-\bmu')\big]< 0$, then  $\overline{\mathrm{gen}}_{\mathrm{SSL}} < \overline{\mathrm{gen}}_{\mathrm{SL}}^{(n+m)}$, which means that SSL with such pseudo-labeling method even has better generalization performance than SL with $n+m$ labeled data. This is the most desirable case in terms of the  generalization error;
    \item If $\bbE\big[ (Y_i\bX_i-\bmu)^\top(\hatY_j\bX_j-\bmu')\big]>\frac{\sigma^2d}{n}$, then $\overline{\mathrm{gen}}_{\mathrm{SSL}}>\overline{\mathrm{gen}}_{\mathrm{SL}}^{(n)}$. This implies that if the cross-covariance between the labeled and pseudo-labeled data is larger than a certain threshold, the pseudo-labeling method does not help to improve the generalization performance;
    \item If $0\leq \bbE\big[ (Y_i\bX_i-\bmu)^\top(\hatY_j\bX_j-\bmu')\big]\leq \frac{\sigma^2d}{n}$, then $\overline{\mathrm{gen}}_{\mathrm{SL}}^{(n+m)} \leq \overline{\mathrm{gen}}_{\mathrm{SSL}} \leq \overline{\mathrm{gen}}_{\mathrm{SL}}^{(n)}$. This implies that if the cross-covariance is sufficiently small,   pseudo-labeling   improves the generalization error.
\end{enumerate}
The value of $\bbE\big[ (Y_i\bX_i-\bmu)^\top(\hatY_j\bX_j-\bmu')\big]$ also depends on the ratio $\lambda$. Next, to study the effect of $\lambda$, let the initial hypothesis learned from the labeled data $S_\rml$ be $\bW_0=\frac{1}{n}\sum_{i=1}^nY_i\bX_i \sim \calN(\bmu,\frac{\sigma^2}{n}\bI_d )$ and the pseudo-label for any $\bX_i\in S_\rmu$ be $\hatY_i=\sgn(\bW_0^\top \bX_i)$. 

Inspired by the derivation techniques in \citet{he2022information}, 
we can rewrite the expected gen-error in \eqref{Eq: mean est gen} as 
{\setlength\abovedisplayskip{5pt}
\setlength\belowdisplayskip{5pt}
\begin{equation}
    \overline{\mathrm{gen}}(P_{\bW|S_\rml,\hatS_\rmu}^\alpha, P_{S_\rml,\hatS_\rmu})
	=\frac{2 \sigma^2 d}{n+m}  +\frac{2m}{n+m}E_n, \label{Eq: mean est gen rewrite}
\end{equation}}
where $E_n=\bbE[\tilg J_\sigma(\gamma_n')+\|\bmu^\perp\|_2 K_\sigma(\gamma_n')] $, $\tilg\sim\calN(0,1)$, $\bmu^\perp\sim\calN(\textbf{0},\bI_d-\bmu\bmu^\top)$, $J_\sigma$ and $K_\sigma$ are functions with domain $[-1,1]$, and $\gamma_n'\in[-1,1]$ is a sequence of correlation coefficients. Details are presented in Appendix \ref{App: pf for Eq: mean est gen rewrite},
where we also prove that $E_n=O(d)$ and $E_n\geq 0$, which means that we always have $\overline{\mathrm{gen}}_{\mathrm{SL}}^{(n+m)}\leq \overline{\mathrm{gen}}_{\mathrm{SSL}}$ here.
Furthermore, we   observe that $E_n$ does not depend on $m$ and thus, \eqref{Eq: mean est gen rewrite} converges to $2E_n$ when we fix $n$, and let $\lambda \to \infty$. 

\begin{figure}[!t]
\centering
\begin{subfigure}{0.45\columnwidth}
    \centering
    \includegraphics[width=\textwidth]{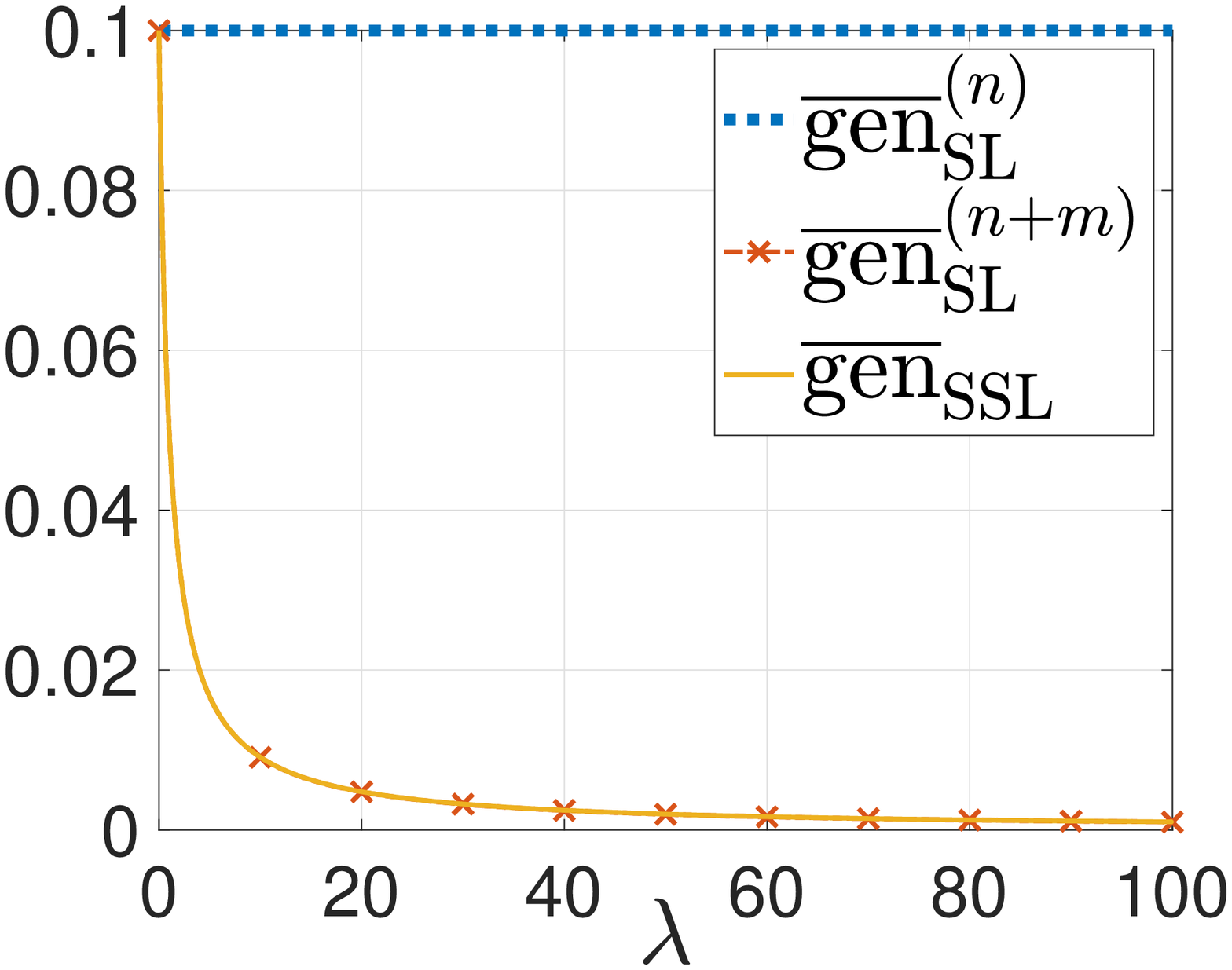}
    \caption{$\sigma=0.5$, $n=5$}
    \label{fig:sig05_n5}
\end{subfigure}
\begin{subfigure}{0.45\columnwidth}
    \centering
    \includegraphics[width=\textwidth]{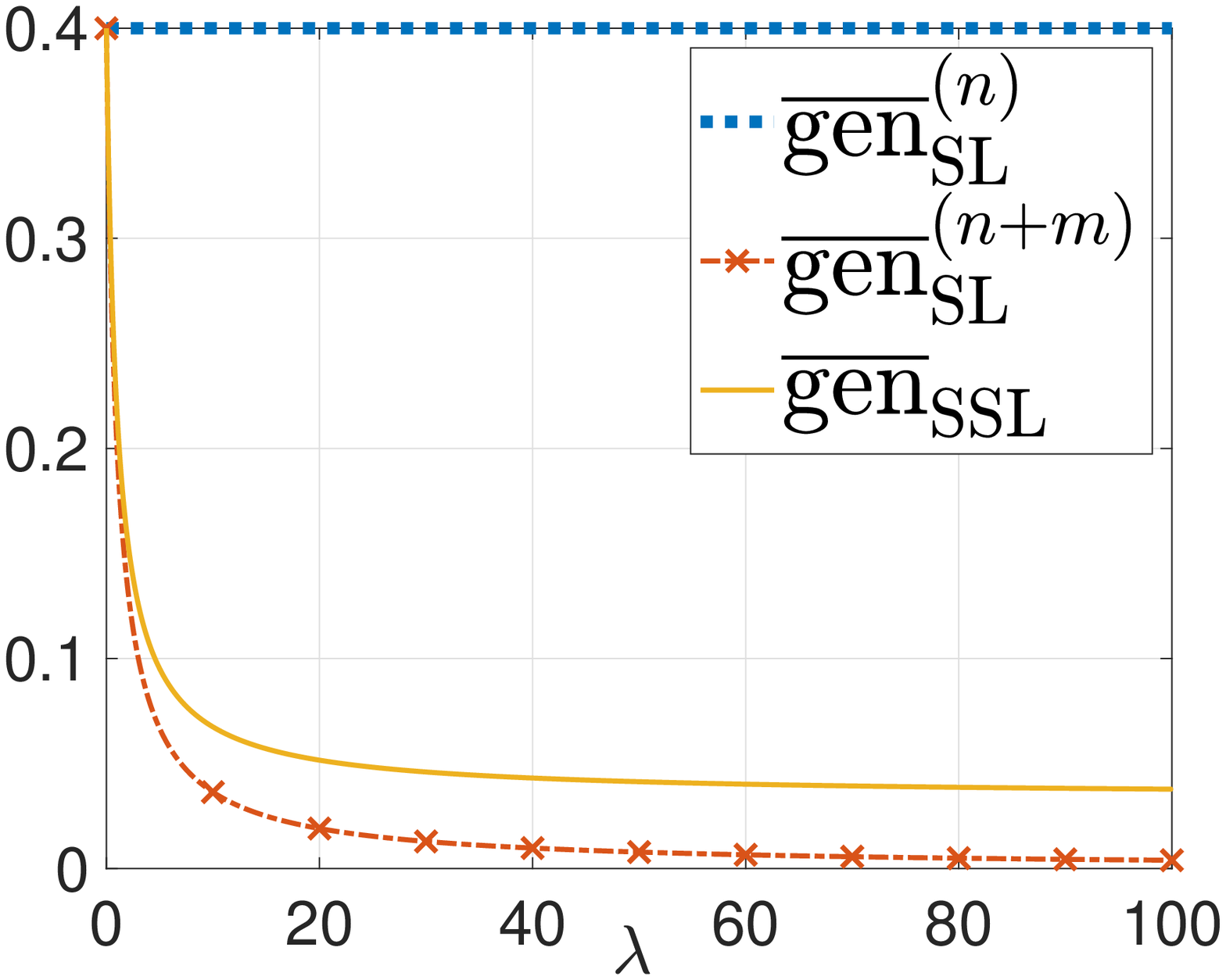}
    \vspace{-15pt}
    \caption{$\sigma=1$, $n=5$}
    \label{fig:sig1_n5}
\end{subfigure}
\begin{subfigure}{0.45\columnwidth}
    \centering
    \includegraphics[width=\textwidth]{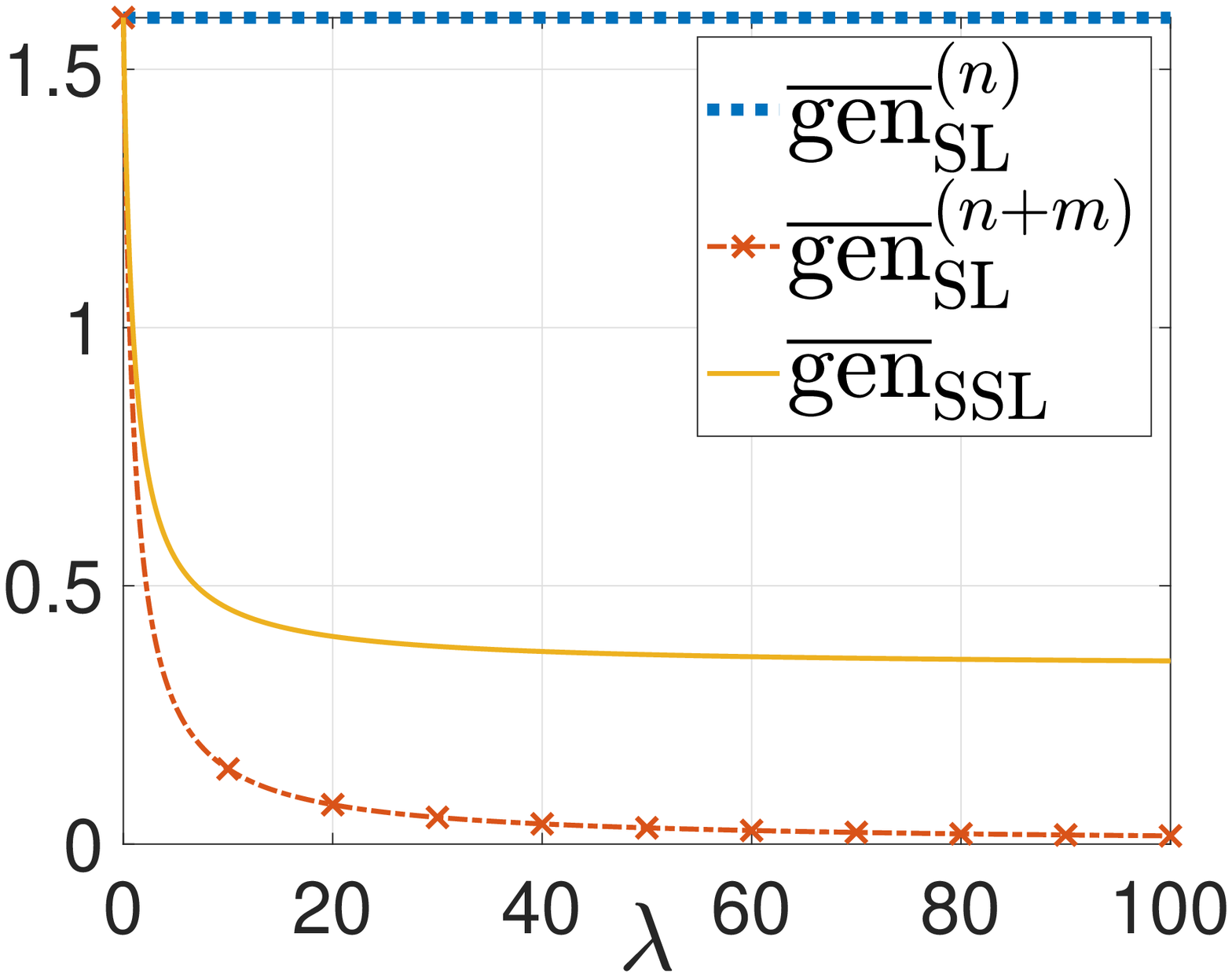}
    \caption{$\sigma=2$, $n=5$}
    \label{fig:sig2_n5}
\end{subfigure}
\begin{subfigure}{0.45\columnwidth}
    \centering
    \includegraphics[width=\textwidth]{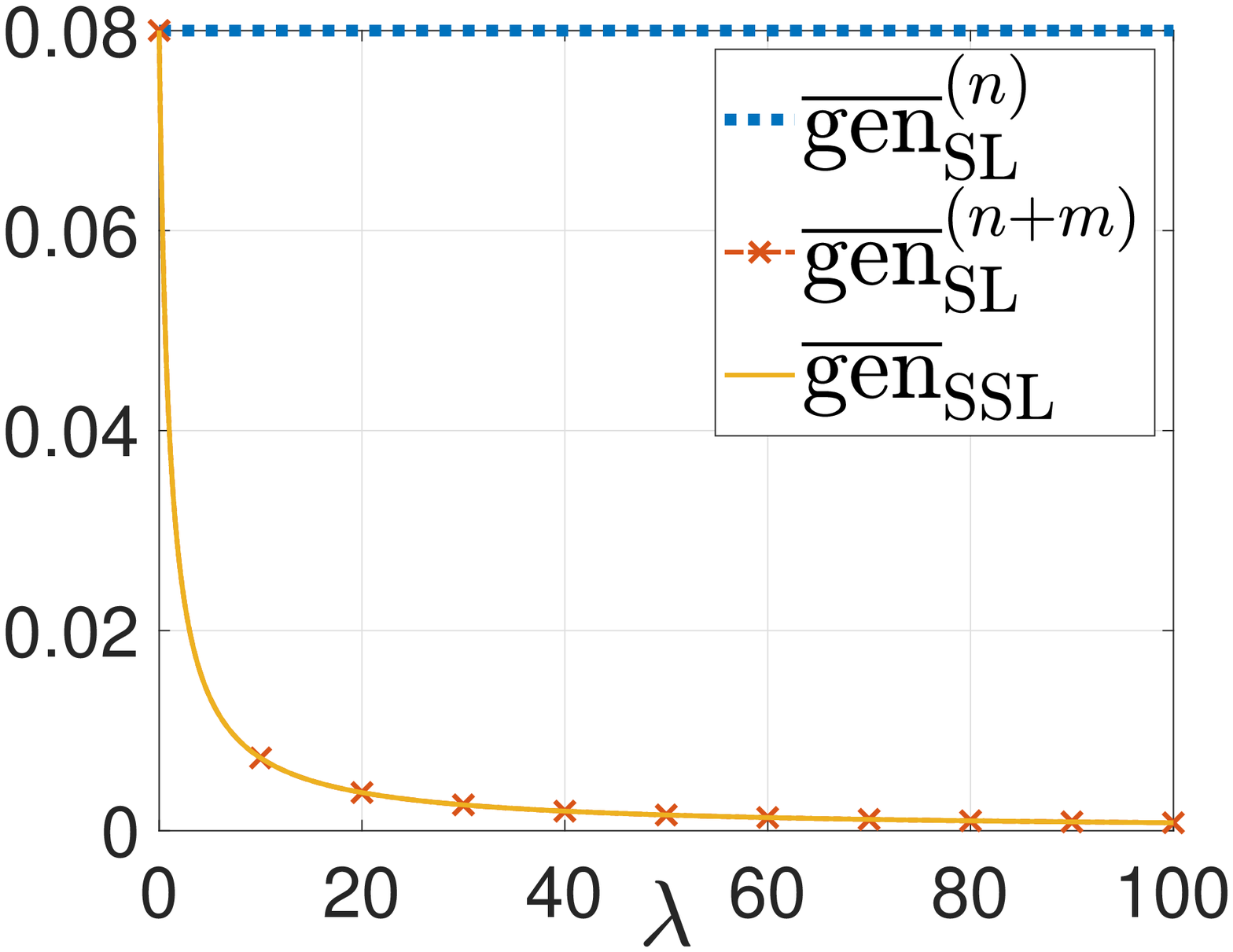}
    \caption{$\sigma=2$, $n=100$}
    \label{fig:sig2_n100}
\end{subfigure}
\vspace{-5pt}
\caption{Gen-error of mean estimation   vs.\  $\lambda=m/n$.}
\vskip -0.05in
\label{Fig: gen vs lambda}
\end{figure}

In Figure \ref{Fig: gen vs lambda}, we numerically plot $\overline{\mathrm{gen}}_{\mathrm{SSL}}$ by varying $\lambda$ and compare it with $\overline{\mathrm{gen}}_{\mathrm{SL}}^{(n)}$ and $\overline{\mathrm{gen}}_{\mathrm{SL}}^{(n+m)}$ for different values of noise level $\sigma$, and number of labeled data samples $n$. Under different choices of $\sigma$, we  observe that  $\overline{\mathrm{gen}}_{\mathrm{SL}}^{(n+m)}\leq \overline{\mathrm{gen}}_{\mathrm{SSL}}\leq \overline{\mathrm{gen}}_{\mathrm{SL}}^{(n)}$. Moreover, the gen-error of SSL  monotonically decreases as $\lambda$ increases, showing obvious improvement compared to $\overline{\mathrm{gen}}_{\mathrm{SL}}^{(n)}$. However, there exists a gap  $\frac{2\lambda E_n}{1+\lambda}$ between $\overline{\mathrm{gen}}_{\mathrm{SSL}}$ and  $\overline{\mathrm{gen}}_{\mathrm{SL}}^{(n+m)}$.
 For $n$ large enough (e.g., $n=100$) or noise level small enough (e.g., $\sigma=0.5$), this gap  is almost negligible, which means   pseudo-labeling yields comparable generalization performance as the SL when {\em  all} the samples are labeled.
 


\subsection{Choosing a  Pseudo-labeling Method from a Given Family of Methods}
\begin{figure}[htbp]
  \begin{center}
\includegraphics[trim={0in .1in .2in 0}, clip, width=0.45\columnwidth]{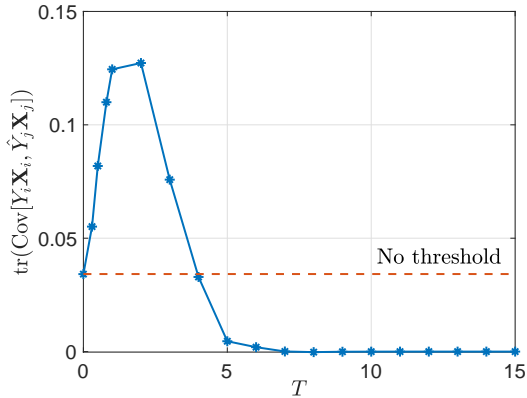}
\caption{The cross-covariance term between labelled and pseudo-labelled data for different confidence threshold $T$.}
\label{Fig: cross-cov vs T}
  \end{center}
  \vspace{-.20in}
\end{figure}

We have shown that the generalization error  decreases as the information shared between the input labeled and pseudo-labeled data (represented by the mutual information $I(\hatS_\rmu;S_\rml)$   or the cross-covariance term $\tr(\Cov[Y_i\bX_i,\hatY_j\bX_j])$ in  the mean estimation example) decreases. This result  serves as a guideline for choosing an appropriate pseudo-labeling method. For example, if we are given a set of pseudo-labeling functions $\{f_i\}_{i=1}^C$ with $f_i: \calX \to \calY$,  we can choose the best one $f_{i^*}$ that yields the minimum  mutual information $I(\hatS_\rmu;S_\rml)$. To make our ideas more concrete, for the mean estimation example that we have been discussing thus far, assume that the pseudo-labeling functions  $\hatY=\sgn(\bW_0^\top \bX)\mathbf{1}_{\{|\bW_0^\top \bX|\geq T\}}$ (where  $\bX\in S_\rmu$) are indexed  by various  ``confidence'' thresholds $T\in\bbR_+$ ($T$ plays the role of the index $i$ in $\{f_i\}_{i=1}^C$). 
For instance, let us consider the case where $n=5$, $\sigma=1$ (cf.\ Figure~\ref{fig:sig1_n5}). As shown in Figure~\ref{Fig: cross-cov vs T}, one can choose the  threshold $T$  that minimizes $\tr(\Cov[Y_i\bX_i,\hatY_j\bX_j])$ for $(\bX_i,Y_i)\in S_\rml$ and $\bX_j\in S_\rmu$. From this figure, we observe that $T\ge 7$ approximately minimizes $\tr(\Cov[Y_i\bX_i,\hatY_j\bX_j])$ and hence, the generalization error. We have thus exhibited a concrete way to choose one pseudo-labeling method from a given family of methods via our characterization of generalization error.  

\section{Particularizing The Gibbs Algorithm To Empirical Risk Minimization}
In this section, we consider the asymptotic behavior of the expected gen-error and excess risk for the Gibbs algorithm under SSL as the ``inverse temperature'' $\alpha\to\infty$. It is known that the Gibbs algorithm converges to empirical risk minimization (ERM) as $\alpha\to\infty$; see Appendix~\ref{App: motivation of Gibbs}.

Given any $S_\rml,\hatS_\rmu$, assume that there exists a unique minimizer of the empirical risk $\barL_{\rmE}(\bW,S_\rml,\hatS_\rmu)$ (also known as  the single-well case), i.e.,
\begin{align}
    \bW^*(S_\rml,\hatS_\rmu)=\argmin_{\bw\in\calW}\barL_{\rmE}(\bw,S_\rml,\hatS_\rmu). \nn
\end{align}
Let the Hessian matrix of the empirical risk $\barL_{\rmE}(\bw,S_\rml,\hatS_\rmu)$ at $\bw=\bW^*(S_\rml,\hatS_\rmu)$ be defined as
\begin{align}
	& H^*(S_\rml,\hatS_\rmu)= \nabla_{\bw}^2\barL_{\rmE}(\bw,S_\rml,\hatS_\rmu)|_{\bw=\bW^*(S_\rml,\hatS_\rmu)}. \nn 
\end{align}
Under the single-well case, we obtain a characterization of the asymptotic expected gen-error in the following theorem. 
\begin{theorem}\label{Thm: gen SSL asym}
Under the $(\alpha,\pi(\bw),\barL_{\rmE}(\bw,S_\rml,\hatS_\rmu))$-Gibbs algorithm, as $\alpha\to\infty$, if the Hessian matrix $H^*(S_\rml,\hatS_u)$ is not singular and $\bW^*(S_\rml,\hatS_\rmu)$ is unique, the expected gen-error converges to
\begin{align}
	&\overline{\mathrm{gen}}(P_{\bW|S_\rml,\hatS_\rmu}^\infty, P_{S_\rml,\hatS_\rmu})\nn\\
	&=
	\bbE_{\hatS_\rmu,S_\rml}\Big[\bW^*(S_\rml,\hatS_\rmu)^\top H^*(S_\rml,\hatS_\rmu)\bW^*(S_\rml,\hatS_\rmu)  \Big]\nn\\
	&\quad +\bbE_{\hatS_\rmu}\bbE_{S_\rml}\Big[ \Big(\bW^*(S_\rml,\hatS_\rmu)-2\bbE_{S_\rml|\hatS_\rmu}[\bW^*(S_\rml,\hatS_\rmu)] \Big)^\top  \cdot H^*(S_\rml,\hatS_\rmu) \bW^*(S_\rml,\hatS_\rmu) \Big]\nn\\
	&\quad -\bbE_{\Delta_{S_\rml,\hatS_\rmu}}\Big[  \barL_{\rmE}(\bW^*(S_\rml,\hatS_\rmu),S_\rml,\hatS_\rmu) \Big]  -\bbE_{\Delta_{(\bW,\hatS_\rmu),S_\rml}}\Big[\bW^\top H^*(S_\rml,\hatS_\rmu)\bW \Big]\bigg), \label{Eq:gen alpha infty}
\end{align}
where the expectations $\bbE_{\Delta_{S_\rml,\hatS_\rmu}}[\cdot]:=\bbE_{S_\rml,\hatS_\rmu}[\cdot]-\bbE_{S_\rml}\bbE_{\hatS_\rmu}[\cdot]$ and $\bbE_{\Delta_{(\bW,\hatS_\rmu),S_\rml}}[\cdot]:=\bbE_{\bW,\hatS_\rmu,S_\rml}[\cdot]-\bbE_{\bW,\hatS_\rmu}\bbE_{S_\rml}[\cdot]$.
\end{theorem}
The proof of Theorem \ref{Thm: gen SSL asym} is provided in Appendix \ref{App: pf of Thm: gen SSL asym}. Theorem~\ref{Thm: gen SSL asym} shows that the gen-error of the Gibbs algorithm under SSL when $\alpha\to\infty$ depends  strongly on the second derivatives of the empirical risk (a.k.a.\ loss landscape) and the relationship between $S_\rml$ and $\hatS_\rmu$. The loss landscape is an important tool to understand the dynamics of the learning process in  deep learning. In the extreme case where $S_\rml$ and $\hatS_\rmu$ are independent, $\bbE_{\Delta_{S_\rml,\hatS_\rmu}}\big[  \barL_{\rmE}(\bW^*(S_\rml,\hatS_\rmu),S_\rml,\hatS_\rmu) \big] =0$ and a simplified expression for the asymptotic gen-error is provided in \cref{App: pf of Coro: asymp gen MLE}.

\vskip -0.5in
\subsection{Semi-Supervised Maximum Likelihood Estimation}\label{Sec:MLE}
In particular, by letting the loss function be the \emph{negative log-likelihood}, this algorithm becomes semi-supervised maximum likelihood estimation (SS-MLE).
In this part, we consider the SS-MLE in the asymptotic regime where $n,m\to\infty$. Throughout this section, we let the mixing weight in \eqref{Eq:normalize emp risk} $\eta=\lambda=m/n$.

We aim to fit the training data with a parametric   family $p(\cdot|\bw)$, where $\bw\in\calW$. Consider the negative log-loss function $l(\bw,\bz)=-\log p(\bz|\bw)$. For the single-well case, the unique minimizer is 
\begin{align}
	&\bW^*(S_\rml,\hatS_\rmu)=\argmin_{\bw\in\calW} \bigg(-\frac{1}{1+\lambda}\cdot \frac{1}{n}\sum_{i=1}^{n}\log p(\bZ_i|\bw)   -\frac{\lambda}{1+\lambda}\cdot \frac{1}{m}\sum_{i=n+1}^{n+m}\log p(\hat{\bZ}_i|\bw)\bigg). \label{Eq: W_ML}
\end{align}
Given any labeled dataset $S_\rml$, we use $P_{\hat{\bZ}|S_\rml}=\bbE_{\bW_0|S_\rml}[P_{\hat{\bZ}|\bW_0}]$ to denote the conditional distribution of the pseudo-labeled data (i.e., $\hat{\bZ}_i\overset{\text{i.i.d.}}{\sim} P_{\hat{\bZ}|S_\rml}$), where $\bW_0$ is the initial hypothesis learned only from $S_\rml$ and used to generate pseudo-labels for $S_\rmu$. Assume that $\bW_0$ is learned from $S_\rml$ using MLE, i.e.,
\begin{align}
    \bW_0=\argmin_{\bw\in\calW} -\frac{1}{n}\sum_{i=1}^n\log p(\bZ_i|\bw).\nn
\end{align}
We have $\bW_0 \xrightarrow{\rmp} \bw^*_{P_\bZ}=\argmin_{\bw\in\calW} D(P_\bZ \| p(\cdot|\bw))$ as $n\to\infty$,
where $\bw^*_{P_\bZ}$ depends only   on the true distribution of the labeled data $P_\bZ$. Then we have $P_{\hat{\bZ}|\bW_0}\xrightarrow{\rmp} P_{\hat{\bZ}|\bw^*_{P_\bZ}}$, $P_{\hat{\bZ}|S_\rml}\xrightarrow{\rmp}P_{\hat{\bZ}|\bw^*_{P_\bZ}}$ and $P_{\hat{\bZ}}=\bbE_{\bW_0}[P_{\hat{\bZ}|\bW_0}]\xrightarrow{\rmp} P_{\hat{\bZ}|\bw^*_{P_\bZ}}$,
which means $\{\hat{\bZ}_i\}_{i=n+1}^{n+m}$ become independent of one another and of $S_\rml$.
We analogously define the minimizer 
\begin{align}
	\bw^*_{\lambda}=&\argmin_{\bw\in\calW} \bigg(\frac{n}{n+m}D\big(P_{\bZ} \| p(\cdot|\bw) \big) +\frac{m}{n+m}D\big(P_{\hat{\bZ}|\bw^*_{P_\bZ}} \| p(\cdot|\bw) \big)\bigg) \label{Eq: w^*_(m/n)}
\end{align}
and   the {\em landscapes} of the loss function
\begin{align*}
	J_\rml(\bw)&:=\bbE_{\bZ\sim P_\bZ}[-\nabla_{\bw}^2\log p(\bZ|\bw)],\\
	J_\rmu(\bw)&:=\bbE_{\hat{\bZ}\sim P_{\hat{\bZ}|\bw^*_{P_\bZ}}}[-\nabla_{\bw}^2\log p(\hat{\bZ}|\bw) ], \\
	\calI_\rml(\bw)&:=\bbE_{\bZ\sim P_\bZ}[\nabla_{\bw}\log p(\bZ|\bw)\nabla_{\bw}\log p(\bZ|\bw)^\top]. 
\end{align*}
Let $J(\bw)=\frac{n}{n+m}J_\rml(\bw)+\frac{m}{n+m}J_\rmu(\bw)$.
Given any ratio $\lambda>0$, as $n,m\to\infty$, by the law of large numbers, the Hessian matrix $H^*(S_\rml,\hatS_\rmu)$ converges as follows
\begin{align}
	H^*(S_\rml,\hatS_\rmu)\xrightarrow{\rmp} J(\bw^*_{\lambda}), \nn
\end{align}
which is independent of  $(S_\rml,\hatS_\rmu)$. 
Leveraging  these asymptotic approximations, from Theorem~\ref{Thm: gen SSL asym}, we obtain the following characterization of the gen-error of SS-MLE.
\begin{corollary}\label{Coro: asymp gen MLE}
In the asymptotic regime where $n,m\to\infty$, the expected gen-error of SS-MLE is given by
\begin{align}
    \overline{\mathrm{gen}}(P_{\bW|S_\rml,\hatS_\rmu}^\infty, P_{S_\rml,\hatS_\rmu})
	=\frac{\tr(J(\bw^*_{\lambda})^{-1}\calI_\rml (\bw^*_{\lambda}) )}{n+m}. \nn
\end{align}
\end{corollary}
The proof of Corollary \ref{Coro: asymp gen MLE} is provided in Appendix \ref{App: pf of Coro: asymp gen MLE}.
For the extreme cases where $\lambda \to 0$, we have
$\bw^*_{\lambda}\to \bw^*_{P_\bZ}$, $J(\bw^*_{\lambda}) \to J_\rml(\bw^*_{P_\bZ})$ and 
$\!	\overline{\mathrm{gen}}(P_{\bW|S_\rml,\hatS_\rmu}^\infty, P_{S_\rml,\hatS_\rmu})\!\! \to\!\! \frac{1}{n}\tr(J_\rml(\bw^*_{P_\bZ})^{-1}\calI_\rml(\bw^*_{P_\bZ}))\!=\! O( \frac{d}{n} ), $
which means the gen-error degenerates to that of the SL case with $n$ labeled data.

For the other   case where $\lambda \to \infty$, from   Appendix~\ref{App: pf of Coro: asymp gen MLE}, we have $\bW_{\mathrm{ML}}(S_\rml,\hatS_\rmu) \to \bbE_{S_\rml}[\bW_{\mathrm{ML}}(S_\rml,\hatS_\rmu)]$ and 
$\overline{\mathrm{gen}}(P_{\bW|S_\rml,\hatS_\rmu}^\infty, P_{S_\rml,\hatS_\rmu})\to 0.$

For $\lambda\!\in\! (0,\infty)$, we have $\overline{\mathrm{gen}}(P_{\bW|S_\rml,\hatS_\rmu}^\infty, P_{S_\rml,\hatS_\rmu})\!=\! O( \frac{d}{n+m} )$,
which is   order-wise the same as the gen-error of SL with $n+m$ labeled data. The intuition is that for large $n$, the pseudo-labeled samples only depend on the labeled data {\em distribution} instead of the  labeled samples. However, the performance of an algorithm  depends not only on the gen-error and but also on the  excess risk. Even when the gen-error is small, the bias of the excess risk may be high. 


\subsubsection{Excess Risk as $\alpha,n,m\to\infty$}\label{Sec: excess risk}
In this section, we discuss   the excess risk of SS-MLE   when $n,m\to\infty$.
The \emph{excess risk} is defined as the gap between the expectation and the minimum of the population risk, i.e., 
\begin{align}
	\calE_{\mathrm{r}}(P_\bW)&:=\bbE_{\bW}[L_\rmP(\bW,P_{S_\rml})]-L_\rmP(\bw_\rml^*,P_{S_\rml})\\
	&=\overline{\mathrm{gen}}(P_{\bW|S_\rml,\hatS_\rmu}^\alpha, P_{S_\rml,\hatS_\rmu})+\bbE_{\bW,S_\rml}[L_{\rmE}(\bW,S_\rml)-L_{\rmE}(\bw_\rml^*,S_\rml)]. \label{Eq: decomp excess risk}
\end{align}
where $\bw_\rml^*=\argmin_{\bw\in\calW} L_\rmP(\bW,P_{S_\rml})$ is the optimal hypothesis. 
The second term in \eqref{Eq: decomp excess risk} is known as the \emph{estimation error}. We  observe that the excess risk depends on both the gen-error and estimation error. When the gen-error is controlled to be sufficiently small, but if the estimation error is large, the excess risk can still be large. 
\begin{corollary}\label{Lem: excess risk MLE}
In the asymptotic regime where $n,m\to\infty$, the excess risk of SS-MLE is given by
\begin{align}
    &\!\!\calE_{\mathrm{r}}(P_\bW)=\frac{1}{2}\tr((\bw^*_{\lambda}-\bw^*_\rml)(\bw^*_{\lambda}-\bw^*_\rml)^\top J_\rml(\bw_\rml^*) ) +\frac{\tr(J_\rml(\bw_\rml^*)J(\bw^*_{\lambda})^{-1}\calI_\rml(\bw^*_{\lambda})J(\bw^*_{\lambda})^{-1} )}{2(1+\lambda)(n+m)}.  \label{Eq:excess risk bias var 2}
\end{align}
\end{corollary}
The proof of \cref{Lem: excess risk MLE} is provided in Appendix \ref{App: pf of Lem: excess risk MLE}.
In~\eqref{Eq:excess risk bias var 2},  the first term represents the bias caused by learning with the mixture of labeled and pseudo-labeled data. When $\lambda \to 0$, the bias  converges to $0$. As $\lambda$ increases, the bias  increases. The second term represents the variance component of the excess risk, which is of order $O(\frac{d}{m+n})$ for $0<\lambda<\infty$, the same as that for the gen-error in \cref{Coro: asymp gen MLE}. 



\subsection{An Application to Logistic Regression}
To re-emphasize, we let the mixing weight in \eqref{Eq:normalize emp risk} $\eta=\lambda=m/n$. We now  apply  SS-MLE to logistic regression to study the effect of $\lambda$ on the gen-error and excess risk.  For any hypothesis $\bw$, let $p(y|\bx,\bw)$ be the conditional likelihood of a label upon seeing a feature sample under $\bw$. Assume that the label $Y\in\{-1,+1\}$. For any $\bz\in\calX\times \{-1,+1\}$, the underlying distribution is $P_\bZ(\bz)=P_{Y|\bX}(y|\bx)P_{\bX}(\bx)$ and the logistic regression model uses $p(y|\bx,\bw)$ to approximate $P_{Y|\bX}(y|\bx)$. Let $P_{\bZ|\bw}(\bz|\bw)=p(y|\bx,\bw)P_\bX(\bx)$. 
To stabilize the solution $\bw$, we consider a   regularized version of the logistic regression model,  where for a fixed $\nu>0$, the objective function can be expressed as
\begin{align}
	l(\bw,\bz)&=-\log p(y|\bx,\bw)+\frac{\nu}{2}\|\bw\|_2^2=\log (1+e^{-y\bw^\top \bx})+\frac{\nu}{2}\|\bw\|_2^2. \nn
\end{align}
We assume that there exists a unique minimizer of the empirical risk in this example. We also assume that the initial hypothesis $\bW_0$ is learned from the labeled dataset $S_\rml$:
\begin{align}
	\bW_0=\argmin_{\bw\in\calW}\bigg(\frac{1}{n}\sum_{i=1}^{n}\log (1+e^{-Y_i\bw^\top \bX_i})+\frac{\nu}{2}\|\bw\|_2^2 \bigg). \nn
\end{align}
Consider the case when $n,m\to\infty$ and $\lambda>0$. Then 
\begin{align*}
	\bW_0 \!\xrightarrow{\rmp}\! \bw^*_0 \!=\! \argmin_{\bw\in\calW} \!\bigg(\! D\bigg(P_\bZ \bigg\| \frac{P_\bX}{1+e^{-Y\bw^\top \bX}} \bigg) \!+\! \frac{\nu}{2}\|\bw\|_2^2 \!\bigg). \nn
\end{align*}
Let the pseudo label for any $\bX_i\in S_\rmu$ be defined as $\hat{Y}_i=\sgn( \bX_i^\top \bW_0)$.
The conditional distribution of the pseudo-labeled data sample given $\bW_0$ converges as follows
\begin{align}
	\!\! P_{\hat{\bZ}|\bW_0}(\hat{\bz}|\bW_0) \!\!\xrightarrow{\rmp}\!\!	P_{\hat{\bZ}|\bw^*_0}(\hat{\bz}|\bw^*_0) \!=\! P_{\bX}(\bx) \! \mathbbm{1}  \{\haty\!=\!\sgn(\bx^\top\bw^*_0 ) \!\}. \nn
\end{align}
Let us rewrite the minimizer $\bw^*_\lambda$ in \eqref{Eq: w^*_(m/n)} as 
\begin{align}
	\bw^*_{\lambda}=&\argmin_{\bw\in\calW} \bigg(\frac{n}{n+m}D\big(P_{\bZ} \| p(\cdot|\bw) \big) +\frac{m}{n+m}D\big(P_{\hat{\bZ}|\bw^*_0} \| p(\cdot|\bw) \big)+\frac{\nu}{2}\|\bw\|_2^2 \bigg).  \nn
\end{align}
Recall the expected gen-error in \cref{Coro: asymp gen MLE}, which can also be rewritten as 
\begin{align}
	&n\cdot \overline{\mathrm{gen}}(P_{\bW|S_\rml,\hatS_\rmu}^\infty, P_{S_\rml,\hatS_\rmu})
	\!=\!\frac{\tr((J(\bw^*_{\lambda})+\nu\bI_d)^{-1}\calI_\rml (\bw^*_{\lambda}) )}{1+\lambda}. \nn
\end{align}
Details of the derivation are provided in Appendix \ref{App: pf of logistic}. We focus on the right-hand side, which depends on the ratio $\lambda$ instead of the individual $m,n$. As mentioned in Section~\ref{Sec:MLE}, when $\lambda \to 0$, $n \cdot \overline{\mathrm{gen}}(P_{\bW|S_\rml,\hatS_\rmu}^\infty, P_{S_\rml,\hatS_\rmu})\to d$.
 On the other hand, the excess risk $\calE_{\mathrm{r}}(P_\bW)$ of this example is given by \cref{Lem: excess risk MLE} where $J(\bw^*_\lambda)$ is replaced by $J(\bw^*_{\lambda})+\nu\bI_d$. Intuitively, as the regularization parameter $\nu$ increases, the gen-error decreases.

\begin{figure}[!t]
\centering
\begin{subfigure}{0.475\columnwidth}
    \includegraphics[width=\columnwidth]{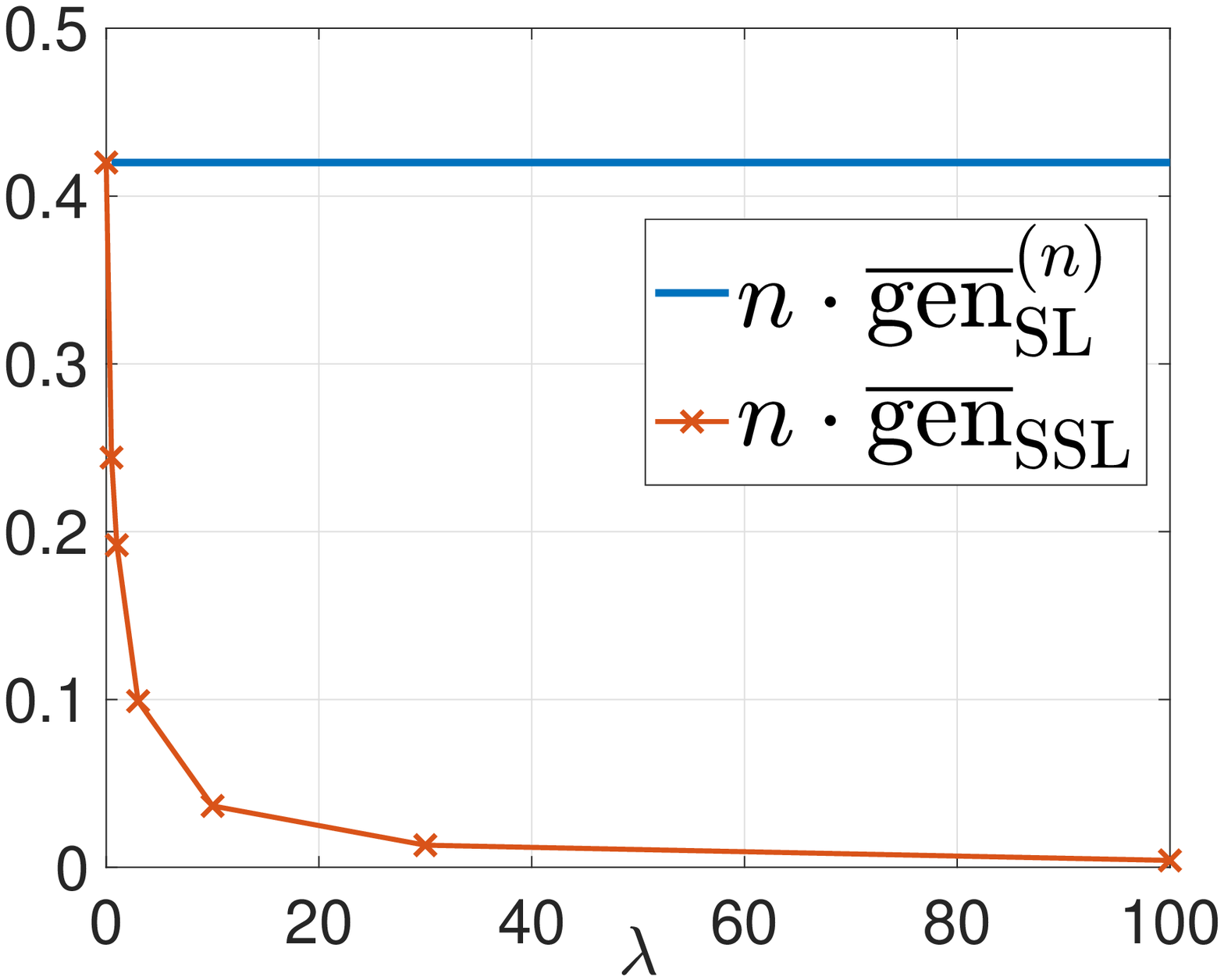}
    \caption{Theoretical gen-error}
\end{subfigure} \hspace{-.1in}
\begin{subfigure}{0.475\columnwidth}
        \includegraphics[width=\columnwidth]{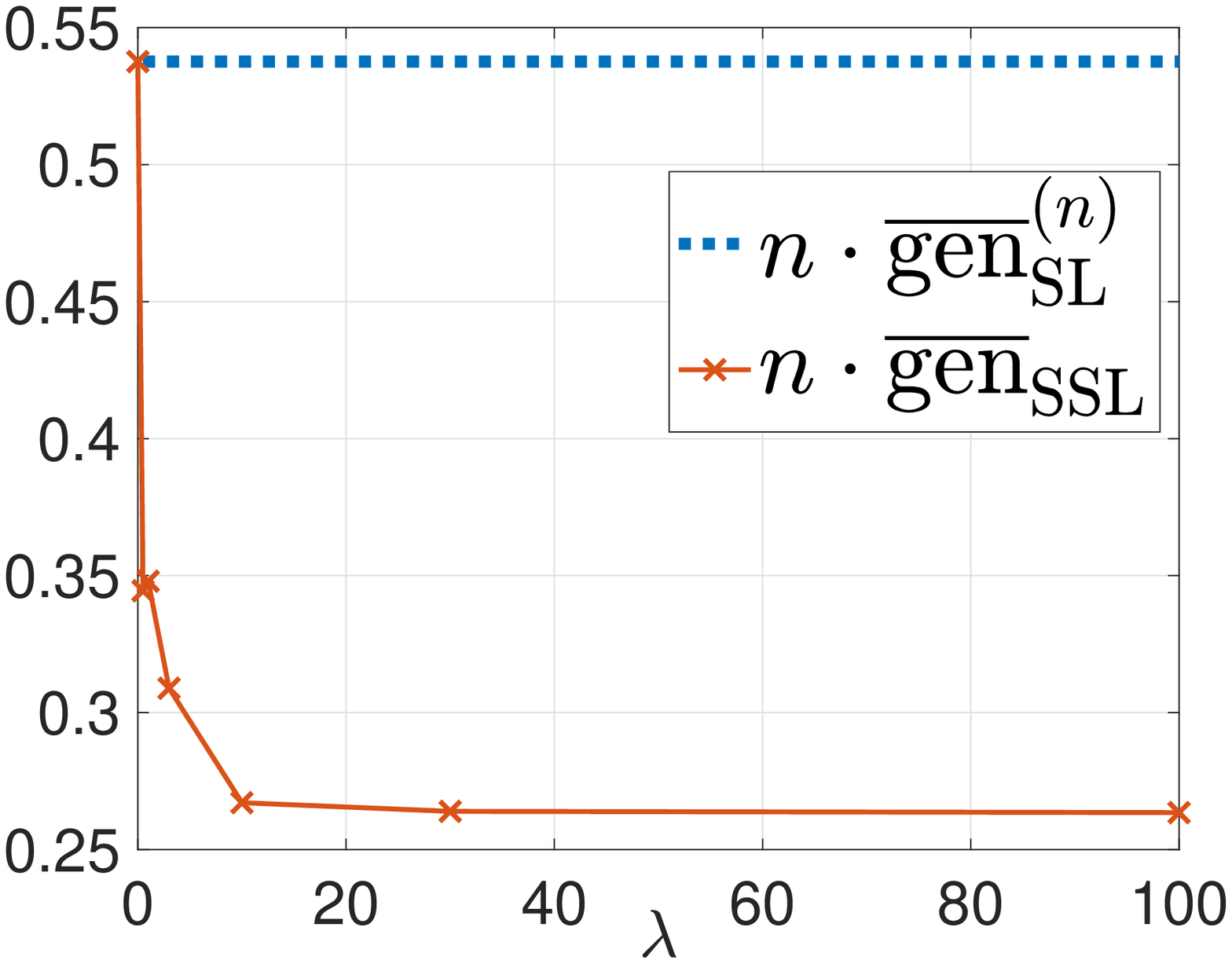}
        \caption{Empirical gen-error}
\end{subfigure}\hspace{-.1in}
 \begin{subfigure}{0.97\columnwidth}
     \includegraphics[width=\columnwidth]{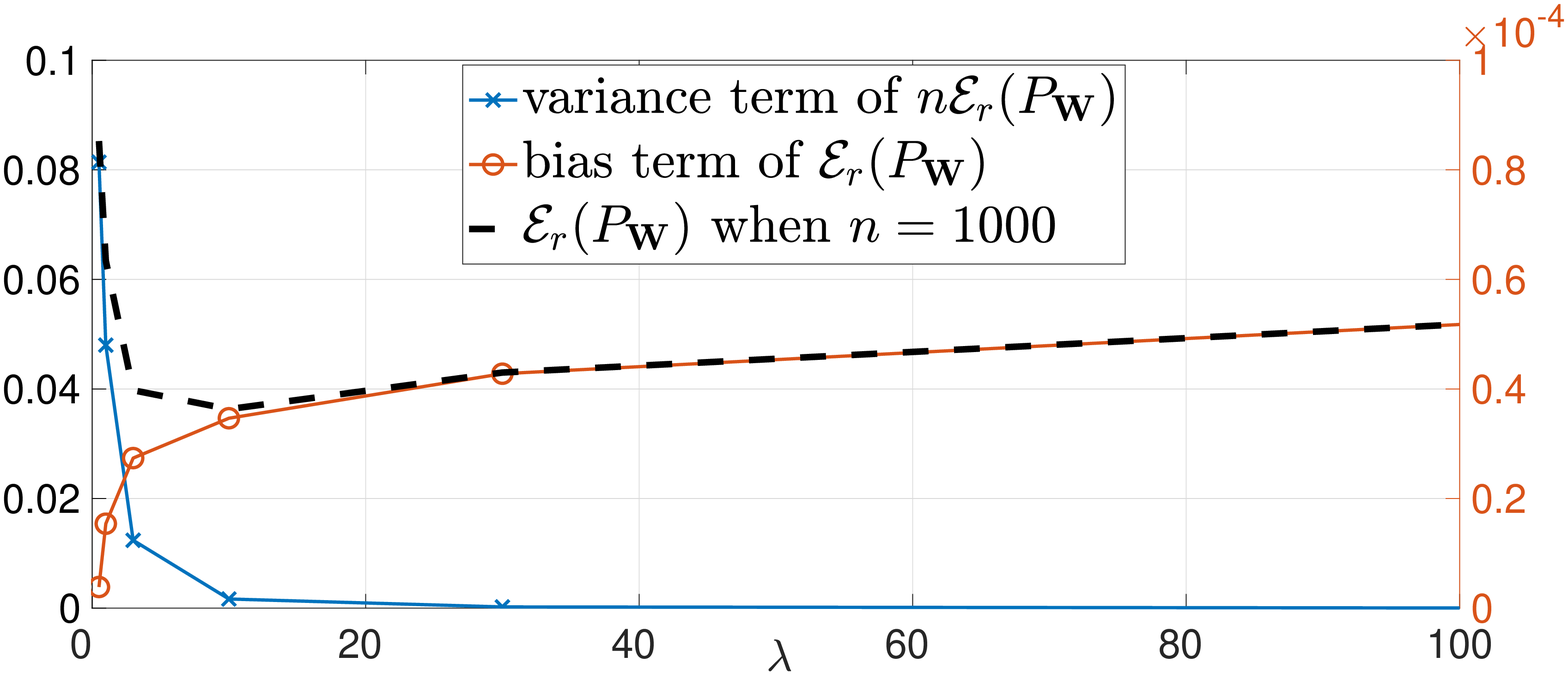}
    \caption{Theoretical excess risk}
 \end{subfigure}
\caption{Theoretical and empirical gen-error and excess risk vs. $\lambda=m/n$ when $\mu=2$ and  $\nu=0.001$.}
\label{Fig: logi plot}
\end{figure}
For different values of $\lambda$, we can numerically calculate the hypothesis $\bw^*_\lambda$, the expected gen-error and the excess risk. Consider the example of a dataset which contains two classes of Gaussian  samples. Let $P_{\bX|Y}=\calN(Y\mu\mathbf{1}_d,\bI_d)$ and $P_Y=\mathrm{Unif}(\{-1,+1 \})$, where $\mu\in\bbR_+$ and $\mathbf{1}_d$  is the   all-ones vector in $\bbR^d$. 
For notational simplicity, let $\overline{\mathrm{gen}}_{\mathrm{SSL}}$ denote $\overline{\mathrm{gen}}(P_{\bW|S_\rml,\hatS_\rmu}^\infty, P_{S_\rml,\hatS_\rmu})$.
In Figure \ref{Fig: logi plot}, we   plot $n\cdot\overline{\mathrm{gen}}_{\mathrm{SSL}}$ and $\calE_{\mathrm{r}}(P_\bW)$ versus  $\lambda$ when $\mu=2$ and $d=2$. We also implement experiments ($n=1,000$) for $40$ times on synthetic data and plot the average empirical gen-error for $\lambda\in\{0,0.5,1,3,10,30,100\}$. In both theoretical and empirical plots, we observe that as the ratio $\lambda$ increases, the gen-error of SSL decreases and is smaller than that of SL with only $n$ labeled data. We also observe similar behaviours of the empirical gen-error of logistic regression on MNIST dataset (see Appendix \ref{App: pf of logistic}). On the other hand, the variance component of the excess risk behaves similarly as the gen-error while the bias component increases. By taking $n=1,000$, we can see the excess risk first decreases and then increases as $\lambda$ increases. In this setting, our results suggest that when  $\lambda$ exceeds some threshold, it may not be beneficial to the performance of the learning algorithm by utilizing any more unlabeled data.

\section{Concluding Remarks}
To develop a comprehensive understanding of SSL, we present an exact characterization of the expected gen-error  of pseudo-labeling-based SSL  via the Gibbs algorithm. Our results reveal that the expected gen-error is influenced by \emph{the information shared between the labeled and pseudo-labeled data} and the ratio of the number of unlabeled data to labeled data. This sheds light in our quest to design good  pseudo-labeling methods, which should  penalize the dependence between the labeled and pseudo-labeled data, e.g., $I(\hatS_\rmu;S_\rml)$, so as to improve the generalization performance. To understand the ERM counterpart of pseudo-labeling-based SSL, we also investigate the asymptotic regime, in which the inverse temperature $\alpha\to\infty$. Finally, we present two examples---mean estimation and logistic regression---as  applications of our theoretical findings. 

\section*{Acknowledgment}
Haiyun He and Vincent Tan are supported by a Singapore National Research Foundation (NRF) Fellowship (Grant Number: A-0005077-01-00). Gholamali Aminian is supported by the UKRI Prosperity Partnership Scheme (FAIR) under the EPSRC Grant EP/V056883/1. M.~R.~D.~Rodrigues and Gholamali Aminian are also supported by the Alan Turing Institute.

\clearpage
{
\bibliographystyle{plainnat}
\bibliography{refs}
}

\newpage
\appendix


\section{Motivations for Gibbs Algorithm}\label{App: motivation of Gibbs}

There are different motivations for the Gibbs algorithm as discussed in \cite{raginsky2017non,kuzborskij2019distribution,asadi2020chaining,aminian2021exact}. Following are an overview of the most prominent motivations for the Gibbs algorithm:

\textbf{Empirical Risk Minimization:}  The $(\alpha,\pi(\bW),\barL_{\rmE}(\bW,S_\rml,\hatS_\rmu))$-Gibbs algorithm can be viewed as a randomized version of empirical risk minimization (ERM). As the inverse temperature $\alpha \to \infty$, then hypothesis generated by the Gibbs algorithm converges to the hypothesis corresponding to standard ERM.

 \textbf{SGLD Algorithm:} 
As discussed in \citet{chiang1987diffusion} and \citet{markowich2000trend}, the Stochastic Gradient Langevin Dynamics (SGLD) algorithm can be viewed as the discrete version of the continuous-time Langevin diffusion, and it is defined as follows:
\begin{equation}
    \bW_{k+1}=\bW_k-\beta\,\nabla \barL_{\rmE}(\bW_k,S_\rml,\hat{S}_\rmu)+\sqrt{\frac{2\beta}{\gamma}}\,\zeta_k, \quad k=0,1,\cdots,
\end{equation}
where $\zeta_k$ is a standard Gaussian random vector  and $\beta>0$ is the step size. In \cite{raginsky2017non}, it is proved that under some conditions on the loss function, the conditional distribution $P_{\bW_k|S_\rml,\hatS_\rmu}$ induced by SGLD algorithm is close to the $(\gamma,\pi(\bW_0),\barL_{\rmE}(\bW_k,S_\rml,\hat{S}_\rmu))$-Gibbs distribution in 2-Wasserstein distance for sufficiently large iterations, $k$.

\section{Proof of Theorem \ref{Thm:gen SSL PL}}\label{App:proof of Thm:gen SSL PL}
Under the $(\alpha,\pi(\bW), \barL_{\rmE}(\bW,S_\rml,\hatS_\rmu))$-Gibbs algorithm, we have
 \begin{align}
	&I_{\mathrm{SKL}}(\bW,\hatS_\rmu;S_\rml) \nn\\
	&=\bbE_{\bW,\hatS_\rmu,S_\rml}[\log P_{\bW|S_\rml,\hatS_\rmu}^\alpha+\log P_{\hatS_\rmu|S_\rml}]-\bbE_{\bW,\hatS_\rmu}\bbE_{S_\rml}[\log P_{\bW|S_\rml,\hatS_\rmu}^\alpha+\log P_{\hatS_\rmu|S_\rml}]\\
	&=\bbE_{\bW,\hatS_\rmu,S_\rml}[-\alpha\barL_{\rmE}(\bW,S_\rml,\hatS_\rmu)-\log \Lambda_{\alpha,\eta}(S_\rml,\hatS_\rmu)]-\bbE_{\bW,\hatS_\rmu}\bbE_{S_\rml}[-\alpha\barL_{\rmE}(\bW,S_\rml,\hatS_\rmu)-\log \Lambda_{\alpha,\eta}(S_\rml,\hatS_\rmu)] \nn\\
	&\quad +\bbE_{\hatS_\rmu,S_\rml}[\log P_{\hatS_\rmu|S_\rml}]-\bbE_{\hatS_\rmu}\bbE_{S_\rml}[\log P_{\hatS_\rmu|S_\rml}]\\
	&=\frac{\alpha}{1+\eta}(\bbE_{\bW}[L_\rmP(\bW,P_{S_\rml})]-\bbE_{\bW,S_\rml}[L_{\rmE}(\bW,S_\rml)]) +\bbE_{\hatS_\rmu,S_\rml}[\log P_{\hatS_\rmu|S_\rml}]-\bbE_{\hatS_\rmu}\bbE_{S_\rml}[\log P_{\hatS_\rmu|S_\rml}]\nn\\
	&\quad -\bbE_{\hatS_\rmu,S_\rml}[\log \Lambda_{\alpha,\eta}(S_\rml,\hatS_\rmu)]+\bbE_{\hatS_\rmu}\bbE_{S_\rml}[\log \Lambda_{\alpha,\eta}(S_\rml,\hatS_\rmu)]\\
	&=\frac{\alpha}{1+\eta}\overline{\mathrm{gen}}(P_{\bW|S_\rml,\hatS_\rmu}, P_{S_\rml,\hatS_\rmu})+I_{\mathrm{SKL}}(\hatS_\rmu;S_\rml)-\bbE_{\Delta_{S_\rml,\hatS_\rmu}}[\log \Lambda_{\alpha,\eta}(S_\rml,\hatS_\rmu)],
\end{align}
where $\Lambda_{\alpha,\eta}(S_\rml,\hatS_\rmu)=\int \pi(\bw)\exp(-\alpha\barL_{\rmE}(\bw,S_\rml,\hatS_\rmu)) \rmd\bw$ and $\bbE_{\Delta_{S_\rml,\hatS_\rmu}}[\cdot]=\bbE_{S_\rml,\hatS_\rmu}[\cdot]-\bbE_{S_\rml}\bbE_{\hatS_\rmu}[\cdot]$.

Thus, the expected gen-error is given by
\begin{align}
    \overline{\mathrm{gen}}(P_{\bW|S_\rml,\hatS_\rmu}, P_{S_\rml,\hatS_\rmu})=\frac{(1+\eta)(I_{\mathrm{SKL}}(\bW,\hatS_\rmu;S_\rml)-I_{\mathrm{SKL}}(\hatS_\rmu;S_\rml)+\bbE_{\Delta_{S_\rml,\hatS_\rmu}}[\log \Lambda_{\alpha,\eta}(S_\rml,\hatS_\rmu)])}{\alpha}.
\end{align}

\begin{remark}(Apply \cref{Thm:gen SSL PL} to SL)
When $\eta\to 0$ and  only the labeled dataset $S_\rml$ is used for learning the output hypothesis~$\bW$, the setup recovers back to SL and we have  $\bbE_{\Delta_{S_\rml,\hatS_\rmu}}[\log\Lambda_{\alpha,\eta}(S_\rml,\hatS_\rmu)]=0$. From \cref{Thm:gen SSL PL} and \eqref{Eq:rewrite I_SKL-I_SKL}, the expected gen-error becomes
		\begin{align}\label{Eq:gen lambda to 0}
		    \overline{\mathrm{gen}}(P_{\bW|S_\rml,\hatS_\rmu}^\alpha, P_{S_\rml,\hatS_\rmu})=\frac{I_{\mathrm{SKL}}(\bW;S_\rml)}{\alpha},
		\end{align}
		reducing to  SL with $n$ labeled data examples as in \citet{aminian2021exact}.
\end{remark}

\section{Exact Characterization of Expected Gen-Error Under Entropy Minimization}\label{Sec: SSL entropy minimization}
Let us recall SSL by entropy minimization in \citet{amini2002semi} and \citet{grandvalet2005semi}. In these  works,  by letting the loss function be the negative log-likelihood $P_{Y|X;\bW}$ under hypothesis $\bW$, the authors considered minimizing the regularized empirical risk as follows: 
\begin{align}
    &L_{\rmE}^{\text{EM}}(\bW;S_\rml,\hatS_\rmu)=\frac{1}{n+m}\bigg(\underbrace{-\sum_{i=1}^n\log P_{Y|X;\bW}(Y_i|X_i;\bW)}_{\text{Empirical risk of $S_\rml$}}
    \underbrace{-\sum_{i=n+1}^{n+m}\sum_{k=1}^K P_{Y|X;\bW}(k|X_i;\bW)\log P_{Y|X;\bW}(k|X_i;\bW)}_{\text{Regularizer: empirical entropy of } S_\rmu} \bigg). \nn
\end{align}
Under the $(\alpha,\pi(\bW),L_{\rmE}^{\text{EM}}(\bW;S_\rml,S_\rmu))$-Gibbs algorithm, the posterior distribution of $\bW$ can be denoted as $P_{\bW|S_\rml,S_\rmu}^\alpha$. The output hypothesis $\bW$ only depends on the labeled data $S_\rml$ and unlabeled data $S_\rmu$ instead of pseudo-labeled data $\hatS_\rmu$.
Since $S_\rml$ and $S_\rmu$ are independent, according to \eqref{Eq:gen indep}, by replacing $\lambda$ with $\frac{m}{n}$, we can characterize the expected gen-error (cf. \eqref{Eq:def gen}) inspired by Theorem~\ref{Thm:gen SSL PL} in the following corollary. 
\begin{corollary}
Under the semi-supervised Gibbs algorithm with entropy minimization (namely, the $(\alpha,\pi(\bW),L_{\rmE}^{\text{EM}}(\bW;S_\rml,S_\rmu))$-Gibbs algorithm), the expected gen-error is
\begin{align}
	\overline{\mathrm{gen}}(P_{\bW|S_\rml,S_\rmu}^\alpha, P_{S_\rml,S_\rmu})=\frac{(n+m)I_{\mathrm{SKL}}(\bW;S_\rml|S_\rmu)}{n\alpha}. \nn
\end{align}
\end{corollary}

\section{Proof of~\cref{Prop: upper and lower bound} }\label{App pf of Prop: upper and lower bound}
We provide the proof of \cref{Prop: upper and lower bound} and a novel lower bound on $\overline{\mathrm{gen}}(P_{\bW|S_\rml,\hatS_\rmu}^\alpha, P_{S_\rml,\hatS_\rmu})$.
\begin{proof}
A {\em $\sigma$-sub-Gaussian random variable}  $L$ is such  that its cumulant generating function $\Lambda_L(s):= \bbE[\exp(s(L-\bbE[L]))]\leq \exp(s^2\sigma^2/2)$ for all $s\in\bbR$ \citep{vershynin2018high}. 

 Assume that the loss function $l(\bW,\bZ)$ is bounded in $[a,b]\subset \bbR_+$ for any $\bW\in\calW$ and $\bZ\in \calZ$. Then we have $\log \Lambda_{\alpha,\eta}(S_\rml,\hatS_\rmu)\in [-\alpha b, -\alpha a]$.
According to Hoeffding's lemma, the loss function $l(\bW,\bZ)$ is $\frac{b-a}{2}$-sub-Gaussian and $\log \Lambda_{\alpha,\eta}(S_\rml,\hatS_\rmu)$ is $\frac{\alpha(b-a)}{2}$-sub-Gaussian. From the Donsker--Varadhan representation, we have
\begin{align}
   & |\bbE_{\Delta_{S_\rml,\hatS_\rmu}}[\log\Lambda_{\alpha,\eta}(S_\rml,\hatS_\rmu)]| 
   \leq\sqrt{\frac{\alpha^2(b-a)^2}{2}I(\hatS_\rmu;S_\rml)},
\end{align}
From \cref{Thm:gen SSL PL}, we can directly obtain
\begin{align}
    \bigg|\overline{\mathrm{gen}}(P_{\bW|S_\rml,\hatS_\rmu}^\alpha, P_{S_\rml,\hatS_\rmu})-\frac{(1+\eta)(I_{\mathrm{SKL}}(\bW,\hatS_\rmu;S_\rml)-I_{\mathrm{SKL}}(\hatS_\rmu;S_\rml))}{\alpha}\bigg| \leq \frac{(1+\eta)(b-a)}{\sqrt{2}}\sqrt{I(\hatS_\rmu;S_\rml)}.
\end{align}
\end{proof}
We can also provide another lower bound on $\overline{\mathrm{gen}}(P_{\bW|S_\rml,\hatS_\rmu}^\alpha, P_{S_\rml,\hatS_\rmu})$. For any  random variable $Z$, from the chain rule of the mutual information, we have $I(X;Y,Z)=I(X;Y)+I(X;Z|Y) \geq I(X;Y)$. 
We can also expand the lautum information $L(X;Y,Z)$ as $L(X;Y,Z)=L(X;Y)+D(P_{Z|Y}\| P_{Z|X,Y}|P_X P_Y) \geq L(X;Y)$.
Thus, we have
\begin{align}\label{Eq:ineq for I_SKL}
    I_{\mathrm{SKL}}(X;Y,Z)\geq I_{\mathrm{SKL}}(X;Y). 
\end{align}
Then the gen-error is lower bounded by
    \begin{align}
        \overline{\mathrm{gen}}(P_{\bW|S_\rml,\hatS_\rmu}^\alpha, P_{S_\rml,\hatS_\rmu}) \geq \frac{1+\eta}{\alpha}\bbE_{\Delta_{S_\rml,\hatS_\rmu}}[\log\Lambda_{\alpha,\eta}(S_\rml,\hatS_\rmu)].
    \end{align}
    The sign of the lower bound depends on the loss function, the prior distribution $\pi(\bW)$ and the data distributions.
    As $\eta\to 0$, or $\eta\to \infty$, or $P_{S_\rml,\hatS_\rmu}=P_{S_\rml}\otimes P_{\hatS_\rmu}$, the lower bound vanishes to $0$.
\section{Proof of \cref{Eq:rewrite I_SKL-I_SKL}}\label{App:rewrite I_SKL-I_SKL}
From the definition of symmetrized KL information in Section \ref{Sec:prob setup}, we have
\begin{align}
    &I_{\mathrm{SKL}}(\bW,\hatS_\rmu;S_\rml)-I_{\mathrm{SKL}}(\hatS_\rmu;S_\rml)\nn\\
    &=I(\bW,\hatS_\rmu;S_\rml)-I(\hatS_\rmu;S_\rml)+L(\bW,\hatS_\rmu;S_\rml)-L(\hatS_\rmu;S_\rml)\\
    &=I(\bW;S_\rml|\hatS_\rmu)+L(\bW,\hatS_\rmu;S_\rml)-L(\hatS_\rmu;S_\rml) \label{Eq:MI chain}\\
    &=I(\bW;S_\rml|\hatS_\rmu)+D(P_{\bW|\hatS_\rmu} \| P_{\bW|S_\rml,\hatS_\rmu} | P_{S_\rml}P_{\hatS_\rmu}), \label{Eq:lautum expand}
\end{align}
where \eqref{Eq:MI chain} follows from the chain rule of mutual information and \eqref{Eq:lautum expand} follows from the expansion of lautum information \citep[Eq. (52)]{palomar2008lautum}.

\section{Proofs for Comparison to SL with $n+m$ Labeled Data}\label{App:proof of Eq:gen_all}
\subsection{Proof of \cref{Eq:gen_all}}
Let $\bar{\bZ}$, $\bar{\bZ}'$ be independent copies of $\bZ_i$ and $\hat{\bZ}_i$. Then $\bar{\bZ}\sim P_{\bZ}$ and  $\bar{\bZ}'\sim P_{\hat{\bZ}}$. Recall that with a perfect pseudo-labeling method, $P_{\bZ}=P_{\hat{\bZ}}$. For $\lambda=\frac{m}{n}$, we have 
	\begin{align}
		&I_{\mathrm{SKL}}(\bW;S_\rml,\hatS_\rmu) =\bbE_{\bW,S_\rml,\hatS_\rmu}[\log 		P_{\bW|S_\rml,\hatS_\rmu}^\alpha]-\bbE_{\bW}\bbE_{S_\rml,\hatS_\rmu}[\log P_{\bW|S_\rml,\hatS_\rmu}^\alpha]\\
		&=\frac{\alpha n}{n+m}\big(\bbE_{\bW}\bbE_{S_\rml,\hatS_\rmu}[L_{\rmE}(\bW,S_\rml)]-\bbE_{\bW,S_\rml,\hatS_\rmu}[L_{\rmE}(\bW,S_\rml)] \big) \nn\\
		&\quad +\frac{\alpha m }{n+m}\big(\bbE_{\bW}\bbE_{S_\rml,\hatS_\rmu}[L_{\rmE}(\bW,\hatS_\rmu)]-\bbE_{\bW,S_\rml^n,\hatS_\rmu}[L_{\rmE}(\bW,\hatS_\rmu)] \big)\\
		&=\frac{\alpha}{n+m}\sum_{i=1}^{n}\big(\bbE_{\bW}\bbE_{\bar{\bZ}}[l(\bW,\bar{\bZ})]-\bbE_{\bW,\bZ_i}[l(\bW,\bZ_i)] \big) \nn\\
		&\quad +\frac{\alpha }{n+m}\sum_{i=n+1}^{n+m}\big(\bbE_{\bW}\bbE_{\bar{\bZ}'}[l(\bW,\bar{\bZ}')]-\bbE_{\bW,\hat{\bZ}_i}[l(\bW,\hat{\bZ}_i)] \big)\\
		&=\alpha \bbE_{\bW}\bbE_{\bar{\bZ}}[l(\bW,\bar{\bZ})]-\frac{\alpha}{n+m}\bigg(\sum_{i=1}^{n}\bbE_{\bW,\bZ_i}[l(\bW,\bZ_i)]  +\sum_{i=n+1}^{n+m}\bbE_{\bW,\hat{\bZ}_i}[l(\bW,\hat{\bZ}_i)] \bigg) \label{Eq: last second line of gen perfect PL}\\
		&=\alpha \overline{\mathrm{gen}}_{\mathrm{all}}(P_{\bW|S_\rml,\hatS_\rmu}^\alpha, P_{S_\rml,\hatS_\rmu}).
	\end{align}
	where \eqref{Eq: last second line of gen perfect PL} follows since $P_{\bZ}=P_{\hat{\bZ}}$.
Thus, \eqref{Eq:gen_all} is proved.

\subsection{Proof for $\overline{\mathrm{gen}}_{\mathrm{all}}$ when the pseudo-labeling is close to perfect}\label{App: gen_all close to perfect}
Without loss of generality, let $S_\rml=\bZ_1=(\bX_1,Y_1)$ and $\hatS_\rmu=\hat{\bZ}_2=(\bX_2,\hatY_2)$. 
With the assumption that $P_{\hat{\bZ}}=P_{\bZ}+\epsilon\Delta$, the joint distributions $P_{\bW,\hat{\bZ}_2}$ and $P_{\bW,\bZ_1}$ are given by
\begin{align}
    P_{\bW,\hat{\bZ}_2}(\cdot,\ast)&=P_{\hat{\bZ}}(\ast)\sum_{\bz}P_{\bZ}(\bz)P_{\bW|\hat{\bZ}_2,\bZ_1}(\cdot|\ast,\bz)=(P_{\bZ}(\ast)+\epsilon\Delta(\ast))\sum_{\bz}P_{\bZ}(\bz)P_{\bW|\hat{\bZ}_2,\bZ_1}(\cdot|\ast,\bz)\\
    P_{\bW,\bZ_1}(\cdot,\ast)&=P_{\bZ}(\ast)\sum_{\bz}P_{\hat{\bZ}}(\bz)P_{\bW|\hat{\bZ}_2,\bZ_1}(\cdot|\bz,\ast)=P_{\bZ}(\ast)\sum_{\bz}(P_{\bZ}(\bz)+\epsilon\Delta(\bz))P_{\bW|\hat{\bZ}_2,\bZ_1}(\cdot|\bz,\ast)\\
    &=P_{\bZ}(\ast)\sum_{\bz}P_{\bZ}(\bz)P_{\bW|\hat{\bZ}_2,\bZ_1}(\cdot|\bz,\ast)+\epsilon P_{\bZ}(\ast)\sum_{\bz}\Delta(\bz)P_{\bW|\hat{\bZ}_2,\bZ_1}(\cdot|\bz,\ast)\\
    &=P_{\bW,\hat{\bZ}_2}(\cdot,\ast)-\epsilon\Delta(\ast)\sum_{\bz}P_{\bZ}(\bz)P_{\bW|\hat{\bZ}_2,\bZ_1}(\cdot|\ast,\bz)+\epsilon P_{\bZ}(\ast)\sum_{\bz}\Delta(\bz)P_{\bW|\hat{\bZ}_2,\bZ_1}(\cdot|\bz,\ast) \label{Eq:symmetry of P_{W|Z_1,Z_2}}\\
    &=P_{\bW,\hat{\bZ}_2}(\cdot,\ast)+\epsilon\Delta'(\cdot,\ast),
\end{align}
where \eqref{Eq:symmetry of P_{W|Z_1,Z_2}} follows since $P_{\bW|\hat{\bZ}_2,\bZ_1}(\cdot|\bz,\ast)=P_{\bW|\hat{\bZ}_2,\bZ_1}(\cdot|\ast,\bz)$, $\Delta'(\cdot,\ast)=-\Delta(\ast)\sum_{\bz}P_{\bZ}(\bz)P_{\bW|\hat{\bZ}_2,\bZ_1}(\cdot|\ast,\bz)+ P_{\bZ}(\ast)\sum_{\bz}\Delta(\bz)P_{\bW|\hat{\bZ}_2,\bZ_1}(\cdot|\bz,\ast)$ and $\sum_{\bw,\bz}\Delta'(\bw,\bz)=0$.

First recall that in the SL setting with $2$ labeled training data samples, the expected gen-error is given by
\begin{align}
    \overline{\mathrm{gen}}(P_{\bW_{\mathrm{SL}}^{(2)}|S_\rml^{(2)}},P_{S_\rml^{(2)}})=\frac{I_{\mathrm{SKL}}(\bW_{\mathrm{SL}}^{(2)};S_\rml^{(2)})}{\alpha}=\bbE_{\bW}\bbE_{\bZ}[l(\bW,\bZ)]-\bbE_{\bW,\bZ_1}[l(\bW,\bZ_1)].
\end{align}
Next, in the SSL setting with this assumption, the expected gen-error is given by
\begin{align}
    &\overline{\mathrm{gen}}_{\mathrm{all}}(P_{\bW|S_\rml,\hatS_\rmu}^\alpha, P_{S_\rml,\hatS_\rmu})\nn\\
    &=\bbE_{\bW}\bbE_{\bZ}[l(\bW,\bZ)]-\frac{1}{2}\big(\bbE_{\bW,\bZ_1}[l(\bW,\bZ_1)]  +\bbE_{\bW,\hat{\bZ}_2}[l(\bW,\hat{\bZ}_2)] \big)\\
    &=\bbE_{\bW}\bbE_{\bZ}[l(\bW,\bZ)]-\bbE_{\bW,\bZ_1}[l(\bW,\bZ_1)]  -\frac{\epsilon}{2}\bbE_{(\bW,\hat{\bZ}_2)\sim \Delta'}[l(\bW,\hat{\bZ}_2)] \\
    &=\frac{I_{\mathrm{SKL}}(\bW_{\mathrm{SL}}^{(2)};S_\rml^{(2)})}{\alpha}-\frac{\epsilon}{2}\bbE_{(\bW,\hat{\bZ}_2)\sim \Delta'}[l(\bW,\hat{\bZ}_2)]\\
    &=\overline{\mathrm{gen}}(P_{\bW_{\mathrm{SL}}^{(2)}|S_\rml^{(2)}},P_{S_\rml^{(2)}})+\Tilde{\epsilon},
\end{align}
where $\tilde{\epsilon}$ is proportional to $\epsilon$ and $\Tilde{\epsilon}\to 0$ as $\epsilon\to 0$. The result can be directly extended to $S_\rml$ with $n$ data samples and $\hatS_\rmu$ with $m$ data samples.

\section{Proof of Proposition \ref{Prop:dis-free ub}}\label{App:pf of Prop:dis-free ub}
The formal version of \cref{Prop:dis-free ub} is stated as follows.
\begin{proposition}[Formal Version]\label{Prop:dis-free ub formal}
    For any $0<\lambda<\infty$, suppose $(I(\bW,\hatS_\rmu;S_\rml)-I(\hatS_\rmu;S_\rml))(1+C_\alpha)\leq  I_{\mathrm{SKL}}(\bW,\hatS_\rmu;S_\rml)-I_{\mathrm{SKL}}(\hatS_\rmu;S_\rml)$ for some constant $C_\alpha \geq 0$ and $I(\hatS_\rmu;S_\rml) =\gamma_{\alpha,\lambda} I(\bW,\hatS_\rmu;S_\rml)$, where $\gamma_{\alpha,\lambda}$ depends on $\lambda$ and $0\leq \gamma_{\alpha,\lambda} \leq 1$.
    If $l(\bW,\bZ)$ is bounded within $[a,b]\subset\bbR_+$ for any $\bW\in\calW$ and $\bZ\in\calZ$, the expected gen-error can be bounded as follows
    \begin{align}
    \frac{-\alpha(b-a)^2\sqrt{\gamma_{\alpha,\lambda}}}{2(1+C_\alpha)(1-\gamma_{\alpha,\lambda})} \bigg(\frac{1}{\sqrt{n}}+(1+\lambda)\sqrt{\gamma_{\alpha,\lambda}} \bigg)&\leq \overline{\mathrm{gen}}(P_{\bW|S_\rml,\hatS_\rmu}^\alpha, P_{S_\rml,\hatS_\rmu})\nn\\
    &\leq \frac{\alpha(b-a)^2\sqrt{\gamma_{\alpha,\lambda}}}{2\sqrt{n}(1+C_\alpha)(1-\gamma_{\alpha,\lambda})} +\frac{\alpha(b-a)^2}{2n(1+\lambda)(1+C_\alpha)(1-\gamma_{\alpha,\lambda})}.
    \end{align}
\end{proposition}
Since $L(\bW,\hatS_\rmu;S_\rml)-L(\hatS_\rmu;S_\rml) =D(P_{\bW|\hatS_\rmu} \| P_{\bW|S_\rml,\hatS_\rmu} | P_{S_\rml}P_{\hatS_\rmu})\geq 0$, 
we always have
\begin{align}
    I(\bW,\hatS_\rmu;S_\rml)-I(\hatS_\rmu;S_\rml)\leq  I(\bW,\hatS_\rmu;S_\rml)-I(\hatS_\rmu;S_\rml)+L(\bW,\hatS_\rmu;S_\rml)-L(\hatS_\rmu;S_\rml)=I_{\mathrm{SKL}}(\bW,\hatS_\rmu;S_\rml)-I_{\mathrm{SKL}}(\hatS_\rmu;S_\rml).
\end{align}
Without loss of generality, we can take $C_\alpha=0$. Recall $\lambda=m/n$. Then we have (i.e., the informal version in \cref{Prop:dis-free ub})
\begin{align}
    &\frac{-\alpha(b-a)^2\sqrt{\gamma_{\alpha,\lambda}}}{2(1-\gamma_{\alpha,\lambda})} \!\bigg(\!\!\frac{1}{\sqrt{n}}\!+\!\frac{n\sqrt{\gamma_{\alpha,\lambda}}}{n+m}\! \bigg)\!\!\leq\! \overline{\mathrm{gen}}(P_{\bW|S_\rml,\hatS_\rmu}^\alpha, P_{S_\rml,\hatS_\rmu}) \leq
    \frac{\alpha(b-a)^2\sqrt{\gamma_{\alpha,\lambda}}}{2\sqrt{n}(1-\gamma_{\alpha,\lambda})}+\frac{\alpha(b-a)^2}{2(n+m)(1-\gamma_{\alpha,\lambda})}.
\end{align}

\begin{proof}
If $l(\bW,\bZ)$ is $\sigma$-sub-Gaussian under $P_{\bZ}$ for every $\bw\in\calW$, from \citep[Theorem 1]{xu2017information}, we have 
\begin{align}
    |\overline{\mathrm{gen}}(P_{\bW|S_\rml,\hatS_\rmu}^\alpha, P_{S_\rml,\hatS_\rmu})|&=|\bbE_{\bW,S_\rml}[L_{\rm P}(\bW,P_{S_\rml})-L_{\rmE}(\bW,S_\rml)]| \nn\\
    &\leq \sqrt{\frac{2\sigma^2 I(\bW;S_\rml)}{n}}\\
    &\leq \sqrt{\frac{2\sigma^2 I(\bW,\hatS_{\rmu};S_\rml)}{n}}.
\end{align}
With the assumption that  the loss function $l(\bW,\bZ)$ is bounded within $[a,b]\subset \bbR_+$ for any $\bW\in\calW$ and $\bZ\in \calZ$, we have $\log \Lambda_{\alpha,\lambda}(S_\rml,\hatS_\rmu)\in [-\alpha b, -\alpha a]$.
According to Hoeffding's lemma, the loss function $l(\bW,\bZ)$ is $\frac{b-a}{2}$-sub-Gaussian and $\log \Lambda_{\alpha,\lambda}(S_\rml,\hatS_\rmu)$ is $\frac{\alpha(b-a)}{2}$-sub-Gaussian.
Thus, from Theorem \ref{Thm:gen SSL PL}, we have
\begin{align}
    &\frac{(1+\lambda)(I_{\mathrm{SKL}}(\bW,\hatS_\rmu;S_\rml)-I_{\mathrm{SKL}}(\hatS_\rmu;S_\rml)+\bbE_{\Delta_{S_\rml,\hatS_\rmu}}[\log\Lambda_{\alpha,\lambda}(S_\rml,\hatS_\rmu)])}{\alpha} \leq \sqrt{\frac{(b-a)^2 I(\bW,\hatS_{\rmu};S_\rml)}{2n}} \\
    \Longrightarrow &\frac{(1+\lambda)(I_{\mathrm{SKL}}(\bW,\hatS_\rmu;S_\rml)-I_{\mathrm{SKL}}(\hatS_\rmu;S_\rml))}{\alpha} \leq \sqrt{\frac{(b-a)^2 I(\bW,\hatS_{\rmu};S_\rml)}{2n}}-\frac{(1+\lambda)\bbE_{\Delta_{S_\rml,\hatS_\rmu}}[\log\Lambda_{\alpha,\lambda}(S_\rml,\hatS_\rmu)]}{\alpha}. \label{Eq: gen distri-free bd 1}
\end{align}
Furthermore, by Donsker--Varadhan representation, $-\bbE_{\Delta_{S_\rml,\hatS_\rmu}}[\log\Lambda_{\alpha,\lambda}(S_\rml,\hatS_\rmu)]$ can be upper bounded as follows
\begin{align}
    -\bbE_{\Delta_{S_\rml,\hatS_\rmu}}[\log\Lambda_{\alpha,\lambda}(S_\rml,\hatS_\rmu)]=\bbE_{S_\rml}\bbE_{\hatS_\rmu}[\log\Lambda_{\alpha,\lambda}(S_\rml,\hatS_\rmu)]-\bbE_{S_\rml,\hatS_\rmu}[\log\Lambda_{\alpha,\lambda}(S_\rml,\hatS_\rmu)]\leq \sqrt{\frac{\alpha^2(b-a)^2I(\hatS_\rmu;S_\rml)}{2} }.
\end{align}
Then \eqref{Eq: gen distri-free bd 1} can be further upper bounded as
\begin{align}
    \frac{(1+\lambda)(I_{\mathrm{SKL}}(\bW,\hatS_\rmu;S_\rml)-I_{\mathrm{SKL}}(\hatS_\rmu;S_\rml))}{\alpha} \leq \sqrt{\frac{(b-a)^2 I(\bW,\hatS_{\rmu};S_\rml)}{2n}}+\sqrt{\frac{(1+\lambda)^2(b-a)^2I(\hatS_\rmu;S_\rml)}{2} }.
\end{align}
Suppose $(I(\bW,\hatS_\rmu;S_\rml)-I(\hatS_\rmu;S_\rml))(1+C_\alpha)\leq  I_{\mathrm{SKL}}(\bW,\hatS_\rmu;S_\rml)-I_{\mathrm{SKL}}(\hatS_\rmu;S_\rml)$ for some constant $C_\alpha \geq 0$ and $I(\hatS_\rmu;S_\rml) =\gamma_{\alpha,\lambda} I(\bW,\hatS_\rmu;S_\rml)$, where $\gamma_{\alpha,\lambda}$ depends on $\lambda$ and $0\leq \gamma_{\alpha,\lambda} \leq 1$. 
Then we have
\begin{align}
    \frac{(1+\lambda)(1+C_\alpha)(I(\bW,\hatS_\rmu;S_\rml)-I(\hatS_\rmu;S_\rml))}{\alpha}&\leq \sqrt{\frac{(b-a)^2 I(\bW,\hatS_{\rmu};S_\rml)}{2n}}+\sqrt{\frac{(1+\lambda)^2(b-a)^2I(\hatS_\rmu;S_\rml)}{2} }\\
    \frac{(1+\lambda)(1+C_\alpha)(1-\gamma_{\alpha,\lambda})I(\bW,\hatS_\rmu;S_\rml)}{\alpha}&\leq \sqrt{\frac{(b-a)^2 I(\bW,\hatS_{\rmu};S_\rml)}{2n}}+\sqrt{\frac{(1+\lambda)^2(b-a)^2\gamma_{\alpha,\lambda}I(\bW,\hatS_\rmu;S_\rml)}{2} }\\
     \sqrt{I(\bW,\hatS_{\rmu};S_\rml)} &\leq \frac{\alpha}{(1+\lambda)(1+C_\alpha)(1-\gamma_{\alpha,\lambda})} \bigg(\sqrt{\frac{(b-a)^2}{2n}}+\frac{(1+\lambda)(b-a)\sqrt{\gamma_{\alpha,\lambda}}}{\sqrt{2}}\bigg).
\end{align}
Thus, we have
\begin{align}
    \overline{\mathrm{gen}}(P_{\bW|S_\rml,\hatS_\rmu}^\alpha, P_{S_\rml,\hatS_\rmu})\leq \frac{\alpha(b-a)^2\sqrt{\gamma_{\alpha,\lambda}}}{2\sqrt{n}(1+C_\alpha)(1-\gamma_{\alpha,\lambda})}+\frac{\alpha(b-a)^2}{2n(1+\lambda)(1+C_\alpha)(1-\gamma_{\alpha,\lambda})}.
\end{align}

On the other hand, in \eqref{Eq: gen UL bd}, since $\text{SKL}\geq 0$, we have
\begin{align}
\overline{\mathrm{gen}}(P_{\bW|S_\rml,\hatS_\rmu}^\alpha, P_{S_\rml,\hatS_\rmu})&\geq -\frac{(1+\lambda)(b-a)}{\sqrt{2}}\sqrt{I(\hatS_\rmu;S_\rml)}\\
&= -\frac{(1+\lambda)(b-a)}{\sqrt{2}}\sqrt{\gamma_{\alpha,\lambda} I(\bW,\hatS_{\rmu};S_\rml)} \\
&\geq \frac{-\alpha(b-a)^2\gamma_{\alpha,\lambda}}{2(1+C_\alpha)(1-\gamma_{\alpha,\lambda})} \bigg(\sqrt{\frac{1}{n}}+(1+\lambda)\sqrt{\gamma_{\alpha,\lambda}} \bigg).
\end{align}
\end{proof}

\section{Proofs of Mean Estimation Example}\label{App:proofs for mean est}
The $(\alpha,\pi(\bW), \barL_{\rmE}(\bW,S_\rml',\hatS_\rmu'))$-Gibbs algorithm is given by the following Gibbs posterior distribution
\begin{align}
	&P_{\bW|S_\rml',\hatS_\rmu'}^\alpha (\bW|(Y_i\bX_i)_{i=1}^n,(\hatY_i\bX_i)_{i=n+1}^{n+m}) \nn\\
	&=\frac{\pi(\bW)}{\Lambda_{\alpha,\lambda}(S_\rml',\hatS_\rmu')} \exp\bigg[-\frac{\alpha}{(1+\lambda)n}\sum_{i=1}^n\big(\bW^\top\bW-2\bW^\top Y_i\bX_i+\bX_i^\top\bX_i \big)  \nn\\
	&\quad \quad  \quad -\frac{\alpha \lambda}{(1+\lambda)m}\sum_{i=n+1}^{n+m}\big(\bW^\top\bW-2\bW^\top\hatY_i\bX_i+\bX_i^\top\bX_i \big) \bigg]\\
	&=\frac{1}{\sqrt{2\pi}\sigma_{\rml,\rmu}} \exp\bigg(-\frac{1}{2\sigma^2_{\rml,\rmu}}\big\| \bW-\bmu_{n,m}\big\|_2^2 \bigg),
\end{align}
where $\sigma^2_{\rml,\rmu}=\frac{1}{2\alpha}$,
\begin{align}
	\bmu_{n,m}=\frac{1}{(1+\lambda)n}\sum_{i=1}^n Y_i\bX_i+\frac{\lambda}{(1+\lambda)m}\sum_{i=n+1}^{n+m} \hatY_i\bX_i,
\end{align}
and
\begin{align}
    \frac{\pi(\bW)}{\Lambda_{\alpha,\lambda}(S_\rml',\hatS_\rmu')}&=\frac{1}{\sqrt{2\pi}\sigma_{\rml,\rmu}}\exp\bigg(-\alpha\bigg(\bmu_{n,m}^\top \bmu_{n,m}-\frac{1}{(1+\lambda)n}\sum_{i=1}^n \bX_i^\top \bX_i-\frac{\lambda}{(1+\lambda)m}\sum_{i=n+1}^{n+m} \bX_i^\top \bX_i \bigg)\bigg) \nn\\
    &=\frac{1}{\sqrt{2\pi}\sigma_{\rml,\rmu}}\exp\bigg(-\alpha\bigg( \frac{1}{(1+\lambda)^2n^2}\sum_{i,j\in[n]^2}\bX_i^\top \bX_i+\frac{\lambda^2}{(1+\lambda)^2 m^2}\sum_{i,j\in[n+1:n+m]^2}\bX_i^\top \bX_i \nn\\
    &\quad + \frac{2\lambda}{(1+\lambda)^2nm}\sum_{i=1}^n\sum_{j=n+1}^{n+m}(Y_i\bX_i)^\top(\hatY_j\bX_j)-\frac{1}{(1+\lambda)n}\sum_{i=1}^n \bX_i^\top \bX_i-\frac{\lambda}{(1+\lambda)m}\sum_{i=n+1}^{n+m} \bX_i^\top \bX_i \bigg)\bigg).
\end{align}
It can be seen that $P_{\bW|S_\rml',\hatS_\rmu'}^\alpha$ is a Gaussian distribution.
Thus, the output hypothesis $\bW$ can be written as
\begin{align}
	\bW&=\frac{1}{(1+\lambda)n}\sum_{i=1}^n Y_i\bX_i+\frac{\lambda}{(1+\lambda)m}\sum_{i=n+1}^{n+m} \hatY_i\bX_i +N \label{Eq:rewrite W} \\
	&= \frac{1}{(1+\lambda)n}\sum_{i=1}^n (Y_i\bX_i-\bmu)+ \frac{\lambda}{(1+\lambda)m}\sum_{i=n+1}^{n+m}( \hatY_i\bX_i-\bmu')+ \frac{\bmu+\lambda \bmu'}{1+\lambda}  +N\\
	&=\bA \bT(S_\rml')+N_G
\end{align}
where $N\sim \calN(0,\sigma_{\rml,\rmu}^2 \bI_d)$ is independent of $(S_\rml,\hatS_\rmu)$, $\bT(S_\rml')=[Y_1\bX_1-\bmu;\ldots;Y_1\bX_1-\bmu]\in\bbR^{nd\times 1}$, $\bA=[\frac{1}{(1+\lambda)n}\bI_d,\ldots,\frac{1}{(1+\lambda)n}\bI_d]\in\bbR^{d\times nd}$, $N_G|\hatS_\rmu \sim \calN(\bmu_{N_G},\sigma^2_{\rml,\rmu}\bI_d)$ and $\bmu_{N_G}=\frac{\lambda}{(1+\lambda)m}\sum_{i=n+1}^{n+m}( \hatY_i\bX_i-\bmu')+ \frac{\bmu+\lambda \bmu'}{1+\lambda}$.

First, let us calculate the expected gen-error according to \eqref{Eq: gen main thm} in Theorem \ref{Thm:gen SSL PL}. Let $\bT=\bT(S_\rml')$ for simplicity. We have
\begin{align}
    &I_{\mathrm{SKL}}(\bW,\hatS_\rmu;S_\rml)-I_{\mathrm{SKL}}(\hatS_\rmu;S_\rml)\nn\\
    &=I(\bW;S_\rml|\hatS_\rmu)+D(P_{\bW|\hatS_\rmu} \| P_{\bW|S_\rml,\hatS_\rmu} | P_{S_\rml}P_{\hatS_\rmu}) \\
    &=\bbE_{S_\rml,\hatS_\rmu}\bbE_{\bW|S_\rml,\hatS_\rmu}[\log P_{\bW|S_\rml',\hatS_\rmu'}^\alpha]-\bbE_{S_\rml}\bbE_{\hatS_\rmu}\bbE_{\bW|\hatS_\rmu}[\log P_{\bW|S_\rml',\hatS_\rmu'}^\alpha]\\
    &=\frac{1}{2\sigma_{\rml,\rmu}^2}\bigg(\bbE_{S_\rml,\hatS_\rmu}\bbE_{\bW|S_\rml,\hatS_\rmu}[-(\bW-\bmu_{n,m})^\top(\bW-\bmu_{n,m})]+\bbE_{S_\rml}\bbE_{\hatS_\rmu}\bbE_{\bW|\hatS_\rmu}[(\bW-\bmu_{n,m})^\top(\bW-\bmu_{n,m})] \bigg)\\
    &=\frac{1}{2\sigma_{\rml,\rmu}^2}\bigg(\bbE_{S_\rml,\hatS_\rmu}\bbE_{\bW|S_\rml,\hatS_\rmu}[\bmu_{n,m}^\top \bmu_{n,m}]+\bbE_{S_\rml}\bbE_{\hatS_\rmu}\bbE_{\bW|\hatS_\rmu}[-2\bW^\top \bmu_{n,m}+\bmu_{n,m}^\top \bmu_{n,m}] \bigg)\\
    &=\frac{1}{2\sigma_{\rml,\rmu}^2}\bigg(\bbE_{S_\rml,\hatS_\rmu}[(\bA \bT)^\top (\bA \bT)+2(\bA\bT)^\top\bmu_{N_G}+\bmu_{N_G}^\top\bmu_{N_G}]\\
    &\quad +\bbE_{S_\rml}\bbE_{\hatS_\rmu}[-2\bbE_{S_\rml|\hatS_\rmu}[\bA\bT]^\top (\bA\bT+\bmu_{N_G})-2\bmu_{N_G}^\top (\bA\bT+\bmu_{N_G})+(\bA\bT+\bmu_{N_G})^\top(\bA\bT+\bmu_{N_G})] \bigg)\\
    &\overset{\text{(a)}}{=}\frac{1}{\sigma_{\rml,\rmu}^2}\bbE[(\bA\bT)^\top(\bA\bT)]\\
    &=\frac{1}{\sigma_{\rml,\rmu}^2}\tr(\bA^\top\bA \bbE[\bT\bT^\top] )\\
    &=\frac{2\alpha \sigma^2 d}{(1+\lambda)^2n} \label{Eq:mean set I-I},
\end{align}
where (a) follows since $\bbE_{S_\rml'}[\bA\bT(S_\rml')]=0$, and
\begin{align}
    &\bbE_{\Delta_{S_\rml',\hatS_\rmu'}}[\log \Lambda_{\alpha,\lambda}(S_\rml',\hatS_\rmu')] \nn\\
    &=\alpha\bbE_{\Delta_{S_\rml',\hatS_\rmu'}}\bigg[ \frac{1}{(1+\lambda)^2n^2}\sum_{i,j\in[n]^2}\bX_i^\top \bX_i+\frac{\lambda^2}{(1+\lambda)^2 m^2}\sum_{i,j\in[n+1:n+m]^2}\bX_i^\top \bX_i \nn\\
    &\quad + \frac{2\lambda}{(1+\lambda)^2nm}\sum_{i=1}^n\sum_{j=n+1}^{n+m}(Y_i\bX_i)^\top(\hatY_j\bX_j)-\frac{1}{(1+\lambda)n}\sum_{i=1}^n \bX_i^\top \bX_i-\frac{\lambda}{(1+\lambda)m}\sum_{i=n+1}^{n+m} \bX_i^\top \bX_i \bigg]\\
    &=\alpha\bbE_{\Delta_{S_\rml',\hatS_\rmu'}}\bigg[ \frac{2\lambda}{(1+\lambda)^2nm}\sum_{i=1}^n\sum_{j=n+1}^{n+m}(Y_i\bX_i)^\top(\hatY_j\bX_j)\bigg]\\
    &= \frac{2\alpha\lambda}{(1+\lambda)^2nm}\sum_{i=1}^n\sum_{j=n+1}^{n+m}\big(\bbE_{S_\rml',\hatS_\rmu'}\big[ (Y_i\bX_i-\bmu)^\top(\hatY_j\bX_j-\bmu')\big]+\bmu^\top\bmu'-\bmu^\top\bmu \big)\\
    &\overset{\text{(b)}}{=} \frac{2\alpha\lambda}{(1+\lambda)^2}\bbE\big[ (Y_i\bX_i-\bmu)^\top(\hatY_j\bX_j-\bmu')\big], \label{Eq:mean est E[Lambda]}
\end{align}
where (b) follows since $\bbE[ (Y_i\bX_i-\bmu)^\top(\hatY_j\bX_j-\bmu')]$ is symmetric for any $i\in[n]$ and $j\in[n+1:n+m]$. By letting $\lambda=\frac{m}{n}$ and combining \eqref{Eq:mean set I-I} and \eqref{Eq:mean est E[Lambda]}, the expected gen-error of this example is given by
\begin{align}
    \overline{\mathrm{gen}}(P_{\bW|S_\rml,\hatS_\rmu}^\alpha, P_{S_\rml,\hatS_\rmu})=\frac{2 \sigma^2 d}{n+m}+\frac{2m}{n+m}\bbE\big[ (Y_i\bX_i-\bmu)^\top(\hatY_j\bX_j-\bmu')\big].
\end{align}

From the definition of gen-error in \eqref{Eq:def gen}, we can also derive the same result as follows.
Let $\barY\bar{\bX}$ be an independent copy of $Y_i\bX_i$ for any $i\in[n]$. 
\begin{align}
	&\overline{\mathrm{gen}}(P_{\bW|S_\rml,\hatS_\rmu}^\alpha, P_{S_\rml,\hatS_\rmu}) \nn\\
	&=\bbE_{\bW}\bbE_{\barY\bar{\bX}}[\|\barY\bar{\bX}-\bW\|_2^2]-\frac{1}{n}\sum_{i=1}^n\bbE_{\bW,Y_i\bX_i}[\|Y_i\bX_i-\bW\|_2^2]\\
	&=\frac{1}{n}\sum_{i=1}^n \bbE[2\bW^\top (Y_i\bX_i-\barY\bar{\bX}) -\bX_i^\top\bX_i+\bar{\bX}^\top\bar{\bX} ]  \\
	&\overset{\text{(c)}}{=}\frac{1}{n}\sum_{i=1}^n \bbE\big[2(\bmu_{n,m}+N)^\top (Y_i\bX_i-\barY\bar{\bX}) \big] \\
	&=\frac{1}{n}\sum_{i=1}^n \bbE\big[2\bmu_{n,m}^\top (Y_i\bX_i-\barY\bar{\bX}) \big] \\
	&=\frac{1}{n}\sum_{i=1}^n 2\bbE\bigg[\bigg(\frac{1}{n+m}\sum_{j=1}^n (Y_j\bX_j-\bmu)+ \frac{1}{n+m}\sum_{j=n+1}^{n+m}( \hatY_j\bX_j-\bmu')+ \frac{\bmu+\lambda \bmu'}{1+\lambda} \bigg)^\top \!\! \big((Y_i\bX_i-\bmu)-(\barY\bar{\bX}-\bmu) \big) \bigg] \\
	&=\frac{1}{n}\sum_{i=1}^n \bigg( \frac{2}{n+m}\bbE[(Y_i\bX_i-\bmu)^\top(Y_i\bX_i-\bmu)]+\frac{2}{n+m}\sum_{j=n+1}^{n+m}\bbE[(\hatY_j\bX_j-\bmu')^\top (Y_i\bX_i-\bmu)] \bigg)\\
	&=\frac{2 \sigma^2 d}{n+m}+\frac{1}{n}\sum_{i=1}^n\frac{2}{n+m}\sum_{j=n+1}^{n+m}\bbE[(\hatY_j\bX_j-\bmu')^\top (Y_i\bX_i-\bmu)]\\
	&\overset{\text{(d)}}{=}\frac{2 \sigma^2 d}{n+m}+\frac{2m}{n+m}\bbE[(\hatY_j\bX_j-\bmu')^\top (Y_i\bX_i-\bmu)], 
\end{align}
where (c) follows from \eqref{Eq:rewrite W}, (d) follows since $\bbE[(\hatY_j\bX_j-\bmu')^\top (Y_i\bX_i-\bmu)]$ is symmetric for any $i\in[n]$ and $j\in[n+1:n+m]$.

\subsection{Mean Estimation under Supervised Learning}\label{App:mean est gen SL}
Under the supervised $(\alpha,\pi(\bW_{\mathrm{SL}}^{(n)}), L_{\rmE}(\bW_{\mathrm{SL}}^{(n)},S_\rml'))$-Gibbs algorithm,  the posterior distribution $P_{\bW_{\mathrm{SL}}^{(n)}|S_\rml' }$ is given by
\begin{align}
    P_{\bW_{\mathrm{SL}}^{(n)}|S_\rml' }= \calN \bigg(\frac{1}{n}\sum_{i=1}^n Y_i\bX_i, \sigma_{\rml,\rmu}^2\bI_d \bigg).
\end{align}
According to \citep[Theorem 1]{aminian2021exact}, with the similar techniques of obtaining \eqref{Eq:mean set I-I},  the expected gen-error is given by
\begin{align}
    \overline{\mathrm{gen}}(P_{\bW_{\mathrm{SL}}^{(n)}|S_\rml'}^{\alpha}, P_{S_\rml'})=\frac{I_{\mathrm{SKL}}(\bW_{\mathrm{SL}}^{(n)};S_\rml')}{\alpha}=2\tr(\bA_1\bbE[\bT(S_\rml')\bT(S_\rml')^\top]\bA_1^\top)=\frac{2\sigma^2 d}{n}.
\end{align}
where $\bT(S_\rml')=[Y_1\bX_1-\bmu;\ldots;Y_1\bX_1-\bmu]\in\bbR^{nd\times 1}$ and $\bA_1=[\frac{1}{n}\bI_d,\ldots,\frac{1}{n}\bI_d]\in\bbR^{d\times nd}$.

Similarly, under the supervised $(\alpha,\pi(\bW_{\mathrm{SL}}^{(n+m)}), L_{\rmE}(\bW_{\mathrm{SL}}^{(n+m)},S_\rml'^{(n+m)}))$-Gibbs algorithm, where $S_\rml'^{(n+m)}$ contains $n+m$ i.i.d.\ $Y_i\bX_i$ samples and $L_{\rmE}(\bW_{\mathrm{SL}}^{(n+m)},S_\rml'^{(n+m)})=\frac{1}{n+m}\sum_{i=1}^{n+m}l(\bW_{\mathrm{SL}}^{(n+m)},\bZ'_i)$, the expected gen-error (cf. \eqref{Eq:gen n add m label}) is given by
\begin{align}
    \overline{\mathrm{gen}}(P_{\bW_{\mathrm{SL}}^{(n+m)}|S_\rml'^{(n+m)}}^{\alpha}, P_{S_\rml'^{(n+m)}})=\frac{I_{\mathrm{SKL}}(\bW_{\mathrm{SL}}^{(n+m)};S_\rml'^{(n
    +m)})}{\alpha}=\frac{2\sigma^2 d}{n+m}.
\end{align}

\subsection{Proofs for \cref{Eq: mean est gen rewrite}}\label{App: pf for Eq: mean est gen rewrite}

Let us rewrite the gen-error in \eqref{Eq: mean est gen} as follows: for any $i\in[n]$ and $j\in[n+1:n+m]$.
\begin{align}
    &\overline{\mathrm{gen}}(P_{\bW|S_\rml,\hatS_\rmu}^\alpha, P_{S_\rml,\hatS_\rmu})=\frac{2 \sigma^2 d}{n+m}+\frac{2m}{n+m}\bbE\big[ (Y_i\bX_i-\bmu)^\top(\hatY_j\bX_j-\bmu')\big]\nn\\
    &=\frac{2 \sigma^2 d}{n+m}+\frac{2m}{n+m}\bbE\big[ (Y_i\bX_i-\bmu)^\top(\sgn(\bW_0^\top\bX_j)\bX_j-\bmu')\big]\\
    &=\frac{2 \sigma^2 d}{n+m}+\frac{2m}{n+m}\big(\bbE\big[ (\sgn(\bW_0^\top\bX_j)\bX_j^\top Y_i\bX_i)\big]-\bbE\big[(\sgn(\bW_0^\top\bX_j)\bX_j^\top)\big]\bmu \big). \label{Eq: mean est gen 1}
\end{align}

Let $\bW_0=\frac{1}{n}\sum_{i=1}^nY_i\bX_i \sim \calN(\bmu,\frac{\sigma^2}{n}\bI_d )$. Using the proof idea from \citep{he2022information}, we can decompose it as
\begin{align}
	\bW_0=\bmu+\frac{\sigma}{\sqrt{n}}\bxi=\bigg(1+\frac{\sigma}{\sqrt{n}}\xi_0\bigg)\bmu+\frac{\sigma}{\sqrt{n}}\bmu^\perp,
\end{align}
where $\bxi\sim\calN(\textbf{0},\bI_d)$, $\xi_0\sim\calN(0,1)$, $\bmu^\perp \sim \calN(\textbf{0},\bI_d-\bmu\bmu^\top)$ and $\bmu^\perp$ is perpendicular to $\bmu$ and independent of $\xi_0$. 
The normalized $\bW_0$ can be written as
\begin{align}
	\overline{\bW}_0=\frac{\bW_0}{\|\bW_0\|_2}=\gamma_n\bmu+\bar{\gamma}_n\bup
\end{align}
where $\bup=\bmu^\perp/\|\bmu^\perp\|_2$, $\gamma_n^2+\bar{\gamma}_n^2=1$ and
\begin{align}
	\gamma_n=\gamma_n(\xi_0,\bmu^\perp):=\frac{1+\frac{\sigma}{\sqrt{n}}\xi_0}{\sqrt{(1+\frac{\sigma}{\sqrt{n}}\xi_0)^2+\frac{\sigma^2}{n}\|\bmu^\perp\|_2^2}}.
\end{align}
For any $i\in[n+m]$, since $Y_i\bX_i\sim \calN(\bmu,\sigma^2\bI_d)$, we have
\begin{align}
	Y_i\bX_i=\bmu+\sigma\bg_i=\bmu+\tilg_i\bmu+\bmu_i^\perp,
\end{align}
where $\bg_i\sim\calN(\textbf{0},\bI_d)$, $\tilg_i\sim \calN(0,1)$, $\bmu_i^\perp\sim\calN(\textbf{0},\bI_d-\bmu\bmu^\top)$ and $\tilg_i$ is independent of $\bmu_i^\perp$.
Given any $Y_i\bX_i$ for $i\in[1:n]$, we have
\begin{align}
	\bW_0|Y_i\bX_i&=\frac{1}{n}Y_i\bX_i+\frac{n-1}{n}\bmu+\frac{\sqrt{n-1}}{n}\sigma\bxi' \\
	&=\frac{1}{n}(\bmu+\sigma \bg_i)+\frac{n-1}{n}\bmu+\frac{\sqrt{n-1}}{n}\sigma(\xi_0'\bmu+\bmu'^\perp)\\
	&=\bigg(1+\frac{\sqrt{n-1}}{n}\sigma\xi_0'+\frac{\sigma}{n}\tilg_i\bigg)\bmu+\bigg(\frac{\sqrt{n-1}}{n}\sigma\|\bmu'^\perp\|_2+\frac{\sigma}{n}\|\bmu_i^\perp\|_2 \bigg)\bup,
\end{align}
where $\bxi'\sim\calN(\textbf{0},\bI_d)$, $\xi_0'\sim\calN(0,1)$, $\bmu'^\perp\sim\calN(\textbf{0},\bI_d-\bmu\bmu^\top)$ and $\bmu'^\perp$ is perpendicular to $\bmu$ and independent of $\xi_0'$. 
The normalized version is given by
\begin{align}
	\overline{\bW}_0|Y_i\bX_i=\frac{\bW_0}{\|\bW_0\|_2} \big|Y_i\bX_i =\gamma_n'\bmu+\bar{\gamma}_n'\bup
\end{align}
where $\gamma_n'^2+\bar{\gamma}_n'^2=1$ and 
\begin{align}
	\gamma_n'=\gamma_n'(\xi_0',\bmu'^\perp,\tilg_i,\bmu_i^\perp):=\frac{1+\frac{\sqrt{n-1}}{n}\sigma\xi_0'+\frac{\sigma}{n}\tilg_i}{\sqrt{(1+\frac{\sqrt{n-1}}{n}\sigma\xi_0'+\frac{\sigma}{n}\tilg_i )^2+(\frac{\sqrt{n-1}}{n}\sigma\|\bmu'^\perp\|_2+\frac{\sigma}{n}\|\bmu_i^\perp\|_2)^2}}.
\end{align}

Define the correlation evolution function $F_{\sigma}:[-1,1]\to [-1,1]$:
\begin{align}
	F_{\sigma}(x):=J_{\sigma}(x)/\sqrt{J_{\sigma}(x)^2+K_{\sigma}(x)^2}, 
\end{align}
where $J_{\sigma}(x):=1-2Q(\frac{x}{\sigma})+\frac{2\sigma x}{\sqrt{2\pi}}\exp(-\frac{x^2}{2\sigma^2})$ and $K_{\sigma}:=\frac{2\sigma\sqrt{1-x^2}}{\sqrt{2\pi}}\exp(-\frac{x^2}{2\sigma^2})$.

For any $j\in[n+1:n+m]$, we decompose the Gaussian random vector $\bg_j\sim\calN(0,\bI_d)$ in another way
	\begin{align}
		\bg_j=\tilg_j \overline{\bW}_0+\tilde{\bg}_j^{\bot}, \label{Eq:new decomp Gaussian vec}
	\end{align}
	where $\tilg_j\sim\calN(0,1)$,  $\tilde{\bg}_j^{\bot}\sim\calN(0,\bI_d-\bar{\bW}_0\bar{\bW}_0^\top)$, $\tilg_j$ and $\tilde{\bg}_j^{\bot}$ are mutually independent and $\tilde{\bg}_j^{\bot} \perp \bar{\bW}_0$.  Then we decompose $\bX_j$ and $\bar{\bW}_0^\top\bX_j$ as
	\begin{align}
		\bX_j&=Y_j\bmu+\sigma\tilg_j \overline{\bW}_0+\sigma\tilde{\bg}_j^{\bot}, \text{~and} \label{Eq:X new decomp} \\ 
		\overline{\bW}_0^\top\bX_j&=Y_j\gamma_n+\sigma \tilg_j. \label{Eq:theta*X new decomp}
	\end{align}		
	Then we have
	\begin{align}
		&\bbE[\sgn(\bar{\bW}_0^\top \bX_j)\bX_j ~|~ \xi_0,\bmu^{\bot}, Y_j=-1] \nn\\
		&=-\bbE[\sgn(-\gamma_n+\sigma \tilg_j)| \xi_0,\bmu^{\bot}]\bmu+ \sigma\bbE[\sgn(-\gamma_n+\sigma \tilg_j)\tilg_j|\xi_0,\bmu^{\bot}]\overline{\bW}_0 \label{Eq:simplify E[theta_1]},
	\end{align}
	where \eqref{Eq:simplify E[theta_1]} follows since $\tilde{\bg}^{\bot}$ is independent of $\tilg_j$ and $\bbE[\tilde{\bg}^{\bot}]=0$.
	Since $\tilg_j\sim \calN(0,1)$, we have
		\begin{align}
		\bbE\big[\sgn\big(\overline{\bW}_0^\top \bX'_j\big)\bX'_j ~|~\xi_0,\bmu^{\bot}, Y_j'=-1 \big]  
		&=\bigg(1-2\rmQ \bigg(\frac{\gamma_n}{\sigma}\bigg) \bigg)\bmu+\frac{2\sigma}{\sqrt{2\pi}}\exp\bigg(-\frac{\gamma_n^2}{2\sigma^2}\bigg)\overline{\bW}_0,
	\end{align}
	and similarly, 
	\begin{align}
		\bbE\big[\sgn\big(\overline{\bW}_0^\top \bX'_j\big)\bX'_j ~|~\xi_0,\bmu^{\bot}, Y_j'=1 \big]  
		&=\bigg(2\rmQ \bigg(-\frac{\gamma_n}{\sigma}\bigg)-1 \bigg)\bmu+\frac{2\sigma}{\sqrt{2\pi}}\exp\bigg(-\frac{\gamma_n^2}{2\sigma^2}\bigg)\overline{\bW}_0.
	\end{align}
Recall the definitions of $J_\sigma$ and $K_\sigma$. Then we have
\begin{align}
	&\bbE[\sgn(\bW_0^\top\bX_j)\bX_j^\top]\bmu=\bbE[\sgn(\overline{\bW}_0^\top\bX_j)\bX_j^\top]\bmu
	=\bbE_{\xi_0,\bmu^\perp}[J_{\sigma}(\gamma_n(\xi_0,\bmu^\perp))] \label{Eq:expand gen example term 1}
\end{align}
and similarly
\begin{align}
	&\bbE[\sgn(\bW_0^\top\bX_j)\bX_j^\top Y_i\bX_i]=\bbE[\sgn(\overline{\bW}_0^\top\bX_j)\bX_j^\top Y_i\bX_i] \nn\\
	&=\bbE\big[\bbE[\sgn(\overline{\bW}_0^\top\bX_j)\bX_j^\top|Y_i\bX_i]  Y_i\bX_i \big]\\
	&=\bbE_{\tilg_i,\bmu_i^\perp }\big[((1+\tilg_i)\bmu+\bmu_i^\perp)^\top \bbE_{\xi_0',\bmu'^\perp}[J_{\sigma}(\gamma_n'(\xi_0',\bmu'^\perp,\tilg_i,\bmu_i^\perp))\bmu+K_{\sigma}(\gamma_n'(\xi_0',\bmu'^\perp,\tilg_i,\bmu_i^\perp))\bup |\tilg_i,\bmu_i^\perp ] \big]\\
	&=\bbE_{\tilg_i,\bmu_i^\perp }\Big[(1+\tilg_i)\bbE_{\xi_0',\bmu'^\perp}\big[J_{\sigma}(\gamma_n'(\xi_0',\bmu'^\perp,\tilg_i,\bmu_i^\perp)) | \tilg_i,\bmu_i^\perp \big] +\|\bmu_i^\perp\|_2\bbE_{\xi_0',\bmu'^\perp}\big[K_{\sigma}(\gamma_n'(\xi_0',\bmu'^\perp,\tilg_i,\bmu_i^\perp)) | \tilg_i,\bmu_i^\perp \big] \Big]\\
	&=\bbE\Big[(1+\tilg_i)J_{\sigma}(\gamma_n'(\xi_0',\bmu'^\perp,\tilg_i,\bmu_i^\perp))+\|\bmu_i^\perp\|_2K_{\sigma}(\gamma_n'(\xi_0',\bmu'^\perp,\tilg_i,\bmu_i^\perp)) \Big].  \label{Eq:expand gen example term 2}
\end{align}
By plugging \eqref{Eq:expand gen example term 1} and \eqref{Eq:expand gen example term 2} back to \eqref{Eq: mean est gen 1}, we have
\begin{align}
	&\overline{\mathrm{gen}}(P_{\bW|S_\rml,\hatS_\rmu}^\alpha, P_{S_\rml,\hatS_\rmu})\nn\\
	&=\frac{2 \sigma^2 d}{n+m} +\frac{2m}{n+m} \bigg(\bbE\Big[(1+\tilg_i)J_{\sigma}(\gamma_n'(\xi_0',\bmu'^\perp,\tilg_i,\bmu_i^\perp))+\|\bmu_i^\perp\|_2K_{\sigma}(\gamma_n'(\xi_0',\bmu'^\perp,\tilg_i,\bmu_i^\perp)) \Big]-\bbE\big[J_{\sigma}(\gamma_n(\xi_0,\bmu^\perp)) \big] \bigg)\\
	&=\frac{2 \sigma^2 d}{n+m} +\frac{2m}{n+m}\bigg( \underbrace{\bbE\Big[\tilg_i J_{\sigma}(\gamma_n'(\xi_0',\bmu'^\perp,\tilg_i,\bmu_i^\perp))+\|\bmu_i^\perp\|_2K_{\sigma}(\gamma_n'(\xi_0',\bmu'^\perp,\tilg_i,\bmu_i^\perp)) \Big]}_{E_n} \bigg), \label{Eq:EJ4=EJ2}
\end{align}
where \eqref{Eq:EJ4=EJ2} follows since $\bbE[J_{\sigma}(\gamma_n'(\xi_0',\bmu'^\perp,\tilg_i,\bmu_i^\perp))]=\bbE[J_{\sigma}(\gamma_n(\xi_0,\bmu^\perp)) ]$. Since $J_{\sigma}(x)\in[J_{\sigma}(-1),J_{\sigma}(1)]$ and $K_{\sigma}(x)\in[0,\frac{2\sigma}{\sqrt{2\pi}}]$ for any $x\in[-1,1]$, we have that $E_n=O(d)$. 

\begin{remark}[Sign of $E_n$]
	First, since $\|\mu_i^\bot\|_2 \geq 0$ and $K_{\sigma}(\gamma_n'(\xi_0',\bmu'^\perp,\tilg_i,\bmu_i^\perp)\geq 0$, we have
	\begin{align}
		\bbE\big[\|\bmu_i^\perp\|_2K_{\sigma}(\gamma_n'(\xi_0',\bmu'^\perp,\tilg_i,\bmu_i^\perp)) \big]\geq 0.
	\end{align}
	Next, for some constant $g\geq 0$, by fixing $\xi_0',\bmu'^\perp,\bmu_i^\perp$, we have
	\begin{align}
		\gamma_n'(\xi_0',\bmu'^\perp,g,\bmu_i^\perp)\geq\gamma_n'(\xi_0',\bmu'^\perp,-g,\bmu_i^\perp).
	\end{align}
	Since $J_{\sigma}(x)$ is an odd increasing function on $[-1,1]$, we have
	\begin{align}
		J_{\sigma}( \gamma_n'(\xi_0',\bmu'^\perp,g,\bmu_i^\perp)) \geq J_{\sigma}( \gamma_n'(\xi_0',\bmu'^\perp,-g,\bmu_i^\perp))
	\end{align}
	and 
	\begin{align}
		\bbE_{\xi_0',\bmu'^\perp,\bmu_i^\perp}[J_{\sigma}( \gamma_n'(\xi_0',\bmu'^\perp,g,\bmu_i^\perp))] \geq \bbE_{\xi_0',\bmu'^\perp,\bmu_i^\perp}[J_{\sigma}( \gamma_n'(\xi_0',\bmu'^\perp,-g,\bmu_i^\perp))].
	\end{align}
	Thus, by recalling that $\tilg_i\sim \calN(0,1)$, we can deduce that
	\begin{align}
		\bbE[\tilg_i J_{\sigma}(\gamma_n'(\xi_0',\bmu'^\perp,\tilg_i,\bmu_i^\perp))]\geq 0.
	\end{align}
	In conclusion, $E_n\geq 0$.
\end{remark}

\section{Proof of \cref{Thm: gen SSL asym}}\label{App: pf of Thm: gen SSL asym}
In this case, there exists a unique minimizer of the empirical risk, i.e.,
\begin{align}
	\bW^*(S_\rml,\hatS_\rmu)=\argmin_{\bw\in\calW} \bigg(\frac{1}{1+\lambda}L_{\rmE}(\bw,S_\rml)+\frac{\lambda}{1+\lambda}L_{\rmE}(\bw,\hatS_\rmu)\bigg).
\end{align}

According to \citep{athreya2010gibbs}, if the following Hessian matrix
\begin{align}
	H^*(S_\rml,\hatS_u)=\nabla_{\bw}^2\bigg(\frac{1}{1+\lambda}L_{\rmE}(\bw,S_\rml)+\frac{\lambda}{1+\lambda}L_{\rmE}(\bw,\hatS_\rmu)\bigg) \bigg|_{\bw=\bW^*(S_\rml,\hatS_\rmu)}
\end{align}
is not singular, then as $\alpha\to\infty$
\begin{align}
	P_{\bW|S_\rml,\hatS_\rmu}^\alpha \xrightarrow{\rmd} \calN\bigg(\bW^*(S_\rml,\hatS_\rmu), \frac{1}{\alpha}H^*(S_\rml,\hatS_u)^{-1} \bigg)
\end{align}
and 
\begin{align}
    \sqrt{\det\bigg(\frac{\alpha H^*(S_\rml,\hatS_\rmu)}{2} \bigg)} e^{\alpha \barL_{\rmE}(\bW^*(S_\rml,\hatS_\rmu),S_\rml,\hatS_\rmu)} \Lambda_{\alpha,\lambda}(S_\rml,\hatS_\rmu) \to \sqrt{\pi^d}. \label{Eq:Lambda asymptotics}
\end{align}
Then we have
\begin{align}
	\bbE_{\bW|S_\rml,\hatS_{\rmu}}[\bW]=\bW^*(S_\rml,\hatS_{\rmu}) \text{~and~} \bbE_{\bW|\hatS_\rmu}[\bW]=\bbE_{S_\rml|\hatS_{\rmu}}[\bW^*(S_\rml,\hatS_{\rmu})].
\end{align}
By applying Theorem \ref{Thm:gen SSL PL}, we use the Gaussian approximation to simplify the symmetrized KL information as follows
\begin{align}
	&I_{\mathrm{SKL}}(\bW,\hatS_\rmu;S_\rml)-I_{\mathrm{SKL}}(\hatS_\rmu;S_\rml) \nn\\
	&=\bbE_{\bW,\hatS_\rmu,S_\rml}[\log P_{\bW|S_\rml,\hatS_\rmu}^\alpha]-\bbE_{\bW,\hatS_\rmu}\bbE_{S_\rml}[\log P_{\bW|S_\rml,\hatS_\rmu}^\alpha]\\
	&=\bbE_{\bW,\hatS_\rmu,S_\rml}\bigg[-\frac{\alpha}{2}(\bW-\bW^*(S_\rml,\hatS_\rmu))^\top H^*(S_\rml,\hatS_\rmu)(\bW-\bW^*(S_\rml,\hatS_\rmu)) \bigg] \nn\\
	&\quad -\bbE_{\bW,\hatS_\rmu}\bbE_{S_\rml}\bigg[-\frac{\alpha}{2}(\bW-\bW^*(S_\rml,\hatS_\rmu))^\top H^*(S_\rml,\hatS_\rmu)(\bW-\bW^*(S_\rml,\hatS_\rmu)) \bigg] \nn\\
	&\quad +\bbE_{S_\rml,\hatS_\rmu}\bigg[\log\frac{ \sqrt{\det(\alpha H^*(S_\rml,\hatS_\rmu))}}{\sqrt{2\pi}} \bigg]-\bbE_{S_\rml}\bbE_{\hatS_\rmu}\bigg[\log\frac{ \sqrt{\det(\alpha H^*(S_\rml,\hatS_\rmu))}}{\sqrt{2\pi}} \bigg] \\
	&=\bbE_{\bW,\hatS_\rmu,S_\rml}\bigg[-\frac{\alpha}{2}\bW^\top H^*(S_\rml,\hatS_\rmu)\bW \bigg]+\bbE_{\bW,\hatS_\rmu}\bbE_{S_\rml}\bigg[\frac{\alpha}{2}\bW^\top H^*(S_\rml,\hatS_\rmu)\bW \bigg] \nn\\
	&\quad +\bbE_{\bW,\hatS_\rmu,S_\rml}\bigg[\frac{\alpha}{2}\tr\Big(H^*(S_\rml,\hatS_\rmu)(\bW^*(S_\rml,\hatS_\rmu)\bW^\top+\bW\bW^*(S_\rml,\hatS_\rmu)^\top-\bW^*(S_\rml,\hatS_\rmu)\bW^*(S_\rml,\hatS_\rmu)^\top  ) \Big) \bigg] \nn\\
	&\quad -\bbE_{\bW,\hatS_\rmu}\bbE_{S_\rml}\bigg[\frac{\alpha}{2}\tr\Big(H^*(S_\rml,\hatS_\rmu)(\bW^*(S_\rml,\hatS_\rmu)\bW^\top+\bW\bW^*(S_\rml,\hatS_\rmu)^\top-\bW^*(S_\rml,\hatS_\rmu)\bW^*(S_\rml,\hatS_\rmu)^\top  ) \Big) \bigg] \nn\\
	&\quad +\bbE_{S_\rml,\hatS_\rmu}\big[\log\sqrt{\det(H^*(S_\rml,\hatS_\rmu))} \big]-\bbE_{S_\rml}\bbE_{\hatS_\rmu}\big[\log \sqrt{\det(H^*(S_\rml,\hatS_\rmu))} \big] \\
	&=\bbE_{\bW,\hatS_\rmu,S_\rml}\bigg[-\frac{\alpha}{2}\bW^\top H^*(S_\rml,\hatS_\rmu)\bW \bigg]+\bbE_{\bW,\hatS_\rmu}\bbE_{S_\rml}\bigg[\frac{\alpha}{2}\bW^\top H^*(S_\rml,\hatS_\rmu)\bW \bigg] \nn\\
	&\quad +\bbE_{\hatS_\rmu,S_\rml}\bigg[\frac{\alpha}{2}\tr\Big(H^*(S_\rml,\hatS_\rmu)(\bW^*(S_\rml,\hatS_\rmu)\bW^*(S_\rml,\hatS_\rmu)^\top ) \Big) \bigg] \nn\\
	&\quad -\bbE_{\hatS_\rmu}\bbE_{S_\rml}\bigg[\frac{\alpha}{2}\tr\Big(H^*(S_\rml,\hatS_\rmu)\big(\bW^*(S_\rml,\hatS_\rmu)\bbE_{S_\rml|\hatS_\rmu}[\bW^*(S_\rml,\hatS_\rmu)]^\top + \bbE_{S_\rml|\hatS_\rmu}[\bW^*(S_\rml,\hatS_\rmu)]\bW^*(S_\rml,\hatS_\rmu)^\top \nn\\
	&\quad - \bW^*(S_\rml,\hatS_\rmu)\bW^*(S_\rml,\hatS_\rmu)^\top \big) \Big) \bigg] \nn\\
	&\quad +\bbE_{S_\rml,\hatS_\rmu}\big[\log \sqrt{\det(H^*(S_\rml,\hatS_\rmu))} \big]-\bbE_{S_\rml}\bbE_{\hatS_\rmu}\big[\log \sqrt{ \det(H^*(S_\rml,\hatS_\rmu))} \big] \\
	&=\bbE_{\bW,\hatS_\rmu,S_\rml}\bigg[-\frac{\alpha}{2}\bW^\top H^*(S_\rml,\hatS_\rmu)\bW \bigg]+\bbE_{\bW,\hatS_\rmu}\bbE_{S_\rml}\bigg[\frac{\alpha}{2}\bW^\top H^*(S_\rml,\hatS_\rmu)\bW \bigg] \nn\\
	&\quad +\bbE_{\hatS_\rmu,S_\rml}\bigg[\frac{\alpha}{2}\bW^*(S_\rml,\hatS_\rmu)^\top H^*(S_\rml,\hatS_\rmu)\bW^*(S_\rml,\hatS_\rmu)  \bigg]\nn\\
	&\quad +\bbE_{\hatS_\rmu}\bbE_{S_\rml}\bigg[\alpha \Big(\frac{1}{2}\bW^*(S_\rml,\hatS_\rmu)-\bbE_{S_\rml|\hatS_\rmu}[\bW^*(S_\rml,\hatS_\rmu)] \Big)^\top H^*(S_\rml,\hatS_\rmu) \bW^*(S_\rml,\hatS_\rmu) \bigg] \nn\\
	&\quad +\bbE_{S_\rml,\hatS_\rmu}\big[\log \sqrt{\det(H^*(S_\rml,\hatS_\rmu))} \big]-\bbE_{S_\rml}\bbE_{\hatS_\rmu}\big[\log \sqrt{ \det(H^*(S_\rml,\hatS_\rmu))} \big].
\end{align}
From \eqref{Eq:Lambda asymptotics}, we have 
\begin{align}
    0 &= \bbE_{\Delta(S_\rml,\hatS_\rmu)}\bigg[\log \sqrt{\det\bigg(\frac{\alpha H^*(S_\rml,\hatS_\rmu)}{2} \bigg)} + \alpha \barL_{\rmE}(\bW^*(S_\rml,\hatS_\rmu)),S_\rml,\hatS_\rmu)+\log\Lambda_{\alpha,\lambda}(S_\rml,\hatS_\rmu) \bigg] \nn\\
    &=\bbE_{\Delta(S_\rml,\hatS_\rmu)}\big[\log \sqrt{ \det(H^*(S_\rml,\hatS_\rmu)}+ \alpha \barL_{\rmE}(\bW^*(S_\rml,\hatS_\rmu),S_\rml,\hatS_\rmu)+\log\Lambda_{\alpha,\lambda}(S_\rml,\hatS_\rmu) \big],
\end{align}
which means
\begin{align}
    \bbE_{\Delta(S_\rml,\hatS_\rmu)}\big[ \log\Lambda_{\alpha,\lambda}(S_\rml,\hatS_\rmu) \big]= -\bbE_{\Delta(S_\rml,\hatS_\rmu)}\Big[\log \sqrt{ \det(H^*(S_\rml,\hatS_\rmu)} \Big] -\bbE_{\Delta(S_\rml,\hatS_\rmu)}\big[ \alpha \barL_{\rmE}(\bW^*(S_\rml,\hatS_\rmu),S_\rml,\hatS_\rmu) \big].
\end{align}
Therefore, by applying Theorem \ref{Thm:gen SSL PL}, the expected gen-error can be rewritten as
\begin{align}
	\overline{\mathrm{gen}}(P_{\bW|S_\rml,\hatS_\rmu}^\alpha, P_{S_\rml,\hatS_\rmu})&=\frac{1+\lambda}{2}\bigg(\bbE_{\bW,\hatS_\rmu,S_\rml}\bigg[-\bW^\top H^*(S_\rml,\hatS_\rmu)\bW \bigg]+\bbE_{\bW,\hatS_\rmu}\bbE_{S_\rml}\bigg[\bW^\top H^*(S_\rml,\hatS_\rmu)\bW \bigg] \nn\\
	&\quad +\bbE_{\hatS_\rmu,S_\rml}\bigg[\bW^*(S_\rml,\hatS_\rmu)^\top H^*(S_\rml,\hatS_\rmu)\bW^*(S_\rml,\hatS_\rmu)  \bigg]\nn\\
	&\quad +\bbE_{\hatS_\rmu}\bbE_{S_\rml}\bigg[ \Big(\bW^*(S_\rml,\hatS_\rmu)-2\bbE_{S_\rml|\hatS_\rmu}[\bW^*(S_\rml,\hatS_\rmu)] \Big)^\top H^*(S_\rml,\hatS_\rmu) \bW^*(S_\rml,\hatS_\rmu) \bigg]\nn\\
	&\quad -\bbE_{\Delta(S_\rml,\hatS_\rmu)}\big[  \barL_{\rmE}(\bW^*(S_\rml,\hatS_\rmu),S_\rml,\hatS_\rmu) \big]  \bigg). 
\end{align}

\section{Proof of \cref{Coro: asymp gen MLE}}\label{App: pf of Coro: asymp gen MLE}
When $S_\rml$ and $\hatS_\rmu$ are independent, we can simplify the asymptotic gen-error as
\begin{align}
    &\overline{\mathrm{gen}}(P_{\bW|S_\rml,\hatS_\rmu}^\infty, P_{S_\rml,\hatS_\rmu})\nn\\
    &=\frac{1+\lambda}{2}\bigg(\bbE_{\bW,\hatS_\rmu,S_\rml}\bigg[-\bW^\top H^*(S_\rml,\hatS_\rmu)\bW \bigg] +\bbE_{\bW,\hatS_\rmu}\bbE_{S_\rml}\bigg[\bW^\top H^*(S_\rml,\hatS_\rmu)\bW \bigg] \nn\\
    &\quad +2\bbE_{\hatS_\rmu}\bbE_{S_\rml}\bigg[ \Big(\bW^*(S_\rml,\hatS_\rmu)-\bbE_{S_\rml}[\bW^*(S_\rml,\hatS_\rmu)] \Big)^\top  H^*(S_\rml,\hatS_\rmu) \bW^*(S_\rml,\hatS_\rmu) \bigg] \bigg). \label{Eq:gen alpha infty indep}
\end{align}
Let the MLE optimizer be denoted as $\hat{\bW}_{\mathrm{ML}}(S_\rml,\hatS_\rmu)=\bW^*(S_\rml,\hatS_\rmu)$.
Recall the definition
\begin{align}
	\bw^*_{\lambda}=\argmin_{\bw\in\calW} \bigg(\frac{n}{n+m}D\big(P_{\bZ} \| p(\cdot|\bw) \big)+\frac{m}{n+m}D\big(P_{\hat{\bZ}|\bw^*_{P_\bZ}} \| p(\cdot|\bw) \big)\bigg)
\end{align}
and then we have
\begin{align}
	\bbE_{\bZ\sim P_\bZ}\bbE_{\hat{\bZ}\sim P_{\hat{\bZ}|\bw^*_{P_\bZ}}}\bigg[\nabla_{\bw}\bigg(-\frac{n}{n+m}\log p(\bZ|\bw)- \frac{m}{n+m}\log p(\hat{\bZ}|\bw) \bigg)\bigg|_{\bw=\bw^*_{\lambda}}  \bigg]=0 .
\end{align}

According to \cref{Thm: gen SSL asym} and \eqref{Eq:gen alpha infty indep}, the asymptotic gen-error of MLE is given by 
\begin{align}
	&\overline{\mathrm{gen}}(P_{\bW|S_\rml,\hatS_\rmu}^\infty, P_{S_\rml,\hatS_\rmu}) \nn\\
	&=\frac{n+m}{2n}\bigg(\bbE_{\bW,\hatS_\rmu,S_\rml}\bigg[-\bW^\top J(\bw^*_{\lambda})\bW \bigg]+\bbE_{\bW,\hatS_\rmu}\bbE_{S_\rml}\bigg[\bW^\top J(\bw^*_{\lambda})\bW \bigg] \nn\\
	&\quad +2\bbE_{\hatS_\rmu}\bbE_{S_\rml}\bigg[\big(\hat{\bW}_{\mathrm{ML}}(S_\rml,\hatS_\rmu)-\bbE_{S_\rml}[\hat{\bW}_{\mathrm{ML}}(S_\rml,\hatS_\rmu)] \big)^\top J(\bw^*_{\lambda})\hat{\bW}_{\mathrm{ML}}(S_\rml,\hatS_\rmu)  \bigg] \bigg) \\
	&=\frac{n+m}{2n}\bigg(\bbE_{\bW}\bigg[-\bW^\top J(\bw^*_{\lambda})\bW \bigg]+\bbE_{\bW}\bigg[\bW^\top J(\bw^*_{\lambda})\bW \bigg] \nn\\
	&\quad +2\tr\bigg( \bbE_{\hatS_\rmu}\bbE_{S_\rml}\bigg[\big(\hat{\bW}_{\mathrm{ML}}(S_\rml,\hatS_\rmu)-\bbE_{S_\rml}[\hat{\bW}_{\mathrm{ML}}(S_\rml,\hatS_\rmu)] \big)\big(\hat{\bW}_{\mathrm{ML}}(S_\rml,\hatS_\rmu)-\bbE_{S_\rml}[\hat{\bW}_{\mathrm{ML}}(S_\rml,\hatS_\rmu)] \big)^\top J(\bw^*_{\lambda}) \bigg] \bigg) \bigg) \label{Eq:gen asymp rewrite 1}\\
	&=\frac{n+m}{n}\tr\bigg( \bbE_{\hatS_\rmu}\bbE_{S_\rml}\bigg[\big(\hat{\bW}_{\mathrm{ML}}(S_\rml,\hatS_\rmu)-\bbE_{S_\rml}[\hat{\bW}_{\mathrm{ML}}(S_\rml,\hatS_\rmu)] \big)\big(\hat{\bW}_{\mathrm{ML}}(S_\rml,\hatS_\rmu)-\bbE_{S_\rml}[\hat{\bW}_{\mathrm{ML}}(S_\rml,\hatS_\rmu)] \big)^\top \bigg] J(\bw^*_{\lambda})  \bigg). \label{Eq:gen asymp rewrite 2}
\end{align}
Fix any pseudo-labeled data set $\hats_\rmu$ and let
\begin{align}
	\hat{\bW}(\hats_\rmu)=\argmin_{\bw\in\calW}\bigg(\frac{n}{n+m}D\big(P_{\bZ} \| p(\cdot|\bw) \big) -\frac{1}{n+m}\sum_{i=n+1}^{n+m}\log p(\hat{\bz}_i|\bw) \bigg). \label{Eq: hatW(S_u)}
\end{align} 
Then we have given any ratio $m/n>0$, as $m\to\infty$,
\begin{align}
	\hat{\bW}(\hatS_\rmu)\xrightarrow{\rmp} \bw^*_{\lambda}, \label{Eq: converge hatW(S_u)}
\end{align}
and
\begin{align}
	\bbE_{\bZ\sim P_{\bZ}}\bigg[\nabla_{\bw}\bigg(-\frac{n}{n+m}\log p(\bZ|\bw)-\frac{1}{n+m}\sum_{i=m+1}^{n+m}\log p(\hat{\bz}_i|\bw)\bigg)\bigg|_{\bw=\hat{\bW}(\hats_\rmu)} \bigg]=0.
\end{align}
As $n\to\infty$, by central limit theorem, $\bbE_{S_\rml}[\hat{\bW}_{\mathrm{ML}}(S_\rml,\hats_\rmu)]=\hat{\bW}(\hats_\rmu)$ (cf. \eqref{Eq: W_ML} and \eqref{Eq: hatW(S_u)}).

By applying Taylor expansion to $\nabla_{\bw}\barL_{\rmE}(\bw,S_\rml,\hats_\rmu)|_{\bw=\hat{\bW}_{\mathrm{ML}}(S_\rml,\hats_\rmu)}$ around $\hat{\bW}_{\mathrm{ML}}(S_\rml,\hats_\rmu)=\hat{\bW}(\hats_\rmu)$, we have
\begin{align}
	0&=\nabla_{\bw}\bigg(-\frac{1}{n+m}\sum_{i=1}^{n}\log p(\bZ_i|\bw)-\frac{1}{n+m}\sum_{i=m+1}^{n+m}\log p(\hat{\bz}_i|\bw)\bigg)\bigg|_{\bw=\hat{\bW}_{\mathrm{ML}}(S_\rml,\hats_\rmu)}\nn\\
	&\approx \nabla_{\bw}\bigg(-\frac{1}{n+m}\sum_{i=1}^{n}\log p(\bZ_i|\bw)-\frac{1}{n+m}\sum_{i=m+1}^{n+m}\log p(\hat{\bz}_i|\bw)\bigg)\bigg|_{\bw=\hat{\bW}(\hats_\rmu)} \nn\\
	&\quad +\nabla^2_{\bw}\bigg(-\frac{1}{n+m}\sum_{i=1}^{n}\log p(\bZ_i|\bw)-\frac{1}{n+m}\sum_{i=m+1}^{n+m}\log p(\hat{\bz}_i|\bw)\bigg)\bigg|_{\bw=\hat{\bW}(\hats_\rmu)} (\hat{\bW}_{\mathrm{ML}}(S_\rml,\hats_\rmu)-\hat{\bW}(\hats_\rmu)). \label{Eq: taylor first derivative of LrmE}
\end{align}
By multivariate central limit theorem, as $n\to\infty$, the first term in \eqref{Eq: taylor first derivative of LrmE} converges as follows
\begin{align}
	\nabla_{\bw}\bigg(-\frac{1}{n+m}\sum_{i=1}^{n}\log p(\bZ_i|\bw)-\frac{1}{n+m}\sum_{i=m+1}^{n+m}\log p(\hat{\bz}_i|\bw)\bigg)\bigg|_{\bw=\hat{\bW}(\hats_\rmu)}\xrightarrow{\rmd} \calN\bigg(0,\frac{n}{(n+m)^2}\calI_\rml(\hat{\bW}(\hats_\rmu)) \bigg).
\end{align}
By the law of large numbers, as $n\to\infty$, the second term in \eqref{Eq: taylor first derivative of LrmE} converges as follows
\begin{align}
	&\nabla^2_{\bw}\bigg(-\frac{1}{n+m}\sum_{i=1}^{n}\log p(\bZ_i|\bw)-\frac{1}{n+m}\sum_{i=m+1}^{n+m}\log p(\hat{\bz}_i|\bw)\bigg)\bigg|_{\bw=\hat{\bW}(\hats_\rmu)}\nn\\
	& \xrightarrow{\text{p}} \frac{n}{n+m}J_{\rml}(\hat{\bW}(\hats_\rmu))-\nabla^2_{\bw} \bigg(\frac{1}{n+m}\sum_{i=m+1}^{n+m}\log p(\hat{\bz}_i|\bw)\bigg)\bigg|_{\bw=\hat{\bW}(\hats_\rmu)}:=\tilJ(\hat{\bW}(\hats_\rmu)).
\end{align}
Given any ratio $m/n>0$, as $n,m\to\infty$, according to \eqref{Eq: converge hatW(S_u)},
\begin{align}
	\calI_\rml(\hat{\bW}(\hatS_\rmu))\xrightarrow{\rmp} \calI_\rml(\bw^*_{\lambda}) ~\text{~and~}~ \tilJ(\hat{\bW}(\hatS_\rmu))\xrightarrow{\rmp} J(\bw^*_{\lambda}).
\end{align}
Thus, we have
\begin{align}
	\hat{\bW}_{\mathrm{ML}}(S_\rml,\hatS_\rmu) &\xrightarrow{\rmd} \calN\bigg(\bw^*_{\lambda}, \frac{nJ(\bw^*_{\lambda})^{-1} \calI_\rml(\bw^*_{\lambda})J(\bw^*_{\lambda})^{-1}}{(n+m)^2} \bigg). \label{Eq:dist conv of W_ML}
\end{align}
Finally, the expected gen-error in \eqref{Eq:gen asymp rewrite 2} can be rewritten as
\begin{align}
	\overline{\mathrm{gen}}(P_{\bW|S_\rml,\hatS_\rmu}^\infty, P_{S_\rml,\hatS_\rmu})&=\frac{n+m}{n}\tr\bigg( \frac{n}{(n+m)^2}J(\bw^*_{\lambda})^{-1}\calI_\rml (\bw^*_{\lambda})  \bigg)\\
	&=\frac{\tr(J(\bw^*_{\lambda})^{-1}\calI_\rml (\bw^*_{\lambda}) )}{n+m} \label{Eq:gen asymp case 1}.
\end{align}

\section{Proof of \cref{Lem: excess risk MLE}} \label{App: pf of Lem: excess risk MLE}

By applying Taylor expansion of $L_\rmP(\bW,P_{S_\rml})$ around $\bW=\bw_\rml^*$, we have the following approximation
\begin{align}
	&L_\rmP(\bW,P_{S_\rml}) \nn\\
	&\approx L_\rmP(\bw_\rml^*,P_{S_\rml})+(\bW-\bw_\rml^*)^\top \nabla_{\bW}L_\rmP(\bW,P_{S_\rml})|_{\bW=\bw_\rml^*}+\frac{1}{2}(\bW-\bw_\rml^*)^\top \nabla_{\bW}^2 L_\rmP(\bW,P_{S_\rml})|_{\bW=\bw_\rml^*}(\bW-\bw_\rml^*)\\
	&=L_\rmP(\bw_\rml^*,P_{S_\rml})+\frac{1}{2}\tr\big((\bW-\bw_\rml^*)(\bW-\bw_\rml^*)^\top J_\rml(\bw_\rml^*)\big).
\end{align}
Thus, the excess risk can be approximated as follows:
\begin{align}
	&\calE_{\mathrm{r}}(P_\bW)=\bbE_{\bW}[L_\rmP(\bW,P_{S_\rml})]-L_\rmP(\bw_\rml^*,P_{S_\rml})\nn\\
	&\approx \frac{1}{2}\tr\big(\bbE_{\bW}[(\bW-\bw_\rml^*)(\bW-\bw_\rml^*)^\top] J_\rml(\bw_\rml^*)\big)\\
	&=\frac{1}{2}\tr\big(\bbE_{S_\rml,\hatS_\rmu}[(\hat{\bW}(S_\rml,\hatS_\rmu)-\bw_\rml^*)(\hat{\bW}(S_\rml,\hatS_\rmu)-\bw_\rml^*)^\top] J_\rml(\bw_\rml^*)\big)+\frac{\tr(J_\rml(\bw_\rml^*) \bbE_{S_{\rml},S_{\rmu}}[\Cov(\bW|S_\rml,\hatS_\rmu)])}{2} \\
	&=\frac{1}{2}\tr((\bw^*_{\lambda}-\bw^*_\rml)(\bw^*_{\lambda}-\bw^*_\rml)^\top J_\rml(\bw_\rml^*) )+\frac{\tr(J_\rml(\bw_\rml^*)\Cov(\hat{\bW}_{\mathrm{ML}}(S_\rml,\hatS_\rmu)) )}{2} \label{Eq:excess risk bias var}\\
	&=\frac{1}{2}\tr((\bw^*_{\lambda}-\bw^*_\rml)(\bw^*_{\lambda}-\bw^*_\rml)^\top J_\rml(\bw_\rml^*) )+\frac{\tr(J_\rml(\bw_\rml^*)J(\bw^*_{\lambda})^{-1}\calI_\rml(\bw^*_{\lambda})J(\bw^*_{\lambda})^{-1} )}{2(1+\lambda)(n+m)} 
\end{align}
where \eqref{Eq:excess risk bias var} follows since when $\alpha\to\infty$, $\Cov(\bW|S_\rml,\hatS_\rmu)=\frac{1}{\alpha}H^*(S_\rml,\hatS_\rmu)^{-1}\to 0$ and from \eqref{Eq:dist conv of W_ML}. 

\section{Proof of Logistic Regression Example}\label{App: pf of logistic}

The first and second derivatives of the loss function are as follows
\begin{align}
	\nabla_{\bw}l(\bw,\bz)&=\nabla_{\bw} \log (1+\exp(-y\bw^\top \bx))+\nu \bw
	=\frac{-y\bx e^{-y\bw^\top \bx}}{1+e^{-y\bw^\top \bx}}+\nu \bw, \text{~and~} \\
	\nabla_{\bw}^2 l(\bw,\bz)&=\nabla_{\bw}^2 \log (1+\exp(-y\bw^\top \bx))+\nu \bI_d
	=\frac{\bx\bx^\top e^{-y\bw^\top \bx}}{(1+e^{-y\bw^\top \bx})^2}+\nu\bI_d.
\end{align}

 The expected Hessian matrices $J_\rml, J_\rmu$ and expected product of the first derivative $\calI_\rml$ are given as follows:
\begin{align}
	J_\rml(\bw)
	=\bbE_{\bZ\sim P_\bZ}\bigg[\frac{\bX\bX^\top e^{-Y\bw^\top \bX}}{(1+e^{-Y\bw^\top \bX})^2}\bigg]=\bbE_{\bX\sim P_{\bX}}\bigg[\frac{\bX\bX^\top}{e^{-\bw^\top\bX}+e^{\bw^\top\bX}+2} \bigg],
\end{align}
\begin{align}
	J_\rmu(\bw)
	&=\bbE_{\bX\sim P_\bX}\bigg[\frac{\bX\bX^\top e^{-\sgn( \bX^\top \bw_0^*)\bw^\top \bX}}{(1+e^{-\sgn(\bX^\top \bw_0^*)\bw^\top \bX})^2}\bigg]=\bbE_{\bX\sim P_{\bX}}\bigg[\frac{\bX\bX^\top}{e^{-\bw^\top\bX}+e^{\bw^\top\bX}+2} \bigg],
\end{align}
\begin{align}
	\calI_\rml(\bw)
	=\bbE_{\bZ\sim P_\bZ}\bigg[\frac{\bX\bX^\top e^{-2Y\bw^\top \bX}}{(1+e^{-Y\bw^\top \bX})^2} \bigg]. 
\end{align}
We can see that $J(\bw)=J_\rml(\bw)=J_\rmu(\bw)$.

\begin{figure}[t!]
    \centering
    \includegraphics[width=0.4\columnwidth]{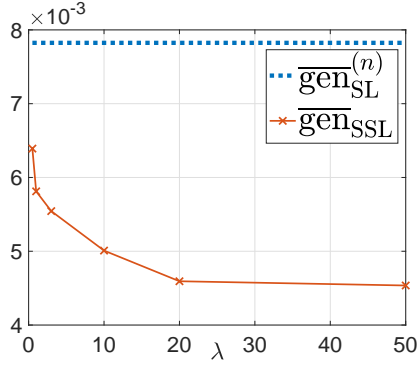}
    \caption{Empirical gen-error of logistic regression on MNIST dataset.}
    \label{fig:logi_MNIST 01}
\end{figure}

Recall the proof of \cref{Coro: asymp gen MLE} in \cref{App: pf of Coro: asymp gen MLE}. In the logistic regression with $l_2$ regularization, the unique minimizer of the empirical risk in \eqref{Eq: W_ML} is rewritten as
\begin{align}
	&\hat{\bW}_{\mathrm{ML}}(S_\rml,\hatS_\rmu)=\argmin_{\bw\in\calW} \bigg(-\frac{1}{1+\lambda}\frac{1}{n}\sum_{i=1}^{n}\log p(\bZ_i|\bw)  -\frac{\lambda}{1+\lambda}\frac{1}{m}\sum_{i=n+1}^{n+m}\log p(\hat{\bZ}_i|\bw) +\frac{\nu}{2}\|\bw\|_2^2 \bigg) \label{Eq: W_ML reg}
\end{align}
and the Hessian matrix of the empirical risk at $\bw=\hat{\bW}_{\mathrm{ML}}(S_\rml,\hatS_\rmu)$ is rewritten as 
\begin{align}
    H^*(S_\rml,\hatS_\rmu)=\nabla_\bw^2\bigg(-\frac{1}{1+\lambda}\frac{1}{n}\sum_{i=1}^{n}\log p(\bZ_i|\bw)  -\frac{\lambda}{1+\lambda}\frac{1}{m}\sum_{i=n+1}^{n+m}\log p(\hat{\bZ}_i|\bw)\bigg)\bigg|_{\bw=\hat{\bW}_{\mathrm{ML}}(S_\rml,\hatS_\rmu)} +\nu \bI_d
    .
\end{align}
Recall the definition of $\bw^*_\lambda$ with regularization
\begin{align}
	\bw^*_{\lambda}=&\argmin_{\bw\in\calW} \bigg(\frac{n}{n+m}D\big(P_{\bZ} \| p(\cdot|\bw) \big) +\frac{m}{n+m}D\big(P_{\hat{\bZ}|\bw^*_0} \| p(\cdot|\bw) \big)+\frac{\nu}{2}\|\bw\|_2^2 \bigg).
\end{align}
Given any ratio $\lambda>0$, as $n,m\to\infty$, the Hessian matrix converges as follows
\begin{align}
    H^*(S_\rml,\hatS_\rmu)\xrightarrow{\rmp} J(\bw^*_\lambda)+\nu \bI_d.
\end{align}
Then the asymptotic expected gen-error in \eqref{Eq:gen asymp rewrite 2} is rewritten as
\begin{align}
&\overline{\mathrm{gen}}(P_{\bW|S_\rml,\hatS_\rmu}^\infty, P_{S_\rml,\hatS_\rmu})\nn\\
&=
    \frac{n+m}{n}\tr\bigg(\! \bbE_{\hatS_\rmu}\bbE_{S_\rml}\! \bigg[\big(\hat{\bW}_{\mathrm{ML}}(S_\rml,\hatS_\rmu)-\bbE_{S_\rml}[\hat{\bW}_{\mathrm{ML}}(S_\rml,\hatS_\rmu)] \big)\big(\hat{\bW}_{\mathrm{ML}}(S_\rml,\hatS_\rmu)-\bbE_{S_\rml}[\hat{\bW}_{\mathrm{ML}}(S_\rml,\hatS_\rmu)] \big)^\top \!\bigg] (J(\bw^*_{\lambda})+\nu \bI_d) \! \bigg). \label{Eq:gen asymp rewrite reg}
\end{align}
By redefining $\hat{\bW}(\hats_\rmu)$ in \eqref{Eq: hatW(S_u)} with the $l_2$ regularization term $\frac{\nu}{2}\|\bw\|_2^2$, we can similarly obtain
\begin{align}
    \hat{\bW}_{\mathrm{ML}}(S_\rml,\hatS_\rmu) &\xrightarrow{\rmd} \calN\bigg(\bw^*_{\lambda}, \frac{n(J(\bw^*_{\lambda})+\nu \bI_d)^{-1} \calI_\rml(\bw^*_{\lambda})(J(\bw^*_{\lambda})+\nu \bI_d)^{-1}}{(n+m)^2} \bigg).
\end{align}
Then the expected gen-error in \eqref{Eq:gen asymp rewrite 2} can be rewritten as
\begin{align}
	\overline{\mathrm{gen}}(P_{\bW|S_\rml,\hatS_\rmu}^\infty, P_{S_\rml,\hatS_\rmu})
	&=\frac{\tr((J(\bw^*_{\lambda})+\nu\bI_d)^{-1}\calI_\rml (\bw^*_{\lambda}) )}{n+m} \label{Eq:gen asymp reg}.
\end{align}
Similarly, the excess risk in \eqref{Eq:excess risk bias var 2} can be rewritten as
\begin{align}
    &\calE_{\mathrm{r}}(P_\bW)=\frac{1}{2}\tr((\bw^*_{\lambda}-\bw^*_\rml)(\bw^*_{\lambda}-\bw^*_\rml)^\top J_\rml(\bw_\rml^*) ) +\frac{\tr(J_\rml(\bw_\rml^*)(J(\bw^*_{\lambda})+\nu\bI_d)^{-1}\calI_\rml(\bw^*_{\lambda})(J(\bw^*_{\lambda})+\nu\bI_d)^{-1} )}{2(1+\lambda)(n+m)},
\end{align}
where $\bw_\rml^*=\argmin_{\bw\in\calW} L_\rmP(\bW,P_{S_\rml})$.

In addition to the experiments on the synthetic datasets, we implement an logistic regression experiment on ``0--1'' digit pair in MNIST dataset by setting $n=200$, $\lambda\in\{0.5,1,3,10,20,50\}$ and $\nu=5$. In Figure \ref{fig:logi_MNIST 01}, we observe that the gen-error decreases as $\lambda$ increases.

\end{document}

%% file: Math-Commands.tex
\newtheorem{theorem}{Theorem}

\newtheorem{proposition}{Proposition}

\newtheorem{corollary}{Corollary}

\newtheorem{remark}{Remark}

\newcommand{\Cov}{\operatorname{Cov}} 

\DeclareMathOperator*{\argmin}{arg\,min}

\DeclareMathOperator{\sgn}{sgn} 

\newcommand{\tr}{\operatorname{tr}} 

\newcommand*{\quotientspace}[2]
{\ensuremath{\raisebox{.25ex}{\ensuremath{#1}} \hspace{-.35ex} \raisebox{-.25ex}{/} 
\hspace{-.05ex} \raisebox{-.85ex}{\ensuremath{#2}}} \hspace{.2ex} }
\newcommand*{\textquotientspace}[2]
{\ensuremath{\raisebox{.05ex}{\ensuremath{#1}} \hspace{-.30ex} \raisebox{-.15ex}{/} 
\hspace{.10ex} \raisebox{-.55ex}{\ensuremath{#2}}}}

\newcommand{\upsum}
{\ensuremath \mkern5.4mu \overline{\vphantom{S}\mkern8mu} \mkern-11mu S}
\newcommand{\lowsum}
{\ensuremath \mkern3.4mu \underline{\vphantom{S}\mkern8mu} \mkern-9mu S}
\newcommand{\upint}
{\ensuremath \mkern10.4mu \overline{\vphantom{\int}\mkern7mu} \mkern-21mu\int}
\newcommand{\upintwdomain}[1]
{\ensuremath \mkern10.4mu \overline{\vphantom{\int}\mkern7mu} \mkern-21mu\int_{#1}}
\newcommand{\textupint}
{\textstyle \mkern5mu \overline{\vphantom{\int}\mkern6mu} \mkern-14mu\intop\mkern0mu}
\newcommand{\textupintwdomain}[1]
{\textstyle \mkern5mu \overline{\vphantom{\int}\mkern6mu} \mkern-14mu\intop_{#1}}
\newcommand{\lowint}
{\ensuremath \underline{\vphantom{\int}\mkern7mu} \mkern-9mu\int}
\newcommand{\lowintwdomain}[1]
{\ensuremath \underline{\vphantom{\int}\mkern7mu} \mkern-9mu\int_{#1}}
\newcommand{\textlowint}
{\textstyle \mkern3mu \underline{\vphantom{\int}\mkern6mu} \mkern-8.5mu\intop}
\newcommand{\textlowintwdomain}[1]
{\textstyle \mkern3mu \underline{\vphantom{\int}\mkern6mu} \mkern-8.5mu\intop_{#1}}

\def\ddefloop#1{\ifx\ddefloop#1\else\ddef{#1}\expandafter\ddefloop\fi}

\def\ddef#1{\expandafter\def\csname c#1\endcsname{\ensuremath{\mathcal{#1}}}}
\ddefloop ABCDEFGHIJKLMNOPQRSTUVWXYZ\ddefloop

\def\ddef#1{\expandafter\def\csname bb#1\endcsname{\ensuremath{\mathbb{#1}}}}
\ddefloop ABCDEFGHIJKLMNOPQRSTUVWXYZ\ddefloop

\def\ddef#1{\expandafter\def\csname fr#1\endcsname{\ensuremath{\mathfrak{#1}}}}
\ddefloop ABCDEFGHIJKLMNOPQRSTUVWXYZabcdefghijklmnopqrstuvwxyz\ddefloop

\def\ddef#1{\expandafter\def\csname ss#1\endcsname{\ensuremath{\mathsf{#1}}}}
\ddefloop ABCDEFGHIJKLMNOPQRSTUVWXYZabcdefghijklmnopqrstuvwxyz\ddefloop

\def\ddef#1{\expandafter\def\csname bf#1\endcsname{\ensuremath{\mathbf{#1}}}}
\ddefloop ABCDEFGHIJKLMNOPQRSTUVWXYZabcdefghijklmnopqrstuvwxyz\ddefloop

\def\ddef#1{\expandafter\def\csname v#1\endcsname{\ensuremath{\boldsymbol{#1}}}}
\ddefloop ABCDEFGHIJKLMNOPQRSTUVWXYZabcdefghijklmnopqrstuvwxyz\ddefloop

\def\ddef#1{\expandafter\def\csname v#1\endcsname{\ensuremath{\boldsymbol{\csname #1\endcsname}}}}
\ddefloop {alpha}{beta}{gamma}{delta}{epsilon}{varepsilon}{zeta}{eta}{theta}{vartheta}{iota}{kappa}{lambda}{mu}{nu}{xi}{pi}{varpi}{rho}{varrho}{sigma}{varsigma}{tau}{upsilon}{phi}{varphi}{chi}{psi}{omega}{Gamma}{Delta}{Theta}{Lambda}{Xi}{Pi}{Sigma}{varSigma}{Upsilon}{Phi}{Psi}{Omega}\ddefloop



%% file: preamble.tex
\usepackage{graphicx,bm,xcolor,url,overpic}
\usepackage{bm}

\newcommand{\nn}{\nonumber} 

\newcommand{\calE}{\mathcal{E}}

\newcommand{\calI}{\mathcal{I}}

\newcommand{\calN}{\mathcal{N}}

\newcommand{\calP}{\mathcal{P}}

\newcommand{\calW}{\mathcal{W}}
\newcommand{\calX}{\mathcal{X}}
\newcommand{\calY}{\mathcal{Y}}
\newcommand{\calZ}{\mathcal{Z}}

\newcommand{\bA}{\mathbf{A}}

\newcommand{\bg}{\mathbf{g}}

\newcommand{\bI}{\mathbf{I}}

\newcommand{\bT}{\mathbf{T}}

\newcommand{\bw}{\mathbf{w}}
\newcommand{\bW}{\mathbf{W}}
\newcommand{\bx}{\mathbf{x}}
\newcommand{\bX}{\mathbf{X}}

\newcommand{\bz}{\mathbf{z}}
\newcommand{\bZ}{\mathbf{Z}}

\newcommand{\rmd}{\mathrm{d}}

\newcommand{\rmE}{\mathrm{E}}

\newcommand{\rml}{\mathrm{l}}

\newcommand{\rmp}{\mathrm{p}}
\newcommand{\rmP}{\mathrm{P}}

\newcommand{\rmQ}{\mathrm{Q}}

\newcommand{\rmu}{\mathrm{u}}

\DeclareMathAlphabet{\mathbsf}{OT1}{cmss}{bx}{n}
\DeclareMathAlphabet{\mathssf}{OT1}{cmss}{m}{sl}

\DeclareSymbolFont{bsfletters}{OT1}{cmss}{bx}{n}  
\DeclareSymbolFont{ssfletters}{OT1}{cmss}{m}{n}
\DeclareMathSymbol{\bsfGamma}{0}{bsfletters}{'000}
\DeclareMathSymbol{\ssfGamma}{0}{ssfletters}{'000}
\DeclareMathSymbol{\bsfDelta}{0}{bsfletters}{'001}
\DeclareMathSymbol{\ssfDelta}{0}{ssfletters}{'001}
\DeclareMathSymbol{\bsfTheta}{0}{bsfletters}{'002}
\DeclareMathSymbol{\ssfTheta}{0}{ssfletters}{'002}
\DeclareMathSymbol{\bsfLambda}{0}{bsfletters}{'003}
\DeclareMathSymbol{\ssfLambda}{0}{ssfletters}{'003}
\DeclareMathSymbol{\bsfXi}{0}{bsfletters}{'004}
\DeclareMathSymbol{\ssfXi}{0}{ssfletters}{'004}
\DeclareMathSymbol{\bsfPi}{0}{bsfletters}{'005}
\DeclareMathSymbol{\ssfPi}{0}{ssfletters}{'005}
\DeclareMathSymbol{\bsfSigma}{0}{bsfletters}{'006}
\DeclareMathSymbol{\ssfSigma}{0}{ssfletters}{'006}
\DeclareMathSymbol{\bsfUpsilon}{0}{bsfletters}{'007}
\DeclareMathSymbol{\ssfUpsilon}{0}{ssfletters}{'007}
\DeclareMathSymbol{\bsfPhi}{0}{bsfletters}{'010}
\DeclareMathSymbol{\ssfPhi}{0}{ssfletters}{'010}
\DeclareMathSymbol{\bsfPsi}{0}{bsfletters}{'011}
\DeclareMathSymbol{\ssfPsi}{0}{ssfletters}{'011}
\DeclareMathSymbol{\bsfOmega}{0}{bsfletters}{'012}
\DeclareMathSymbol{\ssfOmega}{0}{ssfletters}{'012}

\newcommand{\tilg}{\tilde{g}}

\newcommand{\tilJ}{\tilde{J}}

\newcommand{\hats}{\hat{s}}
\newcommand{\hatS}{\hat{S}}

\newcommand{\haty}{\hat{y}}
\newcommand{\hatY}{\hat{Y}}

\newcommand{\barL}{\bar{L}}

\newcommand{\barY}{\bar{Y}}

\newcommand{\bup}{\bm{\upsilon}}

\newcommand{\bxi}{\bm{\xi}}

\newcommand{\bmu}{\bm{\mu}}

%% file: Arxiv-version.bbl
\begin{thebibliography}{55}
\providecommand{\natexlab}[1]{#1}
\providecommand{\url}[1]{\texttt{#1}}
\expandafter\ifx\csname urlstyle\endcsname\relax
  \providecommand{\doi}[1]{doi: #1}\else
  \providecommand{\doi}{doi: \begingroup \urlstyle{rm}\Url}\fi

\bibitem[Amini and Gallinari(2002)]{amini2002semi}
Massih-Reza Amini and Patrick Gallinari.
\newblock Semi-supervised logistic regression.
\newblock In \emph{ECAI}, volume~2, page~11, 2002.

\bibitem[Aminian et~al.(2015)Aminian, Arjmandi, Gohari, Nasiri-Kenari, and
  Mitra]{aminian2015capacity}
Gholamali Aminian, Hamidreza Arjmandi, Amin Gohari, Masoumeh Nasiri-Kenari, and
  Urbashi Mitra.
\newblock Capacity of diffusion-based molecular communication networks over
  {LTI}-{P}oisson channels.
\newblock \emph{IEEE Transactions on Molecular, Biological and Multi-Scale
  Communications}, 1\penalty0 (2):\penalty0 188--201, 2015.

\bibitem[Aminian et~al.(2021{\natexlab{a}})Aminian, Bu, Toni, Rodrigues, and
  Wornell]{aminian2021exact}
Gholamali Aminian, Yuheng Bu, Laura Toni, Miguel Rodrigues, and Gregory
  Wornell.
\newblock An exact characterization of the generalization error for the {Gibbs}
  algorithm.
\newblock \emph{Advances in Neural Information Processing Systems}, 34,
  2021{\natexlab{a}}.

\bibitem[Aminian et~al.(2021{\natexlab{b}})Aminian, Toni, and
  Rodrigues]{aminian2021information}
Gholamali Aminian, Laura Toni, and Miguel~RD Rodrigues.
\newblock Information-theoretic bounds on the moments of the generalization
  error of learning algorithms.
\newblock In \emph{IEEE International Symposium on Information Theory (ISIT)},
  2021{\natexlab{b}}.

\bibitem[Aminian et~al.(2022{\natexlab{a}})Aminian, Abroshan, Khalili, Toni,
  and Rodrigues]{aminian2022information}
Gholamali Aminian, Mahed Abroshan, Mohammad~Mahdi Khalili, Laura Toni, and
  Miguel Rodrigues.
\newblock An information-theoretical approach to semi-supervised learning under
  covariate-shift.
\newblock In \emph{International Conference on Artificial Intelligence and
  Statistics}, pages 7433--7449. PMLR, 2022{\natexlab{a}}.

\bibitem[Aminian et~al.(2022{\natexlab{b}})Aminian, Masiha, Toni, and
  Rodrigues]{aminian2022learning}
Gholamali Aminian, Saeed Masiha, Laura Toni, and Miguel~RD Rodrigues.
\newblock Learning algorithm generalization error bounds via auxiliary
  distributions.
\newblock \emph{arXiv preprint arXiv:2210.00483}, 2022{\natexlab{b}}.

\bibitem[Arazo et~al.(2020)Arazo, Ortego, Albert, O’Connor, and
  McGuinness]{arazo2020pseudo}
Eric Arazo, Diego Ortego, Paul Albert, Noel~E O’Connor, and Kevin McGuinness.
\newblock Pseudo-labeling and confirmation bias in deep semi-supervised
  learning.
\newblock In \emph{2020 International Joint Conference on Neural Networks
  (IJCNN)}, pages 1--8. IEEE, 2020.

\bibitem[Asadi et~al.(2018)Asadi, Abbe, and Verd{\'u}]{asadi2018chaining}
Amir Asadi, Emmanuel Abbe, and Sergio Verd{\'u}.
\newblock Chaining mutual information and tightening generalization bounds.
\newblock In \emph{Advances in Neural Information Processing Systems}, pages
  7234--7243, 2018.

\bibitem[Asadi and Abbe(2020)]{asadi2020chaining}
Amir~R Asadi and Emmanuel Abbe.
\newblock Chaining meets chain rule: Multilevel entropic regularization and
  training of neural networks.
\newblock \emph{Journal of Machine Learning Research}, 21\penalty0
  (139):\penalty0 1--32, 2020.

\bibitem[Athreya and Hwang(2010)]{athreya2010gibbs}
KB~Athreya and Chii-Ruey Hwang.
\newblock Gibbs measures asymptotics.
\newblock \emph{Sankhya A}, 72\penalty0 (1):\penalty0 191--207, 2010.

\bibitem[Berthelot et~al.(2019)Berthelot, Carlini, Goodfellow, Papernot,
  Oliver, and Raffel]{berthelot2019mixmatch}
David Berthelot, Nicholas Carlini, Ian Goodfellow, Nicolas Papernot, Avital
  Oliver, and Colin~A Raffel.
\newblock Mixmatch: A holistic approach to semi-supervised learning.
\newblock \emph{Advances in neural information processing systems}, 32, 2019.

\bibitem[Bu et~al.(2020)Bu, Zou, and Veeravalli]{bu2020tightening}
Yuheng Bu, Shaofeng Zou, and Venugopal~V Veeravalli.
\newblock Tightening mutual information-based bounds on generalization error.
\newblock \emph{IEEE Journal on Selected Areas in Information Theory},
  1\penalty0 (1):\penalty0 121--130, 2020.

\bibitem[Bu et~al.(2022)Bu, Aminian, Toni, Wornell, and
  Rodrigues]{bu2022characterizing}
Yuheng Bu, Gholamali Aminian, Laura Toni, Gregory~W Wornell, and Miguel
  Rodrigues.
\newblock Characterizing and understanding the generalization error of transfer
  learning with gibbs algorithm.
\newblock In \emph{International Conference on Artificial Intelligence and
  Statistics}, pages 8673--8699. PMLR, 2022.

\bibitem[Castelli and Cover(1996)]{castelli1996relative}
Vittorio Castelli and Thomas~M Cover.
\newblock The relative value of labeled and unlabeled samples in pattern
  recognition with an unknown mixing parameter.
\newblock \emph{IEEE Transactions on Information Theory}, 42\penalty0
  (6):\penalty0 2102--2117, 1996.

\bibitem[Catoni(2007)]{catoni2007pac}
Olivier Catoni.
\newblock {PAC}-{B}ayesian supervised classification: the thermodynamics of
  statistical learning.
\newblock \emph{arXiv preprint arXiv:0712.0248}, 2007.

\bibitem[Chapelle et~al.(2003)Chapelle, Weston, and
  Scholkopf]{chapelle2003cluster}
Olivier Chapelle, Jason Weston, and Bernhard Scholkopf.
\newblock Cluster kernels for semi-supervised learning.
\newblock \emph{Advances in neural information processing systems}, pages
  601--608, 2003.

\bibitem[Chapelle et~al.(2006)Chapelle, Sch{\"o}lkopf, and
  Zien]{books/mit/06/CSZ2006}
Olivier Chapelle, Bernhard Sch{\"o}lkopf, and Alexander Zien, editors.
\newblock \emph{Semi-Supervised Learning}.
\newblock The MIT Press, 2006.

\bibitem[Chiang et~al.(1987)Chiang, Hwang, and Sheu]{chiang1987diffusion}
Tzuu-Shuh Chiang, Chii-Ruey Hwang, and Shuenn~Jyi Sheu.
\newblock Diffusion for global optimization in $\mathbb{R}^{n}$.
\newblock \emph{SIAM Journal on Control and Optimization}, 25\penalty0
  (3):\penalty0 737--753, 1987.

\bibitem[Cover(1999)]{cover1999elements}
Thomas~M Cover.
\newblock \emph{Elements of information theory}.
\newblock John Wiley \& Sons, 1999.

\bibitem[Dupre et~al.(2019)Dupre, Fajtl, Argyriou, and
  Remagnino]{dupre2019improving}
Robert Dupre, Jiri Fajtl, Vasileios Argyriou, and Paolo Remagnino.
\newblock Improving dataset volumes and model accuracy with semi-supervised
  iterative self-learning.
\newblock \emph{IEEE Transactions on Image Processing}, 29:\penalty0
  4337--4348, 2019.

\bibitem[Esposito et~al.(2021)Esposito, Gastpar, and
  Issa]{esposito2019generalization}
Amedeo~Roberto Esposito, Michael Gastpar, and Ibrahim Issa.
\newblock Generalization error bounds via {R}{\'e}nyi-, f-divergences and
  maximal leakage.
\newblock \emph{IEEE Transactions on Information Theory}, 2021.

\bibitem[Gibbs(1902)]{gibbs1902elementary}
Josiah~Willard Gibbs.
\newblock Elementary principles of statistical mechanics.
\newblock \emph{Compare}, 289:\penalty0 314, 1902.

\bibitem[G{\"o}pfert et~al.(2019)G{\"o}pfert, Ben-David, Bousquet, Gelly,
  Tolstikhin, and Urner]{gopfert2019can}
Christina G{\"o}pfert, Shai Ben-David, Olivier Bousquet, Sylvain Gelly, Ilya
  Tolstikhin, and Ruth Urner.
\newblock When can unlabeled data improve the learning rate?
\newblock In \emph{Conference on Learning Theory}, pages 1500--1518. PMLR,
  2019.

\bibitem[Grandvalet et~al.(2005)Grandvalet, Bengio, et~al.]{grandvalet2005semi}
Yves Grandvalet, Yoshua Bengio, et~al.
\newblock Semi-supervised learning by entropy minimization.
\newblock \emph{CAP}, 367:\penalty0 281--296, 2005.

\bibitem[Hafez-Kolahi et~al.(2020)Hafez-Kolahi, Golgooni, Kasaei, and
  Soleymani]{hafez2020conditioning}
Hassan Hafez-Kolahi, Zeinab Golgooni, Shohreh Kasaei, and Mahdieh Soleymani.
\newblock Conditioning and processing: Techniques to improve
  information-theoretic generalization bounds.
\newblock \emph{Advances in Neural Information Processing Systems}, 33, 2020.

\bibitem[Haghifam et~al.(2020)Haghifam, Negrea, Khisti, Roy, and
  Dziugaite]{haghifam2020sharpened}
Mahdi Haghifam, Jeffrey Negrea, Ashish Khisti, Daniel~M Roy, and
  Gintare~Karolina Dziugaite.
\newblock Sharpened generalization bounds based on conditional mutual
  information and an application to noisy, iterative algorithms.
\newblock \emph{Advances in Neural Information Processing Systems}, 2020.

\bibitem[He et~al.(2022)He, Yan, and Tan]{he2022information}
Haiyun He, Hanshu Yan, and Vincent~YF Tan.
\newblock Information-theoretic characterization of the generalization error
  for iterative semi-supervised learning.
\newblock \emph{Journal of Machine Learning Research}, 23:\penalty0 1--52,
  2022.

\bibitem[Iscen et~al.(2019)Iscen, Tolias, Avrithis, and Chum]{iscen2019label}
Ahmet Iscen, Giorgos Tolias, Yannis Avrithis, and Ondrej Chum.
\newblock Label propagation for deep semi-supervised learning.
\newblock In \emph{Proceedings of the IEEE/CVF Conference on Computer Vision
  and Pattern Recognition}, pages 5070--5079, 2019.

\bibitem[Jaynes(1957)]{jaynes1957information}
Edwin~T Jaynes.
\newblock Information theory and statistical mechanics.
\newblock \emph{Physical review}, 106\penalty0 (4):\penalty0 620, 1957.

\bibitem[Jeffreys(1946)]{jeffreys1946invariant}
Harold Jeffreys.
\newblock An invariant form for the prior probability in estimation problems.
\newblock \emph{Proceedings of the Royal Society of London. Series A.
  Mathematical and Physical Sciences}, 186\penalty0 (1007):\penalty0 453--461,
  1946.

\bibitem[Ji et~al.(2012)Ji, Yang, Lin, Jin, and Han]{ji2012simple}
Ming Ji, Tianbao Yang, Binbin Lin, Rong Jin, and Jiawei Han.
\newblock A simple algorithm for semi-supervised learning with improved
  generalization error bound.
\newblock In \emph{Proceedings of the 29th International Coference on
  International Conference on Machine Learning}, pages 835--842, 2012.

\bibitem[Kuzborskij et~al.(2019)Kuzborskij, Cesa-Bianchi, and
  Szepesv{\'a}ri]{kuzborskij2019distribution}
Ilja Kuzborskij, Nicol{\`o} Cesa-Bianchi, and Csaba Szepesv{\'a}ri.
\newblock Distribution-dependent analysis of {G}ibbs-{ERM} principle.
\newblock In \emph{Conference on Learning Theory}, pages 2028--2054. PMLR,
  2019.

\bibitem[Lee et~al.(2013)]{lee2013pseudo}
Dong-Hyun Lee et~al.
\newblock Pseudo-label: The simple and efficient semi-supervised learning
  method for deep neural networks.
\newblock In \emph{Workshop on Challenges in Representation Learning, ICML},
  2013.

\bibitem[Li et~al.(2019)Li, Liu, Yin, and Wang]{li2019multi}
Jian Li, Yong Liu, Rong Yin, and Weiping Wang.
\newblock Multi-class learning using unlabeled samples: Theory and algorithm.
\newblock In \emph{International Joint Conference on Artificial Intelligence
  (IJCAI)}, pages 2880--2886, 2019.

\bibitem[Markowich and Villani(2000)]{markowich2000trend}
Peter~A Markowich and C{\'e}dric Villani.
\newblock On the trend to equilibrium for the {F}okker-{P}lanck equation: an
  interplay between physics and functional analysis.
\newblock \emph{Mat. Contemp}, 19:\penalty0 1--29, 2000.

\bibitem[McLachlan(2005)]{mclachlan2005discriminant}
Geoffrey~J McLachlan.
\newblock \emph{Discriminant Analysis and Statistical Pattern Recognition}.
\newblock John Wiley \& Sons, 2005.

\bibitem[Moon(1996)]{moon1996expectation}
Todd~K Moon.
\newblock The expectation-maximization algorithm.
\newblock \emph{IEEE Signal processing magazine}, 13\penalty0 (6):\penalty0
  47--60, 1996.

\bibitem[Niu et~al.(2013)Niu, Jitkrittum, Dai, Hachiya, and
  Sugiyama]{niu2013squared}
Gang Niu, Wittawat Jitkrittum, Bo~Dai, Hirotaka Hachiya, and Masashi Sugiyama.
\newblock Squared-loss mutual information regularization: A novel
  information-theoretic approach to semi-supervised learning.
\newblock In \emph{International Conference on Machine Learning}, pages 10--18.
  PMLR, 2013.

\bibitem[Ouali et~al.(2020)Ouali, Hudelot, and Tami]{ouali2020overview}
Yassine Ouali, C{\'e}line Hudelot, and Myriam Tami.
\newblock An overview of deep semi-supervised learning.
\newblock \emph{arXiv preprint arXiv:2006.05278}, 2020.

\bibitem[Palomar and Verd{\'u}(2008)]{palomar2008lautum}
Daniel~P Palomar and Sergio Verd{\'u}.
\newblock Lautum information.
\newblock \emph{IEEE transactions on information theory}, 54\penalty0
  (3):\penalty0 964--975, 2008.

\bibitem[Raginsky et~al.(2017)Raginsky, Rakhlin, and
  Telgarsky]{raginsky2017non}
Maxim Raginsky, Alexander Rakhlin, and Matus Telgarsky.
\newblock Non-convex learning via stochastic gradient {L}angevin dynamics: a
  nonasymptotic analysis.
\newblock In \emph{Conference on Learning Theory}, pages 1674--1703. PMLR,
  2017.

\bibitem[Rigollet(2007)]{rigollet2007generalization}
Philippe Rigollet.
\newblock Generalization error bounds in semi-supervised classification under
  the cluster assumption.
\newblock \emph{Journal of Machine Learning Research}, 8\penalty0 (7), 2007.

\bibitem[Rizve et~al.(2020)Rizve, Duarte, Rawat, and Shah]{rizve2020defense}
Mamshad~Nayeem Rizve, Kevin Duarte, Yogesh~S Rawat, and Mubarak Shah.
\newblock In defense of pseudo-labeling: An uncertainty-aware pseudo-label
  selection framework for semi-supervised learning.
\newblock In \emph{International Conference on Learning Representations}, 2020.

\bibitem[Russo and Zou(2019)]{russo2019much}
Daniel Russo and James Zou.
\newblock How much does your data exploration overfit? controlling bias via
  information usage.
\newblock \emph{IEEE Transactions on Information Theory}, 66\penalty0
  (1):\penalty0 302--323, 2019.

\bibitem[Seeger(2000)]{seeger2000learning}
Matthias Seeger.
\newblock Learning with labeled and unlabeled data.
\newblock 2000.
\newblock URL \url{http://infoscience.epfl.ch/record/161327}.

\bibitem[Singh et~al.(2008)Singh, Nowak, and Zhu]{singh2008unlabeled}
Aarti Singh, Robert Nowak, and Jerry Zhu.
\newblock Unlabeled data: Now it helps, now it doesn't.
\newblock \emph{{Advances in Neural Information Processing Systems}},
  21:\penalty0 1513--1520, 2008.

\bibitem[Sohn et~al.(2020)Sohn, Berthelot, Carlini, Zhang, Zhang, Raffel,
  Cubuk, Kurakin, and Li]{sohn2020fixmatch}
Kihyuk Sohn, David Berthelot, Nicholas Carlini, Zizhao Zhang, Han Zhang,
  Colin~A Raffel, Ekin~Dogus Cubuk, Alexey Kurakin, and Chun-Liang Li.
\newblock Fixmatch: Simplifying semi-supervised learning with consistency and
  confidence.
\newblock \emph{Advances in neural information processing systems},
  33:\penalty0 596--608, 2020.

\bibitem[Steinke and Zakynthinou(2020)]{steinke2020reasoning}
Thomas Steinke and Lydia Zakynthinou.
\newblock Reasoning about generalization via conditional mutual information.
\newblock In \emph{Conference on Learning Theory}, pages 3437--3452. PMLR,
  2020.

\bibitem[Szummer and Jaakkola(2002)]{szummer2002information}
Martin Szummer and Tommi Jaakkola.
\newblock Information regularization with partially labeled data.
\newblock \emph{Advances in Neural Information processing systems}, 15, 2002.

\bibitem[Vershynin(2018)]{vershynin2018high}
Roman Vershynin.
\newblock \emph{High-dimensional probability: An introduction with applications
  in data science}, volume~47.
\newblock Cambridge university press, 2018.

\bibitem[Wei et~al.(2020)Wei, Shen, Chen, and Ma]{wei2020theoretical}
Colin Wei, Kendrick Shen, Yining Chen, and Tengyu Ma.
\newblock Theoretical analysis of self-training with deep networks on unlabeled
  data.
\newblock In \emph{International Conference on Learning Representations}, 2020.

\bibitem[Xu and Raginsky(2017)]{xu2017information}
Aolin Xu and Maxim Raginsky.
\newblock Information-theoretic analysis of generalization capability of
  learning algorithms.
\newblock In \emph{Advances in Neural Information Processing Systems}, pages
  2524--2533, 2017.

\bibitem[Zhu(2020)]{zhu2020semi}
Jingge Zhu.
\newblock Semi-supervised learning: the case when unlabeled data is equally
  useful.
\newblock In \emph{Conference on Uncertainty in Artificial Intelligence}, pages
  709--718. PMLR, 2020.

\bibitem[Zhu and Goldberg(2009)]{zhu2009introduction}
Xiaojin Zhu and Andrew~B Goldberg.
\newblock Introduction to semi-supervised learning.
\newblock \emph{Synthesis Lectures on Artificial Intelligence and Machine
  Learning}, 3\penalty0 (1):\penalty0 1--130, 2009.

\bibitem[Zhu(2008)]{zhu2008semi}
Xiaojin~Jerry Zhu.
\newblock Semi-supervised learning literature survey.
\newblock Technical report, University of Wisconsin-Madison Department of
  Computer Sciences, 2008.

\end{thebibliography}
